%% file: main.tex
\documentclass[10pt]{article} %

\usepackage[preprint]{tmlr}

\usepackage{url}            %
\usepackage{tikz}
\usepackage{booktabs}       %
\usepackage{amsfonts}       %
\usepackage{amsmath}
\usepackage{mathtools}
\usepackage{amssymb}
\usepackage{graphicx}
\usepackage[ruled,vlined]{algorithm2e}
\usepackage{setspace}
\graphicspath{ {./images/} }
\usepackage{xcolor}
\usepackage[normalem]{ulem}
\usepackage{cancel}

\usepackage{tabularx}
\usepackage{array}

\usepackage{hyperref}

\usepackage[disable,textsize=tiny]{todonotes}
\usepackage{hyperref}       %
\usepackage[capitalize,noabbrev]{cleveref}

\hypersetup{pdfauthor={},pdftitle={A friendly introduction to triangular transport}}

\title{A friendly introduction to triangular transport}

\author{%
    \name Maximilian Ramgraber \email m.ramgraber@tudelft.nl \\
    \addr Department of Geoscience \& Engineering \\
    Delft University of Technology%
    \AND
    \name Daniel Sharp \email dannys4@mit.edu \\
    \addr Center for Computational Science and Engineering \\
    Massachusetts Institute of Technology%
    \AND
    \name Mathieu Le Provost\\
    \addr Department of Aeronautics \& Astronautics \\
    Massachusetts Institute of Technology%
    \AND
    \name Youssef Marzouk \email ymarz@mit.edu \\
    \addr Center for Computational Science and Engineering \\
    Massachusetts Institute of Technology%
}

\input{macros}

\def\month{09}  %
\def\year{2026} %
\def\openreview{\url{https://openreview.net/forum?id=PL0vAVlmAH}} %

\begin{document}

\maketitle

\begin{abstract}
Decision making under uncertainty is a cross-cutting challenge in science and engineering.
Most approaches to this challenge employ probabilistic representations of uncertainty.
In complicated systems accessible only via data or black-box models, however, these representations are rarely known. We discuss how to characterize and manipulate such representations using \textit{triangular transport maps}, which approximate any complex probability distribution as a transformation of a simple, well-understood distribution. %
The particular structure of triangular transport guarantees many desirable mathematical and computational properties that translate well into solving practical problems. Triangular maps are actively used for density estimation, (conditional) generative modelling, Bayesian inference, data assimilation, optimal experimental design, and related tasks. %
While there is ample literature on the development and theory of triangular transport methods, this manuscript provides a detailed introduction for scientists interested in employing measure transport without assuming a formal mathematical background. We build intuition for the key foundations of triangular transport, discuss many aspects of its practical implementation, and outline the frontiers of this field.
\end{abstract}

\section{Motivation}\label{sec:introduction}
\input{sections/introduction}

\section{Theory}\label{sec:theory}
\input{sections/theory}

\section{Implementation}\label{sec:implementation}
\input{sections/implementation}

\section{Summary \& Outlook}\label{sec:outlook}
\input{sections/outlook}

\subsubsection*{Acknowledgements} \label{acknowledgements}

We would like to acknowledge and express our deep gratitude to Mathieu Le Provost, whose invaluable contributions and insights helped shape this work. Mathieu sadly passed away before the completion of this article, and his presence is greatly missed.

The research of MR leading to these results has received funding from the Swiss National Science Foundation under the Early PostDoc Mobility grant P2NEP2 191663 and the Dutch Research Council under the Veni grant VI.Veni.232.140. MR, MLP, and YM also acknowledge support from the Office of Naval Research Multidisciplinary University Research Initiative on Integrated Foundations of Sensing, Modeling, and Data Assimilation for Sea Ice Prediction under award number N00014-20-1-2595. MLP and YM acknowledge support from the National Science Foundation (award PHY-2028125). DS and YM acknowledge support from the US Department of Energy (DOE), Office of Advanced Scientific Computing Research, under grants DE-SC0021226 (FASTMath SciDAC Institute) and DE-SC0023188. DS and YM also acknowledge support from the US Department of Energy (DOE), Office of Science, Office of Advanced Scientific
Computing Research (ASCR), via the FASTMath6 SciDAC Institute via contract number DE-AC5207NA27344, under a subcontract from Lawrence Livermore National Laboratory, and via the M2dt
MMICC center under award number DE-SC0023187.

\bibliography{references,references_daniel}
\bibliographystyle{tmlr}
\clearpage

\appendix
\input{sections/supplementary1}

\end{document}

%% file: macros.tex
\newcommand{\ba}{\mathbf{a}}
\newcommand{\bb}{\mathbf{b}}

\newcommand{\x}{\mathbf{x}}
\newcommand{\y}{\mathbf{y}}
\newcommand{\z}{\mathbf{z}}

\newcommand{\0}{\mathbf{0}}
\newcommand{\R}{\mathbf{R}}
\renewcommand{\S}{\mathbf{S}}

\definecolor{orange_custom}{HTML}{FF5000}
\definecolor{green_custom}{HTML}{01A109} %
\definecolor{blue_custom}{HTML}{1988B8}
\definecolor{crimson_custom}{HTML}{980002}
\newcommand{\co}[1]{\textcolor{orange_custom}{#1}}
\newcommand{\cg}[1]{\textcolor{green_custom}{#1}}
\newcommand{\cb}[1]{\textcolor{blue_custom}{#1}}

\newcommand{\BB}[1]{\mathbf{#1}}
\newcommand{\smap}{\BB{S}}
\newcommand{\rmap}{\BB{R}}

\newcommand{\real}[1]{\mathbb{R}^{#1}}

\DeclareMathOperator*{\argmin}{arg\,min}

\newcommand{\tmlr}{}

%% file: sections/introduction.tex
\paragraph{Who is this tutorial for?} 
This manuscript is an accessible introduction to \textit{triangular transport}, a powerful and versatile method for generative modelling and Bayesian inference. In particular, triangular transport underpins effective algorithms for data assimilation, solving inverse problems, and performing simulation-based inference, with applications across myriad scientific disciplines.

This tutorial targets researchers with an interest in applied statistical methods but without a formal background in mathematics. %
Consequently, we will focus more on intuition, general concepts, and implementation, referring the reader to other relevant articles for more formal exposition and theory.

\paragraph{How does triangular transport work?} Like other measure transport methods, triangular (measure) transport is a framework to transform one probability distribution into another. This operation is highly useful, as it allows us to characterize a complex \co{target distribution} $\co{\pi}$ by transforming a simpler, known \cg{reference distribution} $\cg{\eta}$. Throughout this manuscript, we use \co{orange} to denote variables associated with the (problem-specific) \co{target} distribution and \cg{green} to denote variables associated with the (user-defined) \cg{reference} distribution, e.g., a standard Gaussian. The idea of coupling two distributions is used in a wide range of applications. In \textit{generative modelling}, for example, we are usually interested in creating samples of a target distribution $\co{\pi}$, such as the distribution of $400 \times 400$ pixel images of cats. Measure transport methods approach this challenge by first learning a transport map 
$\mathbf{S}$ that transforms $\cg{\eta}$ to $\co{\pi}$, %
and then use %
$\mathbf{S}$ to convert reference samples $\cg{\z}\sim\cg{\eta}$ (here: $400 \times 400$ pixel \cg{images of Gaussian white noise}) into samples from the target distribution $\co{\x} = \mathbf{S}(\cg{\z}) \sim\co{\pi}$ (here: $400 \times 400$ pixel \co{images of cats}).

\begin{figure}[t!]
  \centering
  \includegraphics[width=\textwidth]{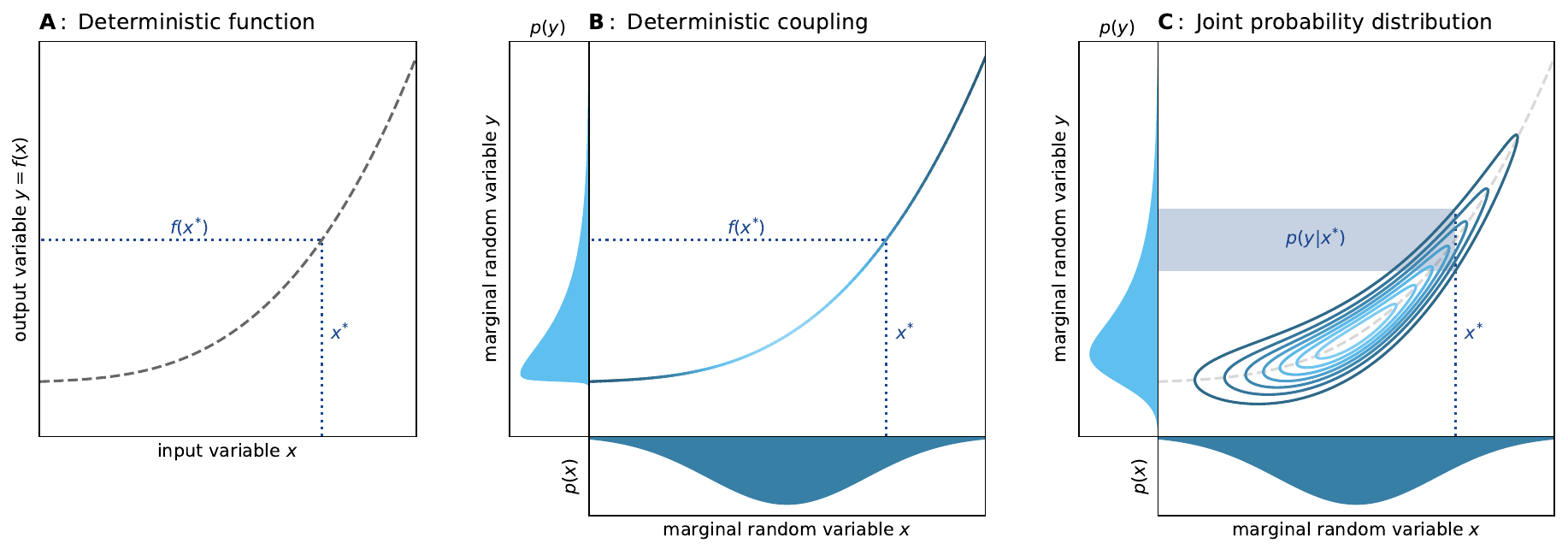}
  \caption{Progression from a fully deterministic to a fully stochastic system. (A) Numerical models are usually represented as deterministic functions. (B) In the presence of input uncertainty, deterministic functions encode a deterministic coupling which yields uncertain output (see \cref{subsec:change_of_variables}). (C) If the function itself is uncertain, this coupling ``blurs'' into a joint probability distribution. Function evaluation now corresponds to characterizing a conditional distribution. Mind that (A) and (B) can also be parsed as degenerate joint probability distributions.}
  \label{fig:deterministic_vs_stochastic}
\end{figure}

Some of these methods---among them triangular transport---can also characterize \textit{conditionals} $\co{\pi}(\co{\boldsymbol{a}}|\cb{\boldsymbol{b}^{*}})$ of the \textit{joint} target distribution $\co{\pi}(\co{\boldsymbol{a}},\co{\boldsymbol{b}})$ of two random variables $\co{\boldsymbol{a}}$ and $\co{\boldsymbol{b}}$. Here \cb{blue} denotes the fact that $\cb{\boldsymbol{b}}^{*}$ is a deterministic, fixed value. 
Conditioning operations often arise as stochastic generalizations of evaluating deterministic processes (see \cref{fig:deterministic_vs_stochastic}). As we will describe in Section~\ref{sec:theory}, conditioning is also central to the Bayesian approach to statistical inference, which is an important tool across many scientific disciplines. We distinguish here between generating from $\pi(\co{\boldsymbol{a}} | \cb{\boldsymbol{b}^{*}})$ (``generate an image of a \cb{grey} \co{cat}'') and $\pi(\co{\boldsymbol{a}},\co{\boldsymbol{b}})$ (``generate a \co{color} and \co{a cat of that color}'') by assuming that $\cb{\boldsymbol{b}^{*}}$ is determined outside of our control, either by user or application.

\paragraph{What makes triangular transport special?} At the heart of triangular transport is their eponymous \textit{triangular} structure.%
\footnote{\tmlr{A triangular map is a vector-valued function whose $k$-th output depends only on the first $k$ input variables. The name derives from the fact that the Jacobian of such a map is lower triangular.}} 
This structure sets them apart from other measure transport methods such as normalizing flows~\citep[e.g., ][]{rezende2015variational,Kobyzev2021NormalizingMethods}, %
which compose together many simpler but somewhat ad hoc transformations, often interleaved with permutations, and even from conditional normalizing flows \citep[e.g., ][]{VanDenOord2016ConditionalDecoders}, which parameterize normalizing flows in order to represent block triangular (rather than strictly triangular) maps. 
Triangular structure---discussed in greater detail in subsequent sections---has many important practical properties:
\begin{itemize}
    \item \textbf{Parsimony}: The parameterization of the map function is at the user's discretion (see \cref{subsec:map_components}). This means we can adjust the map's overall complexity, down to the complexity with which it resolves individual variables and variable dependencies. This allows us to implement nonlinear maps that are as complex as necessary, and yet as simple as possible (see \cref{subsec:adaptation}).
    \item \textbf{Sparsity}: Triangular maps have a natural ability to exploit \textit{conditional independence}. This improves their computational efficiency, which enables these maps to scale to high-dimensional settings. Further, such structure makes them highly robust to spurious correlations and smaller sample sizes (see \cref{subsubsec:conditional_independence}).
    \item \textbf{Numerical convenience}: Constructing triangular maps boils down to parameterizing simple one-dimensional monotone functions, a task with a rich body of supporting literature. Because of this, these maps are easy to optimize and invert, which we investigate in detail.
    \item \textbf{Explainability}: Triangular maps have a clear correspondence between their constituent elements and the statistical features they represent (see \cref{sec:triangular_maps}). We can then readily describe different factorizations of the target distribution using the elements of a triangular map, with a particular focus on combinations of various conditional and marginals of the target. %
\end{itemize}

\paragraph{In what applications has triangular transport been successful?} Triangular transport has been applied to a wide range of statistical problems in many different disciplines. Examples of such applications include: %
\begin{itemize}
    \item \textbf{Bayesian inference}: Triangular transport lends itself exceptionally well to the sampling of conditional distributions. As such, it has found application in both variational~\citep{ElMoselhy2012BayesianMaps} and simulation-based 
    inference~\citep{Marzouk2017SamplingIntroductionPUB,rubio2023transport,baptista2024bayesian}, for large-scale inverse problems~\citep{brennan2020greedy} and in applications with multiscale structure~\citep{parno2016multiscale}.
    
    \item \textbf{Data assimilation}: Triangular transport provides true nonlinear generalizations of popular filtering~\citep{Spantini2022CouplingFiltering} and smoothing~\citep{Ramgraber2023EnsembleFramework,Ramgraber2023EnsembleUpdates} algorithms such as the ensemble Kalman filter and smoother, and their many variants~\citep{grange2024DistributedFiltering}.

    \item \textbf{Density estimation and generative modelling}: The coupling learned by triangular transport is highly useful for the estimation~\citep{wang2022minimax,Martinez-Sanchez2024DecomposingComponents,Lopez-Marrero2024DensitySciences} and sampling~\citep{pmlr-v151-irons22a} of non-Gaussian probability distributions, even in high dimensions~\tmlr{\citep{katzfuss2024scalable,bryutkinNeuralTriangularTransport2025}}.
    
    \item \textbf{Optimal experimental design}: Due to the close connections between conditional densities and expected information gain or mutual information, triangular transport maps are useful for estimating common objectives in Bayesian optimal experimental design~\tmlr{\citep{Huan2024OptimalComputations,kovalTractableOptimalExperimental2024,li2026JUQ}}. %

\end{itemize}

Further applications of triangular transport include methods for joint state-parameter inference in state-space models~\citep{Spantini2018InferenceCouplings,grashorn2024transport,zhao2024tensor}, solving Fokker--Planck equations \citep{zeng2023bounded}, stochastic programming \citep{backhoff2017causal}, and even the discovery of causal models from data~\citep{akbari2023learningcausalgraphsmonotone,xi2023triangular}.

In many of these applications, there are a variety of reasons 
one might prefer to
use triangular transport over other transport methods, e.g., (continuous) normalizing flows, diffusion models, or variational auto-encoders. Some of these reasons include the interpretability of the resulting triangular map, which closely aligns with the structure of graphical~\citep{baptista2024learning} or causal models~\citep{xi2023triangular}; 
the inherent parallelizability of training given a set of samples and computational tractability of the map derivatives;%
\footnote{In particular, the determinant of the transport map's Jacobian---essential for calculating the pdf induced by the transport map---is significantly easier to calculate than in, e.g., continuous normalizing flows.}
the ability to both sample the distribution and evaluate its (normalized) pdf; and the ability to evaluate and sample from conditional distributions~\citep{Marzouk2017SamplingIntroductionPUB}. Other reasons will be discussed throughout this work. We also note that triangular transport has been benchmarked against comparable methods in applications by many referenced works, e.g.,~\citet{Baptista2023OnMaps,bryutkinNeuralTriangularTransport2025,katzfuss2024scalable}.

\paragraph{How is this tutorial structured?} In the following, we will provide an intuition-focused introduction to the theoretical basics of triangular transport (\cref{sec:theory}), discuss practical aspects related to their implementation in code (\cref{sec:implementation}), and conclude with a brief overview of interesting research directions (\cref{sec:outlook}). For researchers interested chiefly in practical implementation, a flow chart of the most important steps in the construction and application of triangular transport is provided in \cref{fig:flowchart}. First, we define the target $\co{\pi}$ and reference $\cg{\eta}$ (\cref{sec:triangular_maps}). Then, we structure, parameterize (\cref{subsec:map_components}), and optimize (\cref{subsec:optimization}) the triangular map. Finally, we can deploy the map in the application of our choice, with some practical heuristics listed in \cref{sec:practice}.
The tutorial will involve several recurring variables, which are summarized in \cref{tab:nomenclature}. 
\begin{figure}
    \centering
    \resizebox{\linewidth}{!}{\input{figures/flowchart}}
    \caption{Flowchart of the main steps required to define, build, optimize, and apply triangular maps, with links to the relevant sections.}
    \label{fig:flowchart}
\end{figure}
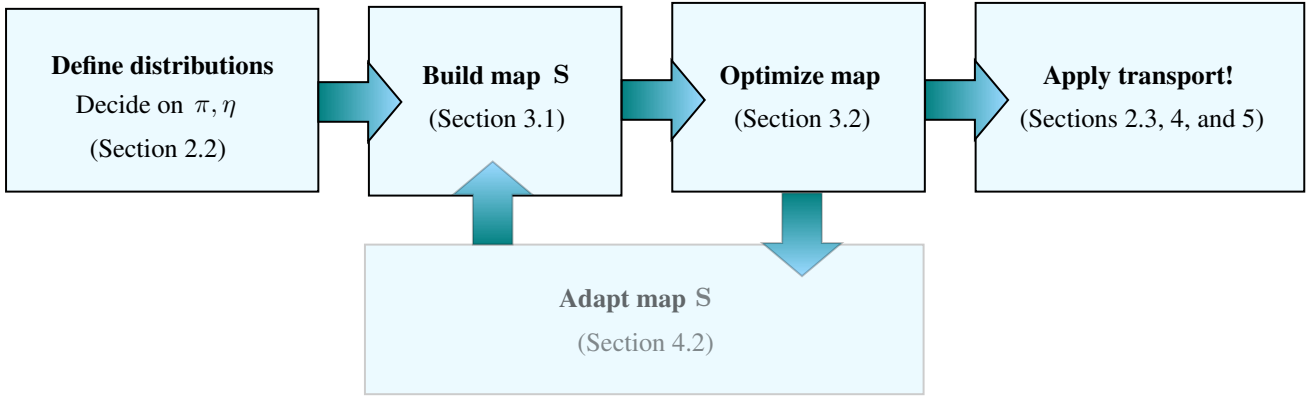
\begin{table}[!ht]
    \caption{Recurring notation and variables.}
    \begin{center}
    \begin{tabularx}{\linewidth}{@{}r >{\raggedright\arraybackslash}X r >{\raggedright\arraybackslash}X@{}} %
          \textbf{bold font} & vector-valued variable or function  & \text{Roman font} & scalar-valued variable or function \\  
     $\smap$ & Target-to-reference map & $S_{k}$ & $k$-th map component function \\
     $\mathbf{R}$ &  Reference-to-target map (see \cref{subsubsec:maps_from_densities}) & $\tau$ & twice Archimedes' constant, i.e., $6.283185...$\\%$\mathbf{T}$ & composite map \\
     $\co{\pi}$ & target distribution of interest & $\cg{\eta}$ & reference distribution, often standard Gaussian \\
     $\co{\text{orange}}$ & variable associated with $\co{\pi}$ & $\cg{\text{green}}$ & variable associated with $\cg{\eta}$ \\
     $\co{\x} \sim \co{\pi}$ & target random variable & $\cg{\z} \sim \cg{\eta}$ & reference random variable \\
     $\co{\boldsymbol{\mathsf{X}}^i}$ & $i$-th realization of $\co{\x}\sim\co{\pi}$ & $\cg{\boldsymbol{\mathsf{Z}}^i}$ & $i$-th realization of $\cg{\z}\sim\cg{\eta}$ \\
     $\co{S^{\sharp}\eta}$ & pullback distribution & $\cg{S_{\sharp}\pi}$ & pushforward distribution \\
     $K$ & number of target dimensions & $N$ & ensemble size \\
     $p$ & general probability density function (pdf) & $\mathbf{a},\mathbf{b}$ & general random variables \\
     $\cb{\y^{*}}$ & conditioning variable & $\co{\x^{*}}$ & conditioned variable $\co{\x^{*}}\sim p(\co{\x}|\cb{\y^{*}})$ \\
     $c$ & basis function coefficient & $r$ & rectifier ($r: \mathbb{R} \to\mathbb{R}^{+}$) \\
     $f$ & monotone function & $g$ & nonmonotone function \\ 
    \end{tabularx}%
    \end{center}
    \label{tab:nomenclature}
\end{table}

%% file: figures/flowchart.tex
\tikzset {_brnzclcwp/.code = {\pgfsetadditionalshadetransform{ \pgftransformshift{\pgfpoint{0 bp } { 0 bp }  }  \pgftransformrotate{-180 }  \pgftransformscale{2 }  }}}
\pgfdeclarehorizontalshading{_qczoichd6}{150bp}{rgb(0bp)=(0.6,0.85,1);
rgb(37.5bp)=(0.6,0.85,1);
rgb(62.5bp)=(0,0.5,0.5);
rgb(100bp)=(0,0.5,0.5)}

\tikzset {_gcs4bff9p/.code = {\pgfsetadditionalshadetransform{ \pgftransformshift{\pgfpoint{0 bp } { 0 bp }  }  \pgftransformrotate{-180 }  \pgftransformscale{2 }  }}}
\pgfdeclarehorizontalshading{_pp57rcskb}{150bp}{rgb(0bp)=(0.6,0.85,1);
rgb(37.5bp)=(0.6,0.85,1);
rgb(62.5bp)=(0,0.5,0.5);
rgb(100bp)=(0,0.5,0.5)}

\tikzset {_ehigoh1wy/.code = {\pgfsetadditionalshadetransform{ \pgftransformshift{\pgfpoint{0 bp } { 0 bp }  }  \pgftransformrotate{-180 }  \pgftransformscale{2 }  }}}
\pgfdeclarehorizontalshading{_9qshgj8gm}{150bp}{rgb(0bp)=(0.6,0.85,1);
rgb(37.5bp)=(0.6,0.85,1);
rgb(62.5bp)=(0,0.5,0.5);
rgb(100bp)=(0,0.5,0.5)}

\tikzset {_fvlvrwhiz/.code = {\pgfsetadditionalshadetransform{ \pgftransformshift{\pgfpoint{0 bp } { 0 bp }  }  \pgftransformrotate{-90 }  \pgftransformscale{2 }  }}}
\pgfdeclarehorizontalshading{_rm88ritm8}{150bp}{rgb(0bp)=(0.6,0.85,1);
rgb(37.5bp)=(0.6,0.85,1);
rgb(62.5bp)=(0,0.5,0.5);
rgb(100bp)=(0,0.5,0.5)}

\tikzset {_bgjf49vby/.code = {\pgfsetadditionalshadetransform{ \pgftransformshift{\pgfpoint{0 bp } { 0 bp }  }  \pgftransformrotate{-270 }  \pgftransformscale{2 }  }}}
\pgfdeclarehorizontalshading{_ezjtyhw1y}{150bp}{rgb(0bp)=(0.6,0.85,1);
rgb(37.5bp)=(0.6,0.85,1);
rgb(62.5bp)=(0,0.5,0.5);
rgb(100bp)=(0,0.5,0.5)}
\tikzset{every picture/.style={line width=0.75pt}} %

\begin{tikzpicture}[x=0.75pt,y=0.75pt,yscale=-1,xscale=1]

\draw  [fill={rgb, 255:red, 236; green, 250; blue, 255 }  ,fill opacity=1 ] (1.5,93) -- (159,93) -- (159,185.25) -- (1.5,185.25) -- cycle ;
\draw  [fill={rgb, 255:red, 236; green, 250; blue, 255 }  ,fill opacity=1 ] (184.5,92) -- (311.88,92) -- (311.88,187.25) -- (184.5,187.25) -- cycle ;
\path  [shading=_qczoichd6,_brnzclcwp] (159,130) -- (184.13,130) -- (184.13,120) -- (200.88,140) -- (184.13,160) -- (184.13,150) -- (159,150) -- cycle ; %
 \draw   (159,130) -- (184.13,130) -- (184.13,120) -- (200.88,140) -- (184.13,160) -- (184.13,150) -- (159,150) -- cycle ; %

\draw  [fill={rgb, 255:red, 236; green, 250; blue, 255 }  ,fill opacity=1 ] (337.5,90) -- (464.88,90) -- (464.88,185.25) -- (337.5,185.25) -- cycle ;
\path  [shading=_pp57rcskb,_gcs4bff9p] (312,129) -- (337.13,129) -- (337.13,119) -- (353.88,139) -- (337.13,159) -- (337.13,149) -- (312,149) -- cycle ; %
 \draw   (312,129) -- (337.13,129) -- (337.13,119) -- (353.88,139) -- (337.13,159) -- (337.13,149) -- (312,149) -- cycle ; %

\draw  [fill={rgb, 255:red, 236; green, 250; blue, 255 }  ,fill opacity=1 ] (490.5,90) -- (656.25,90) -- (656.25,185.25) -- (490.5,185.25) -- cycle ;
\path  [shading=_9qshgj8gm,_ehigoh1wy] (465,129) -- (490.13,129) -- (490.13,119) -- (506.88,139) -- (490.13,159) -- (490.13,149) -- (465,149) -- cycle ; %
 \draw   (465,129) -- (490.13,129) -- (490.13,119) -- (506.88,139) -- (490.13,159) -- (490.13,149) -- (465,149) -- cycle ; %

\draw  [color={rgb, 255:red, 0; green, 0; blue, 0 }  ,draw opacity=0.2 ][fill={rgb, 255:red, 236; green, 250; blue, 255 }  ,fill opacity=1 ] (182.5,212) -- (464.25,212) -- (464.25,287.5) -- (182.5,287.5) -- cycle ;
\path  [shading=_rm88ritm8,_fvlvrwhiz] (236.94,211.94) -- (236.94,186.81) -- (226.94,186.81) -- (246.94,170.06) -- (266.94,186.81) -- (256.94,186.81) -- (256.94,211.94) -- cycle ; %
 \draw  [color={rgb, 255:red, 0; green, 0; blue, 0 }  ,draw opacity=0.25 ] (236.94,211.94) -- (236.94,186.81) -- (226.94,186.81) -- (246.94,170.06) -- (266.94,186.81) -- (256.94,186.81) -- (256.94,211.94) -- cycle ; %

\path  [shading=_ezjtyhw1y,_bgjf49vby] (392.94,186.06) -- (392.94,211.19) -- (382.94,211.19) -- (402.94,227.94) -- (422.94,211.19) -- (412.94,211.19) -- (412.94,186.06) -- cycle ; %
 \draw  [color={rgb, 255:red, 0; green, 0; blue, 0 }  ,draw opacity=0.25 ] (392.94,186.06) -- (392.94,211.19) -- (382.94,211.19) -- (402.94,227.94) -- (422.94,211.19) -- (412.94,211.19) -- (412.94,186.06) -- cycle ; %

\draw (15,115) node [anchor=north west][inner sep=0.75pt]   [align=center] {\textbf{Define distributions }};
\draw (30,135) node [anchor=north west][inner sep=0.75pt]   [align=center] {\begin{minipage}[lt]{48.68pt}\setlength\topsep{0pt}
\begin{center}
Decide on
\end{center}

\end{minipage}};
\draw (95,138) node [anchor=north west][inner sep=0.75pt]    {$\pi ,\eta $};
\draw (205,120) node [anchor=north west][inner sep=0.75pt]   [align=left] {\textbf{Build map} $\mathbf{{\displaystyle S}}$};
\draw (207,142) node [anchor=north west][inner sep=0.75pt]   [align=left] {\begin{minipage}[lt]{60.58pt}\setlength\topsep{0pt}
\begin{center}
(Section~\ref{subsec:map_components})
\end{center}

\end{minipage}};
\draw (357,120) node [anchor=north west][inner sep=0.75pt]   [align=left] {\textbf{Optimize map}};
\draw (361,142) node [anchor=north west][inner sep=0.75pt]   [align=left] {\begin{minipage}[lt]{60.58pt}\setlength\topsep{0pt}
\begin{center}
(Section~\ref{subsec:optimization})
\end{center}

\end{minipage}};
\draw (515,120) node [anchor=north west][inner sep=0.75pt]   [align=left] {\textbf{Apply transport!}};
\draw (506,142) node [anchor=north west][inner sep=0.75pt]   [align=left] {\begin{minipage}[lt]{100pt}\setlength\topsep{0pt}
\begin{center}
(Sections~\ref{sec:triangular_maps}, \ref{sec:practice}, and \ref{sec:outlook})
\end{center}

\end{minipage}};
\draw (36,157) node [anchor=north west][inner sep=0.75pt]   [align=left] {\begin{minipage}[lt]{60.58pt}\setlength\topsep{0pt}
\begin{center}
(Section~\ref{subsec:change_of_variables})
\end{center}

\end{minipage}};
\draw (279,233) node [anchor=north west][inner sep=0.75pt]  [color={rgb, 255:red, 0; green, 0; blue, 0 }  ,opacity=0.59 ] [align=left] {\textbf{Adapt map} $\mathbf{{\displaystyle S}}$};
\draw (282,255) node [anchor=north west][inner sep=0.75pt]  [color={rgb, 255:red, 0; green, 0; blue, 0 }  ,opacity=0.46 ] [align=left] {\begin{minipage}[lt]{60.58pt}\setlength\topsep{0pt}
\begin{center}
(Section~\ref{subsec:adaptation})
\end{center}

\end{minipage}};

\end{tikzpicture}

%% file: sections/theory.tex
\subsection{Bayesian inference}\label{subsec:Bayesian_inference}

To begin, let us briefly revisit some basic concepts of Bayesian inference which will serve to motivate the operations explored in the following sections. In short, Bayesian inference is based on \textit{Bayes' theorem}: given two random variables (RVs) $\mathbf{a}$, $\mathbf{b}$ with joint probability density function (pdf) $p(\mathbf{a},\mathbf{b})$, we see:

\begin{equation}
    p\left(\mathbf{a}|\mathbf{b}^{*}\right) = \frac{p\left(\mathbf{a}\right)p\left(\mathbf{b}^{*}|\mathbf{a}\right)}{p\left(\mathbf{b}^{*}\right)},
\label{eq:Bayes_theorem}
\end{equation}
\cref{eq:Bayes_theorem} %
subsumes three sequential operations \citep[see \cref{fig:Bayes_theorem}; e.g., ][]{Gelman2013BayesianAnalysis}:

\begin{figure}
  \centering
  \includegraphics[width=\textwidth]{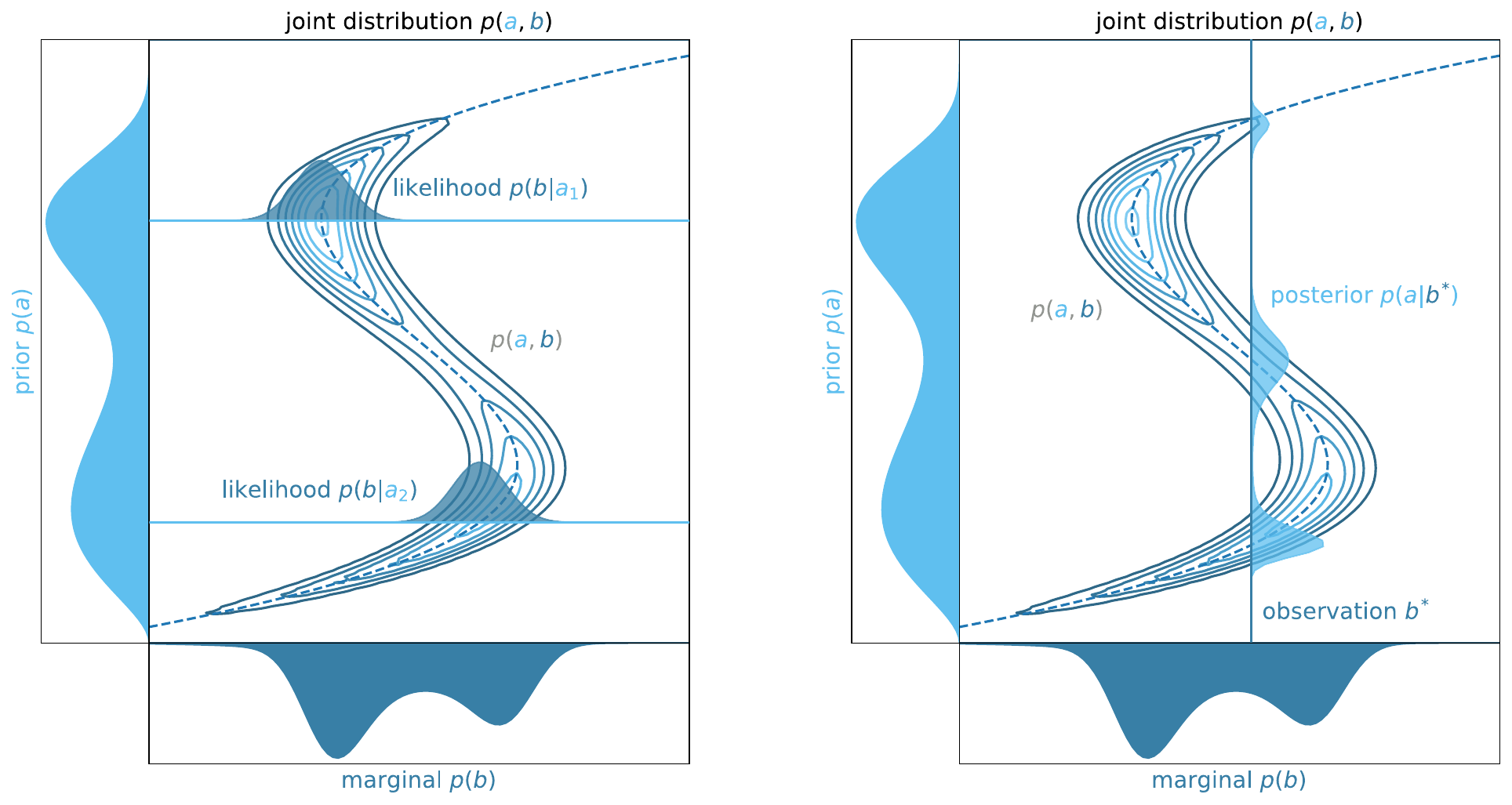}
  \caption{Schematic illustration of Bayes theorem, using a Beta mixture prior $p(a)$ and an observation model $p(b|a)=\mathcal{N}\left(\mu=2a^3 - a,\sigma=0.075\right)$. Left: We can create a joint distribution $p(a,b)$ from a prior $p(a)$ and the observation model $p(b|a)$. Right: conditioning this joint distribution on a specific value $b^{*}$ retrieves a posterior $p(a|b^{*})$.}
  \label{fig:Bayes_theorem}
\end{figure}

\begin{enumerate}
    \item First, the marginal prior $p\left(\mathbf{a}\right)$ is combined with a conditional observation model (sometimes also called \textit{likelihood model}) $p\left(\mathbf{b}|\mathbf{a}\right)$, yielding a joint probability distribution $p\left(\mathbf{a},\mathbf{b}\right)=p\left(\mathbf{a}\right)p\left(\mathbf{b}|\mathbf{a}\right)$ over all possible combinations of $\mathbf{a}$ and $\mathbf{b}$. This joint distribution describes how the variable of interest $\mathbf{a}$ and the predicted observations $\mathbf{b}$ relate to each other.
    \item Next, this joint distribution is conditioned on a specific observation $\mathbf{b}^{*}$. In practice, this means evaluating $p\left(\mathbf{a},\mathbf{b}\right)$ for all possible values $\mathbf{a}$ while keeping $\mathbf{b}$ fixed at the value of $\mathbf{b}^{*}$. This extracts a slice $p\left(\mathbf{a},\mathbf{b}^{*}\right)$ of this joint distribution at $\mathbf{b}^{*}$ along different values of $\mathbf{a}$.
    \item Since the probability densities along this slice do not generally integrate to $1$, this slice does not constitute a valid pdf. The final step thus normalizes the probability densities against the slice's probability mass $p\left(\mathbf{b}^{*}\right) = \int p\left(\mathbf{t},\mathbf{b}^{*}\right) d\mathbf{t}$, yielding the posterior pdf $p\left(\mathbf{a}|\mathbf{b}^{*}\right)$.
\end{enumerate}

In summary, Bayes' theorem first constructs a joint distribution $p\left(\mathbf{a},\mathbf{b}\right)$ from a prior $p(\mathbf{a})$ and an observation model $p(\mathbf{b}|\mathbf{a})$, then conditions it on a specific observation value $\mathbf{b}^{*}$ and re-normalizes. In consequence, one could reformulate \cref{eq:Bayes_theorem} equivalently as:
\begin{equation}
    p\left(\mathbf{a}|\mathbf{b}^{*}\right) = \frac{p\left(\mathbf{a},\mathbf{b}^{*}\right)}{\int p\left(\mathbf{t},\mathbf{b}^{*}\right) d \mathbf{t}},
\label{eq:Bayes_theorem_better}
\end{equation}
where $p\left(\mathbf{a},\mathbf{b}^{*}\right)$ evaluates the joint pdf $p\left(\mathbf{a},\mathbf{b}\right)$ for any desired $\mathbf{a}$ while keeping $\mathbf{b}$ fixed at $\mathbf{b}^{*}$, and the denominator acts as a normalizing constant. This equation, or reformulation thereof, lie at the heart of all Bayesian inference algorithms. \tmlr{Unfortunately, it is generally intractable to evaluate the normalized posterior density $p\left(\mathbf{a}|\mathbf{b}^{*}\right)$. In some cases, it is also impossible even to evaluate the joint density. 
To overcome these challenges}, different Bayesian inference methods use different strategies. As we shall see in the following, 
triangular transport addresses this challenge by first approximating 
the joint pdf $p(\mathbf{a},\mathbf{b})$ using the concept of measure transport (\cref{subsec:connection_to_transport}). Crucially, the structure of this transformation then allows us to evaluate any of its conditionals $p\left(\mathbf{a}|\mathbf{b}^{*}\right)$ (\cref{subsubsec:conditioning}).

The remainder of this tutorial drops this generic notation for two RVs $\ba$ and $\bb$ jointly distributed as $p(\ba,\bb)$ in favour of a single RV $\co{\x}$ distributed according to a distribution $\co{\pi}$. You can think of $\co{\x}$ as an augmented RV $\co{\x}=[\bb,\ba]$, and of the joint density as $\co{\pi}(\co{\x})=p(\ba,\bb)$. Transport methods generally operate on this joint distribution $p(\ba,\bb)$; we focus primarily on methods that will also allow us to characterize the desired conditionals $p(\ba | \bb)$. 

\subsection{The change-of-variables formula}\label{subsec:change_of_variables}

The key to understand transport methods is the \textit{change-of-variables formula}. This formula allows us to relate a RV $\co{\x} \in \mathbb{R}^K$ associated with a complicated target pdf $\co{\pi}$, known only to proportionality or through samples, to a second RV $\cg{\z} \in \mathbb{R}^K$, associated with a much simpler, user-specified reference pdf $\cg{\eta}$ through an invertible, differentiable transformation $\smap$. In essence, the change-of-variables formula describes what happens to pdfs when they are subjected to specific transformations. For scalar-valued RVs $\co{x}$ and $\cg{z}$, the change-of-variables formula is defined as:

\begin{equation}
\begin{aligned}
\co{\pi}(\co{x}) &= \co{S^{{\sharp}}\eta}(\co{x}) = \cg{\eta}(S(\co{x})) \left| \frac{\partial S(\co{x})}{\partial \co{x}} \right|, \\
\cg{\eta}(\cg{z}) &= \cg{S_{{\sharp}}\pi}(\cg{z}) = \co{\pi}(S^{-1}(\cg{z})) \left| \frac{\partial S^{-1}(\cg{z})}{\partial \cg{z}} \right|, \\
\end{aligned}
\label{eq:change_of_variables}
\end{equation}

where $\cg{z}=S(\co{x})$ and $\co{x}=S^{-1}(\cg{z})$, and the \textit{pullback density}%
\footnote{The \textit{pullback} density $\co{S^{{\sharp}}\eta}(\co{x})$ is the resulting pdf when applying of the \textit{inverse} map $S^{-1}$ to a RV $\cg{z}\sim\cg{\eta}$.}
$\co{S^{{\sharp}}\eta}(\co{x})$ obtains $\co{\pi}$ by applying the inverse map $S^{-1}$ to the reference distribution $\cg{\eta}$~\tmlr{\citep{bogachevMeasureTheory2007}}. %
The alternate form in the second line of \cref{eq:change_of_variables} reflects the fact that since $S$ is invertible, each distribution $\co{\pi}$ and $\cg{\eta}$ can be expressed in terms of the other through either the (forward) map $S$ or its inverse $S^{-1}$. Consequently, the \textit{pushforward pdf}%
\footnote{The \textit{pushforward} density $\cg{S_{{\sharp}}\pi}(\cg{z})$ is the resulting pdf when applying the \textit{forward} map $S$ to a RV $\co{x}\sim\co{\pi}$.}
$\cg{S_{{\sharp}}\pi}(\cg{z})$ likewise approximates $\cg{\eta}$ by applying the forward map $S$ to the target $\co{\pi}$.%
\footnote{A mnemonic bridge to remember the $\sharp$ notation is that the pull\textit{back} relies on the \textit{inverse} map $\smap^{-1}$ to sample, and has the $\sharp$ in the superscript where the $-1$ would be.}
For multivariate RVs $\co{\x}$ and $\cg{\z}$, the same principle applies.
Given a multivariate monotone function $\smap$, \cref{eq:change_of_variables} generalizes to:

\begin{equation}
\begin{aligned}
    \co{\pi}(\co{\x}) &= \co{\smap^{{\sharp}}\eta}(\co{\x}) = \cg{\eta}(\smap(\co{\x})) \left| \det \co{\nabla_{\x}} \smap(\co{\x}) \right|, \\
    \cg{\eta}(\cg{\z}) &= \cg{\smap_{{\sharp}}\pi}(\cg{\z}) = \co{\pi}(\smap^{-1}(\cg{\z})) \left| \det \cg{\nabla_{\z}} \smap^{-1}(\cg{\z}) \right|.
\end{aligned}
\label{eq:change_of_variables_multidimensional}
\end{equation}

The change-of-variables formula in \cref{eq:change_of_variables} has a surprisingly intuitive interpretation. It states that the probability density $\cg{\eta}(\cg{z})$ at a point of interest $\cg{z}$ after the transformation equals the original probability density $\co{\pi}(\co{x})$ at the pre-transformation point $\co{x} = S^{-1}(\cg{z})$, adjusted for any deformation %
the transformation might have induced at this location $\left| \partial S(\co{x}) / \partial \co{x} \right|$. \cref{eq:change_of_variables_multidimensional} extends this notion to multi-dimensional systems. Intuitively, the absolute value of the determinant of the Jacobian of $\smap$, namely $\left| \det \co{\nabla_{\x}} \smap(\co{\x}) \right|$, measures the inflation/deflation of an infinitesimal volume centered about $\x$ by the map $\smap$. If $\left| \det \co{\nabla_{\x}} \smap(\co{\x}) \right| = 1$, the map $\smap$ preserves infinitesimal volumes about $\x$. The compensation term, similar to `$u$-substitution' in calculus, is necessary because transformations $\smap$ ``stretch'' or ``squeeze'' the spaces they are applied to. Accounting for this spatial distortion ensures that the probability mass is preserved and thus the transformed density $\cg{\eta}(\cg{\z})$ remains a valid pdf. 

\begin{figure}
  \centering
  \includegraphics[width=\textwidth]{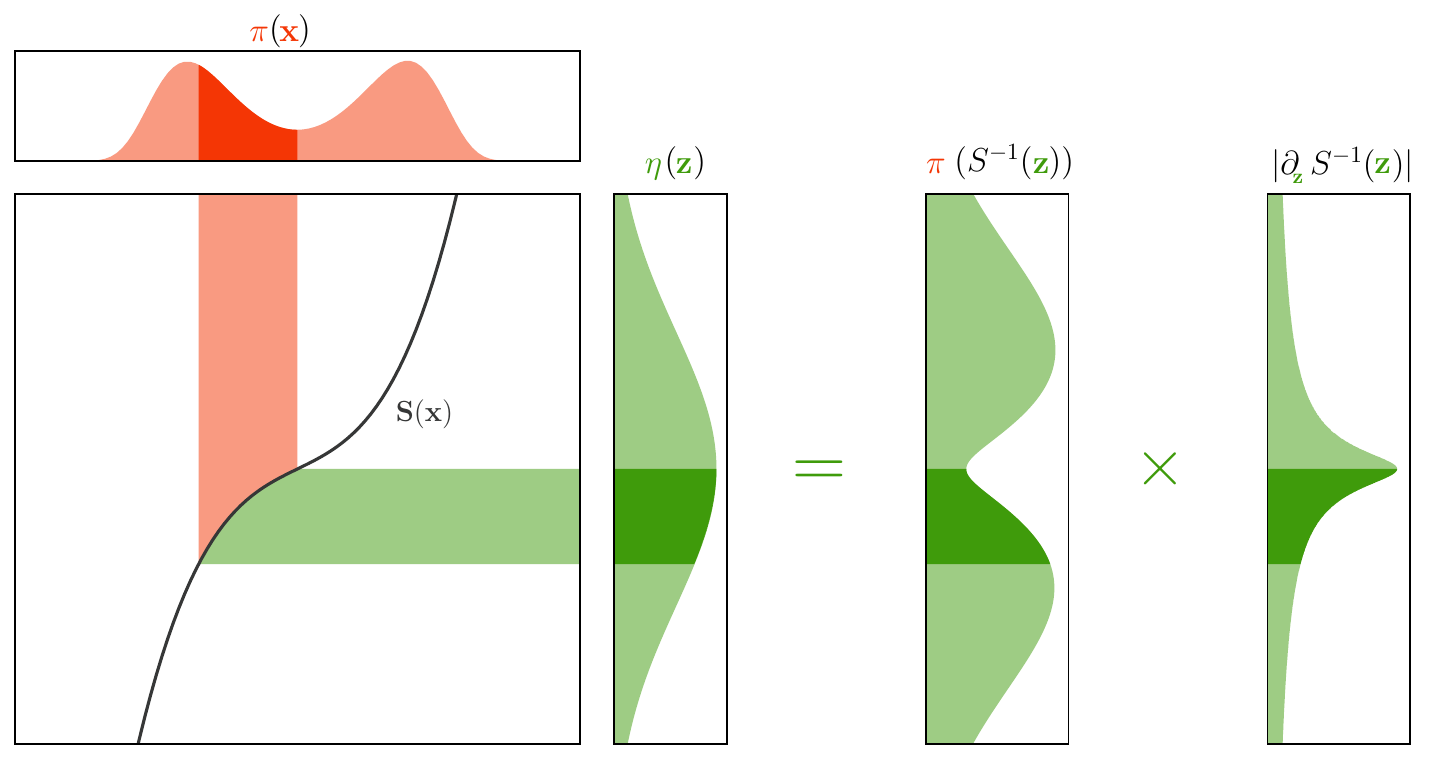}
  \caption{Illustration of the change-of-variables formula. A monotone function $S$ allows us to relate a RV $\co{x}$ associated with a pdf $\co{\pi}$ to a RV $\cg{z}$ associated with a pdf $\cg{\eta}$. We can evaluate $\cg{\eta}(\cg{z})$ by evaluating the target $\co{\pi}$ at the pre-image of $\cg{z}$, that is to say $\co{\pi}(S^{-1}(\cg{z}))$, then multiplying it with the inverse map's absolute gradient $|\partial_{\cg{z}}S^{-1}(\cg{z})|$.}
  \label{fig:change_of_variables}
\end{figure}

The change-of-variables formula allows us to describe how a probability density $\co{\pi}$ changes when subjected to a specific transformation $\smap$. An example is provided in \cref{fig:change_of_variables}, which shows how a non-Gaussian target pdf $\co{\pi}$ can be related to a Gaussian reference pdf $\cg{\eta}$ through an invertible transformation; this invertibility is equivalent to monotonicity in the scalar case.

\tmlr{Given the ability to change variables in this way, one can estimate many quantities of interest. Consider an arbitrary function $\BB{Q}:\mathbb{R}^K\to\mathbb{R}^M$. Then, one can use \textit{Monte Carlo} samples, $\cg{\boldsymbol{\mathsf{Z}}^{i}}\sim\cg{\eta}$ for $i=1,\ldots,N$, to estimate the average value of $\BB{Q}$, i.e.,
\begin{equation}
    \int \BB{Q}(\co\x)\,\mathrm{d}\co{\pi}(\co{\x}) = \int \BB{Q}(\co\x)\,\mathrm{d} \co{\smap^\sharp\eta}(\co{\x}) = \int \BB{Q}(\smap^{-1}(\cg{\z}))\,\mathrm{d}\cg{\eta}(\cg{\z}) \approx \frac{1}{N}\sum_{i=1}^{N} \BB{Q}(\smap^{-1}(\cg{\boldsymbol{\mathsf{Z}}^{i}})).
\end{equation}
Applications of this estimation scheme include, for example, the mean ($\BB{Q}(\co{\x}) = \co{\x}$), the sum of marginal second moments ($Q(\co{\x}) = \|\co{\x}\|^2$), or even the entropy ($Q(\co{\x}) = \log\co\pi(\co{\x}) = \log \co{\smap^\sharp\eta}(\co{\x})$).}

\subsubsection{Connection to transport methods}\label{subsec:connection_to_transport}

The change-of-variables formula has three knobs to tune: the original density $\co{\pi}$, the map $\smap$, and its transformed output $\cg{\eta}$. When considering the change-of-variables in elementary calculus courses, $\co{\pi}$ and $\smap$ are often assumed known, and we seek the transformed density $\cg{\eta}$. Transport methods choose a slightly different approach. We assume $\co{\pi}$ and $\cg{\eta}$ to be known (at least partially), and instead seek a map $\smap$ which relates the two distributions to each other. 

\begin{figure}
  \centering
  \includegraphics[width=\textwidth]{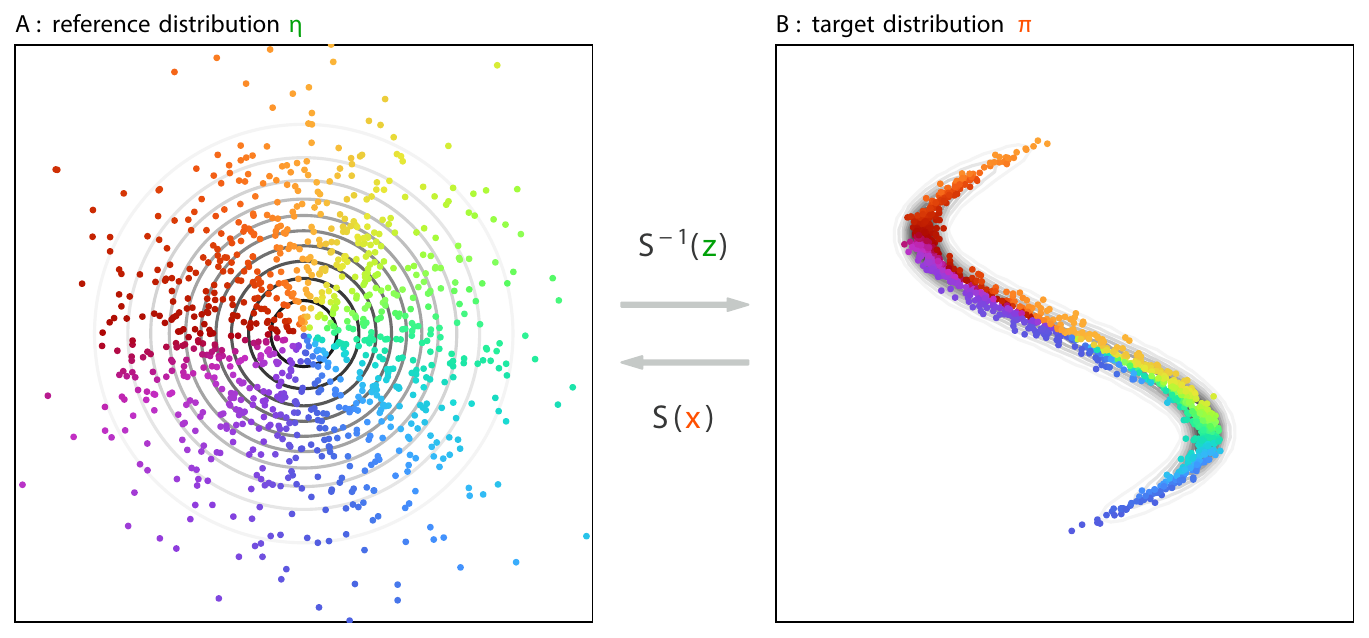}
  \caption{\tmlr{A transport map $\smap$ relates a RV $\co{\x}$ associated with the target $\co{\pi}$ to a RV $\cg{\z}$ associated with the reference $\cg{\eta}$. If the map is invertible, it can be applied both ways. This figure is adapted from \citet[Figure 3]{Ramgraber2023EnsembleFramework}.}}
  \label{fig:map_example}
\end{figure}

As discussed in \cref{sec:introduction}, we generally do not have access to the target pdf $\co{\pi}$ in closed form. 
Rather, we might only have samples\footnote{Sample approximations are common if $\co{\pi}$ is only known from a dataset
or if the prior can be sampled but not evaluated.} $\co{\boldsymbol{\mathsf{X}}}\sim\co{\pi}$, or we might only be able to evaluate $\co{\pi}$ up to a constant of proportionality (i.e., $\co{\tilde{\pi}}=\co{m}\co{\pi}$, where $\co{m} > 0$ is an unknown constant).%
\footnote{Unnormalized densities $\co{\tilde{\pi}}$ often arise in Bayesian statistics, when the prior and likelihood can be evaluated at least point-wise, but it is infeasible to quantify the model evidence (see \cref{eq:Bayes_theorem}).}
On the other hand, the reference %
$\cg{\eta}$ is chosen to be a simple, well-known distribution, often a standard Gaussian $\cg{\mathcal{N}}\left(\mathbf{0},\boldsymbol{I}\right)$ (\cref{fig:map_example}). 
The map $\smap$ is identified by minimizing an objective function over a specified class of functions; see the discussion in \cref{subsec:optimization}. By finding this map, we learn how to construct the unknown target distribution $\co{\pi}$ by transforming a well-defined reference distribution $\cg{\eta}$.

Among its other uses, learning the map $\smap$ allows us to cheaply draw new samples from the target distribution $\co{\pi}$. This is achieved by first sampling the reference $\cg{\eta}$, then applying the inverse map $\smap^{-1}$ to the resulting reference samples $\cg{\z}$. This is especially useful in applications where sampling the target conventionally would involve computationally expensive simulations of, e.g., partial differential equations (\textit{emulation}), or systems in which the sample-generating process is not known exactly \citep[\textit{generative modelling}: e.g.,][]{doi:10.1137/23M1581546}. %

\subsection{Triangular maps and their uses}\label{sec:triangular_maps}

Many classes of functions are viable choices for the transport map $\smap$. \tmlr{Outside of the constructions shown here, common examples of transport maps include normalizing flows \citep{papamakarios2021normalizing}, optimal transport maps \citep{peyre2019computational}, and the transformations induced by flow-based generative models~\citep{albergo2023stochastic,lipman2023flow}.
By contrast, we consider} \textit{triangular} transport maps. In addition to sampling the target $\co{\pi}$, triangular maps are unique in allowing us to sample \textit{conditionals} of $\co{\pi}$, which makes them a flexible and versatile tool for %
Bayesian inference.
Triangular maps are structured as follows:

\begin{equation}
\smap(\co{\x})=\begin{bmatrix*}[l]
    S_{1}(\co{x_{1}}) \\
    S_{2}(\co{x_{1}},\co{x_{2}}) \\
    S_{3}(\co{x_{1}},\co{x_{2}},\co{x_{3}}) \\
    \vdots \\
    S_{K}(\co{x_{1}},\dots,\co{x_{K}})
\end{bmatrix*}=\begin{bmatrix*}[l]
    \cg{z_{1}} \\
    \cg{z_{2}} \\
    \cg{z_{3}} \\
    \vdots \\
    \cg{z_{K}}
\end{bmatrix*} = \cg{\z},
\label{eq:decomposed_S}
\end{equation}

where the full map $\smap:\real{K} \to \real{K}$ is comprised of map components $S_{k}:\real{k} \to \real{}$, $k=1,\ldots,K$, each of which depends only on the first $k$ entries of the target RV $\co{\x}=\left[\co{x_{1}},\dots,\co{x_{K}}\right]^{\top}$ and we \textit{enforce} that $\partial_{\co{x_k}}S_k(\co{x_1},\ldots,\co{x_k}) > 0$ for any feasible choice of $\co{x_1},\ldots,\co{x_k}$. When all of $S_1,\ldots,S_K$ satisfy this condition, we call the map $\smap$ ``monotone.'' The eponymous \textit{triangular} nature of $\smap$ refers to the fact that the map's partial derivatives with regards to $\co{\x}$ are lower-triangular; that is to say, the Jacobian matrix $\nabla\smap$ has all zeros above its diagonal. This structure 
has a number of highly desirable properties: 

\begin{enumerate}

    \item \tmlr{For any pair of target and reference pdfs,}%
    \footnote{\tmlr{The existence of a transport map, triangular or not, is more delicate if we do not assume that both distributions have densities. For example, if $\co\pi$ is a discrete distribution, such that $X\sim\co\pi$ takes on only a finite number of values, and $\cg\eta$ is a Gaussian distribution, there is no map achieving $\cg{\smap_\sharp \pi} = \cg{\eta}$~\citep{Villani2007OptimalNew}. The failure of existence is due to the fact that a deterministic map cannot split mass.}}
    \tmlr{$\co\pi$ and $\cg\eta$, there exists a unique function $\smap$ satisfying this structure (i.e., a monotone triangular map) that achieves $\cg{\smap_\sharp \pi} = \cg{\eta}$. %
    This map is known as the \textit{Knothe--Rosenblatt rearrangement} \citep{Knothe1957ContributionsBodies,Rosenblatt1952RemarksTransformation} or KR map.}

    \item This triangular structure allows us to evaluate the \textbf{determinant of the map's Jacobian} $\det \nabla \smap(\co{\x})$ efficiently as the product of its diagonal entries \citep{Marzouk2017SamplingIntroductionPUB}, which proves highly useful for the map's optimization (see \cref{subsec:optimization}), not to mention when performing the density estimation itself for the changed variables:

    \begin{equation}
        \det \nabla \smap(\co{\x}) = \prod_{k=1}^{K} \frac{\partial S_{k}(\co{x_{1}},\ldots,\co{x_{k}})}{\partial \co{x_k}}.
        \label{eq:triangular_determinant}
    \end{equation} 
    
    \item Triangular maps are \textbf{easily invertible}. In particular, we pick a class of functions that are monotone in their last input $\co{x_{k}}$ (with no requirements on the other inputs $\co{x_{1}},\dots,\co{x_{k-1}}$), i.e. $\partial_{\co{x_{k}}} S_{k} > 0$ everywhere; this guarantees the monotonicity and thus eases our inversion computation. We will discuss ways to guarantee this property in \cref{subsec:monotonicity}. 
    
    \item Perhaps most importantly, triangular maps naturally \textbf{factorize the target distribution} into a product of marginal conditional pdfs. We will investigate this property in greater detail in the following sections.

\end{enumerate}

\tmlr{Why does a unique triangular map exist? In the scalar-valued case $K=1$, it is well known that the cumulative distribution function (cdf) of $\co{\pi}$, defined as $F_{\co{\pi}}(\co{x}) =\int_{-\infty}^{\co{x}} \co{\pi}(t)\,\mathrm{d}t$, is the unique monotone increasing function that pushes forward $\co{\pi}$ to the uniform distribution on the unit interval, $\mathcal{U}(0,1)$~\citep{devroyeNonUniformRandomVariate1986}.}
\tmlr{Using this property, it is straightforward to couple two univariate distributions, $\co\pi$ and $\cg\eta$, by thinking of the uniform distribution as an intermediate ``bridge'' between the two: concretely, the map $S = F_{\cg{\eta}}^{-1} \circ F_{\co{\pi}}$ pushes forward $\co\pi$ to $\cg\eta$. Since $S$ is the composition of monotone functions, it is also monotone. This construction establishes existence; to guarantee uniqueness, we can notice that replacing the intermediate uniform distribution with any other distribution does not change the final map $S$. In particular, if we write the intermediate distribution as $B_\sharp \,  \mathcal{U}(0,1)$ for some invertible $B$ (which is possible for any absolutely continuous distribution), we can easily see that $S = F_{\cg{\eta}}^{-1} \circ B^{-1} \circ B \circ F_{\co{\pi}} =  F_{\cg{\eta}}^{-1} \circ F_{\co{\pi}}$.}

\tmlr{In the more general multivariate case, a similar principle is applied \textit{recursively}. As we shall see in Section~\ref{subsubsec:conditioning}, a triangular map corresponds to a factorization of a multivariate pdf $\co\pi$ into a product of univariate \textit{marginal conditional} pdfs, i.e., $\co{\pi}(\co{x_k} \vert \co{x_{1:k-1}})$ 
for any integer $k \leq K$. The corresponding marginal conditional cdfs, $F_{\co{\pi}}(\co{x_k}|\co{x_{1:k-1}}) = \int_{-\infty}^{\co{x_k}}\co{\pi}(t|\co{x_{1:k-1}})\,\mathrm{d}t$, along with the corresponding marginal conditional cdfs of the reference distribution $\cg \eta$, then uniquely yield the monotone component functions $S_k$ of the full KR map $\smap$. For more details on the construction of the KR map from marginal conditional distributions, we refer to, e.g.,~\citet{wang2022minimax,Santambrogio2015OptimalMathematicians,figalliInvitationOptimalTransport2021,zech2022sparseI}.}%

\subsubsection{Map inversion}\label{subsubsec:inversion}

In the forward map evaluation (\cref{eq:decomposed_S}), each of the map's component functions $S_{k}$ can be evaluated independently, even in parallel, and then assembled into the full reference vector $\cg{\z}$. However, the same does not hold for the inverse map:

\begin{equation}
\smap^{-1}(\cg{\mathbf{z}})=\begin{bmatrix*}[l]
    S_{1}^{-1}(\cg{z_{1}}) \\
    S_{2}^{-1}(\cg{z_{2}};\co{x_{1}}) \\
    S_{3}^{-1}(\cg{z_{3}};\co{x_{1}},\co{x_{2}}) \\
    \hfill\vdots\hfill \\
    S_{K}^{-1}(\cg{z_{K}};\co{x_{1}},\dots,\co{x_{K-1}})\end{bmatrix*}=\begin{bmatrix*}[l]
    \co{x_{1}} \\
    \co{x_{2}} \\
    \co{x_{3}} \\
    \hfill\vdots\hfill \\
    \co{x_{K}}
\end{bmatrix*}  = \co{\x}.
\label{eq:inverse_decomposed_S}
\end{equation}

Here, the inverse map's {component functions} $S_{k}^{-1}$ must be evaluated in sequence and cannot be evaluated independently. This process begins by inverting the first map component $S_{1}^{-1}(\cg{z_{1}})$, a trivial one-dimensional root-finding problem,%
\footnote{This root finding problem has a single unique solution within the domain of $S_{k}^{-1}$, since we require $S_{k}$ be monotone in $\co{x_{k}}$.}
which yields $\co{x_{1}}$. This output $\co{x_{1}}$ serves as auxiliary input for the second map component's inversion $S_{2}^{-1}(\cg{z_{2}};\co{x_{1}})$, yielding another one-dimensional root-finding problem, which provides $\co{x_{2}}$. All subsequent map component inversions are similar one-dimensional root-finding problems that likewise depend on the outcomes of previous inversions. This dependence of each inversion $S_{k}^{-1}$ on each of $\co{x_{1}},\ldots,\co{x_{k-1}}$, the outcomes of previous inversions, effectively factorizes the target distribution as a product of marginal conditionals:%

\begin{equation}
\pi\left(\co{\x}\right) = \underbrace{\pi\left(\co{x_{1}}\right)}_{S_{1}^{-1}\left(\cg{z_{1}}\right)}\underbrace{ \pi\left(\co{x_{2}}|\co{x_{1}}\right)}_{S_{2}^{-1}\left(\cg{z_{2}};\co{x_{1}}\right)}\underbrace{\pi\left(\co{x_{3}}|\co{x_{1}},\co{x_{2}}\right)}_{S_{3}^{-1}\left(\cg{z_{3}};\co{x_{1}},\co{x_{2}}\right)}\dots \underbrace{\pi\left(\co{x_{K}}|\co{x_{1}},\dots,\co{x_{K-1}}\right)}_{S_{K}^{-1}\left(\cg{z_{K}};\co{x_{1}},\dots,\co{x_{K-1}}\right)},
\label{eq:factorization}
\end{equation}
where each term corresponds to, and is in turn sampled by, one of the inverse map components indicated in the underbraces. \tmlr{We also note that the variable ordering defines the factorization, and through it, the map structure; we will explore the consequences of different orderings in more detail in \cref{subsubsec:conditional_independence}.} Due to the chain rule of probability, such a triangular factorization always exists. For a particular sample $i$, each row of \cref{eq:inverse_decomposed_S} can be used to generate a sample $\co{\mathsf{X}_k^i}$ from a particular marginal distribution conditioned on $ \co{\mathsf{X}_{1}},\dots,\co{\mathsf{X}_{k-1}}$. Given a reference sample $\cg{\boldsymbol{\mathsf{Z}}^{i}}\sim\cg{\eta}$, we get:

\begin{equation}
\co{\mathsf{X}_{k}^{i}} = S_{k}^{-1}(\cg{\mathsf{Z}_{k}^{i}};\co{\mathsf{X}_{1}^{i}},\dots,\co{\mathsf{X}_{k-1}^{i}}) \sim
\co{\pi}\left(\co{x_{k}}|\co{x_{1}},\dots,\co{x_{k-1}}\right).
\label{eq:marginal_conditional}
\end{equation}

\tmlr{The direct correspondence between $S_{k}^{-1}\left(\cg{z_{k}};\co{\x_{1:k-1}}\right)$ and $\co{\pi}\left(\co{x_{k}}|\co{\x_{1:k-1}}\right)$ in Equation~\eqref{eq:factorization} only applies if the reference distribution $\cg\eta$ has no dependence among any of its marginals~\citep{Rosenblatt1952RemarksTransformation, Spantini2022CouplingFiltering}, i.e., $\cg\eta(\cg{\z}) = \cg{\eta_1}(\cg{z_1})\cg{\eta_2}(\cg{z_2})\cdots\cg{\eta_K}(\cg{z_K})$. A standard Gaussian reference fulfills this property.
An intuitive explanation is that this tensor-product reference distribution, because it introduces no dependence among the random variables $\cg{z_1, \ldots, z_K}$, introduces no extraneous information about dependencies in the target distribution.
The absence of dependence in the reference measure forces the map $\smap$ itself to learn \textit{all} dependencies among the random target variables $\co{x_1, \ldots, x_K}$.
Conditions for this correspondence can also be deduced from the explicit construction of the KR map via cdfs in the general case where $\cg \eta$ is not a product measure; see, e.g., \citet[Sec.\ 2]{wang2022minimax}.}

\subsubsection{Sampling conditionals}\label{subsubsec:conditioning}

As it turns out, the factorization of the target distribution $\co{\pi}$ in \cref{eq:factorization,eq:marginal_conditional} also allows us to sample \textit{conditionals} of $\co{\pi}$, including the Bayesian posterior $p(\mathbf{a}|\mathbf{b})$ (assuming $p:=\co{\pi}$ and $[\mathbf{b},\mathbf{a}]:=[\co{\x_{1:k}},\co{\x_{k+1:K}}]$). This can be achieved by manipulating the inversion process. First, observe that the factorization in \cref{eq:factorization} can be aggregated into two blocks:

\begin{equation}
\pi\left(\co{\x}\right) = \underbrace{\pi\left(\co{\x_{1:k}}\right)}_{\smap_{1:k}^{-1}\left(\cg{\z_{1:k}}\right)}\underbrace{\pi\left(\co{\x_{k+1:K}}|\co{\x_{1:k}}\right)}_{\;\;\smap_{k+1:K}^{-1}\left(\cg{\z_{k+1:K}};\,\co{\x_{1:k}}\right)}\hspace{-0.8em}.
\label{eq:factorization_blocks}
\end{equation}

Similarly, we can aggregate the map component functions into two blocks:

\begin{equation}
\smap^{-1}(\cg{\mathbf{z}})=\begin{bmatrix*}[l]
    \smap_{1:k}^{-1}(\cg{\z_{1:k}}) \\[0.3em]
    \smap_{k+1:K}^{-1}(\cg{\z_{k+1:K}};\co{\x_{1:k}}) \end{bmatrix*}=\begin{bmatrix*}[l]
    \co{\x_{1:k}} \\
    \co{\x_{k+1:K}} \\
\end{bmatrix*}  = \co{\x}.
\label{eq:inverse_decomposed_S_blocks}
\end{equation}

Instead of evaluating of \cref{eq:inverse_decomposed_S} from top to bottom ($S_{1}^{-1}$ to $S_{K}^{-1}$), if we are interested in sampling conditionals, we may skip the upper map block $\smap_{1:k}^{-1}$ and replace its corresponding output $\co{\x_{1:k}}=[\co{x_1},\dots,\co{x_{k}}]$ with arbitrary, user-specified values $\cb{\x_{1:k}^{*}}=[\cb{x_1^{*}},\dots,\cb{x_{k}^{*}}]$. This results in the following truncated inversion for the lower map block $\smap_{k+1:K}^{-1}$:

\begin{equation}
\smap^{-1}_{k+1:K}(\cg{\mathbf{z}_{k+1:K}};\cb{\x_{1:k}^{*}})=\begin{bmatrix*}[l]
    S_{k+1}^{-1}(\cg{z_{k+1}};\cb{\x_{1:k}^{*}}) \\
    S_{k+2}^{-1}(\cg{z_{k+2}};\cb{\x_{1:k}^{*}},\co{x_{k+1}^{*}}) \\
    \vdots \\
    S_{K}^{-1}(\cg{z_{K}};\cb{\x_{1:k}^{*}},\co{x_{k+1}^{*}},\dots,\co{x_{K-1}^{*}})\end{bmatrix*}=\begin{bmatrix*}[l]
    \co{x_{k+1}^{*}} \\
    \co{x_{k+2}^{*}} \\
    \vdots \\
    \co{x_{K}^{*}}
\end{bmatrix*}  = \co{\x_{k+1:K}^{*}}.
\label{eq:inverse_decomposed_S_manipulated}
\end{equation}

Resuming the inversion starting with the map component inverse $S_{k+1}^{-1}$ thus yields samples $\co{\x_{k+1:K}^{*}}$ from the conditional $\co{\pi}\left(\co{\x_{k+1:K}}|\cb{\x_{1:k}^{*}}\right)$ instead of the target $\co{\pi}\left(\co{\x}\right)$. Equivalent to \cref{eq:factorization}, the corresponding conditional distribution now factorizes as:

\begin{equation}
\pi\left(\co{\x_{k+1:K}}|\cb{\x_{1:k}^{*}}\right) = \underbrace{\pi\left(\co{x_{k+1}}|\cb{\x_{1:k}^{*}}\right)}_{S_{k+1}^{-1}(\cg{z_{k+1}};\,\cb{\x_{1:k}^{*}})}\ \underbrace{\pi\left(\co{x_{k+2}}|\cb{\x_{1:k}^{*}},\co{x_{k+1}}\right)}_{S_{k+2}^{-1}(\cg{z_{k+2}};\,\cb{\x_{1:k}^{*}},\co{x_{k+1}^{*}})}\dots \underbrace{\pi\left(\co{x_{K}}|\cb{\x_{1:k}^{*}},\co{\x_{k+1:K-1}}\right)}_{S_{K}^{-1}(\cg{z_{K}};\,\cb{\x_{1:k}^{*}},\co{x_{k+1}^{*}},\dots,\co{x_{K-1}^{*}})},
\label{eq:factorization_conditioned}
\end{equation}

\begin{figure}[ht!]
  \centering
  \includegraphics[width=\textwidth]{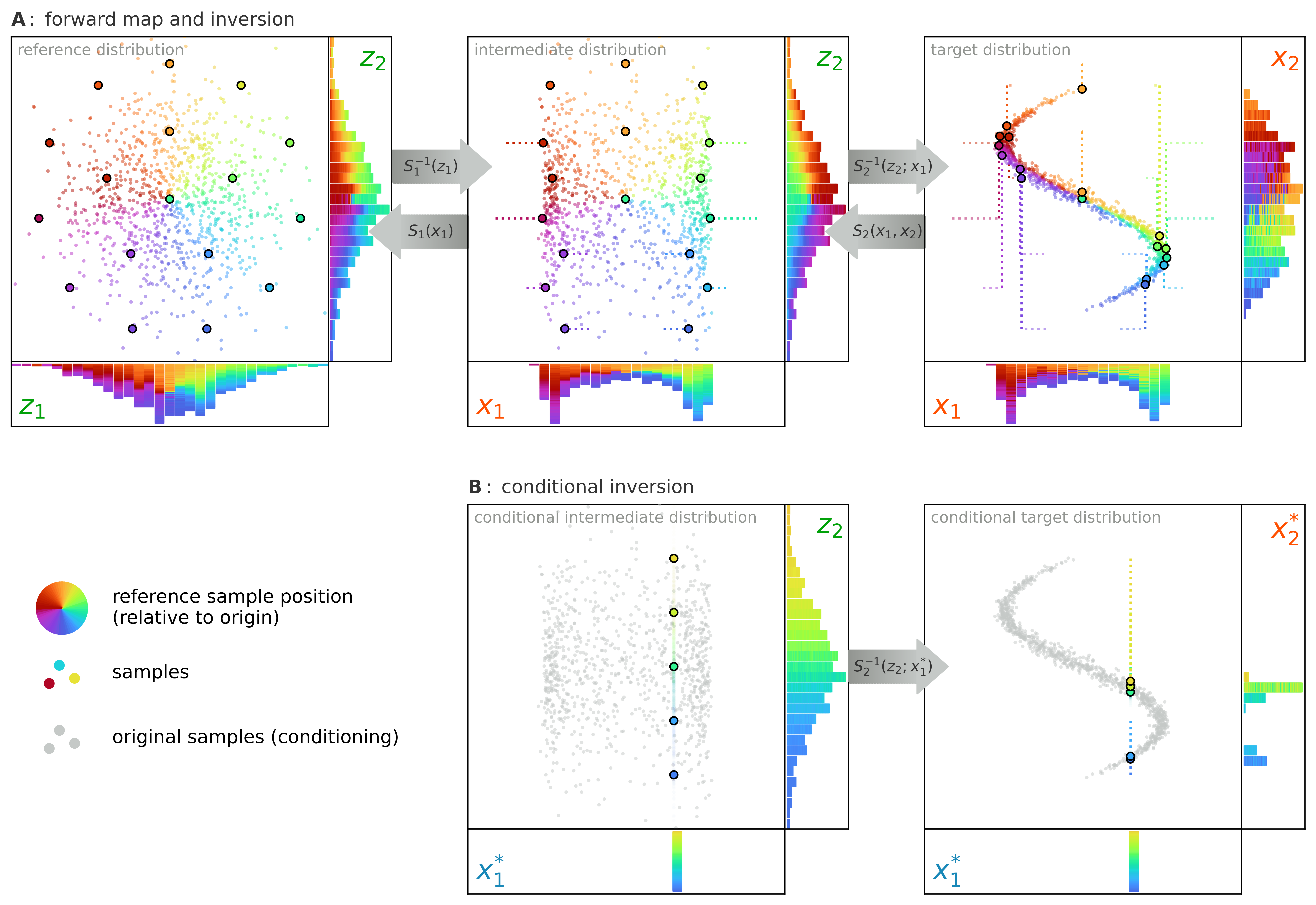}
  \caption{(A) The forward map (top row, from right) and its inverse (top row, from left) operate via implicit intermediate distributions (center), as the map components transform the distributions one marginal at a time. (B) Supplying the inverse map with a manipulated intermediate distribution (bottom row, center) as a starting point will instead sample conditionals of the target distribution. This figure is adapted from \citet[Figure 1]{Ramgraber2023EnsembleUpdates}.}
  \label{fig:conditioning}
\end{figure}

This means that the manipulated triangular map inversion in \cref{eq:inverse_decomposed_S} allows us to sample conditionals of the target distribution $\co{\pi}$ for arbitrary $\cb{\x_{1:k}^{*}}$, or to estimate the density of this conditional distribution. Recalling \cref{subsec:Bayesian_inference}, it is plain to see why this operation proves extremely useful in Bayesian inference: If we define our target pdf $\co{\pi}$ as the joint distribution $p\left(\co{\ba},\co{\bb}\right)$ (using the notation of \cref{subsec:Bayesian_inference}) between the RV of interest $\co{\ba}$ and the observation predictions $\co{\bb}$, and consider the manipulated samples $\cb{\cb{\x_{1:k}^{*}}}$ to be the observations $\cb{\bb^{*}}$ of $\co{\bb}$, then the manipulated inversion in \cref{eq:inverse_decomposed_S_manipulated} samples the posterior $p\left(\co{\ba}|\cb{\bb^{*}}\right)$. An illustration of the forward, inverse, and conditional mapping operations is provided in \cref{fig:conditioning}.

\subsubsection{Conditional independence}\label{subsubsec:conditional_independence}

A related, very useful property of triangular transport maps is that they naturally allow for the exploitation of \textit{conditional independence}
by construction.%
\footnote{By default, many statistical methods assume that all RVs directly influence each other. \textit{Conditional independence} arises whenever two RVs $\boldsymbol{a}$ and $\boldsymbol{c}$ only affect each other indirectly via a third RV $\boldsymbol{b}$. In this case, we say that $\boldsymbol{a}$ is conditionally independent of $\boldsymbol{c}$ (and vice versa) given $\boldsymbol{b}$, represented symbolically as $\boldsymbol{a} \perp \!\!\! \perp \boldsymbol{c} \; | \; \boldsymbol{b}$. 
As an example, consider $a$ as the chance of rain, $b$ as the chance of wet ground (which can, but does not have to be caused by rain), and $c$ as the chance of slipping. As rain ($a$) only affects the chance of slipping ($c$) via wet ground ($b$), we have $a \perp \!\!\! \perp c \; | \; b$.
}
Recalling the map's generic factorization of the conditionals of target $\co{\pi}$ in \cref{eq:factorization}, we might ask ourselves what happens in systems in which we can exploit conditional independence. For example, if we have a target distribution $\co{\pi}\left(\co{\x_{1:4}}\right)$ and conditional independence properties $\co{x_{3}} \perp \!\!\! \perp \co{x_{1}} | \co{x_{2}}$ and $\co{x_{4}} \perp \!\!\! \perp \co{x_{1}},\co{x_{2}} | \co{x_{3}}$ (corresponding to \textit{Markov structure}), the map's factorization could be reduced as follows:

\begin{equation}
\begin{aligned}
\co{\pi}\left(\co{\x_{1:4}}\right) &= \co{\pi}\left(\co{x_{1}}\right)\co{\pi}\left(\co{x_{2}}|\co{x_{1}}\right)\co{\pi}\left(\co{x_{3}}|\cancel{\co{x_{1}}},\co{x_{2}}\right)\co{\pi}\left(\co{x_{4}}|\cancel{\co{x_{1}}},\cancel{\co{x_{2}}},\co{x_{3}}\right) \\
&= \co{\pi}\left(\co{x_{1}}\right)\co{\pi}\left(\co{x_{2}}|\co{x_{1}}\right)\co{\pi}\left(\co{x_{3}}|\co{x_{2}}\right)\co{\pi}\left(\co{x_{4}}|\co{x_{3}}\right).
\end{aligned}
\label{eq:factorization_sparse}
\end{equation}

We refer to this reduction as \textit{sparsification}. Triangular transport maps allows us to leverage these conditional independence properties by simply dropping the corresponding arguments from the map components $S_{k}$:

\begin{equation}
\smap(\co{\x_{1:4}})=\begin{bmatrix*}[l]
    S_{1}(\co{x_{1}}) \\
    S_{2}(\co{x_{1}},\co{x_{2}}) \\
    S_{3}(\cancel{\co{x_{1}}},\co{x_{2}},\co{x_{3}}) \\
    S_{4}(\cancel{\co{x_{1}}},\cancel{\co{x_{2}}},\co{x_{3}},\co{x_{4}})
\end{bmatrix*}=\begin{bmatrix*}[l]
    S_{1}(\co{x_{1}}) \\
    S_{2}(\co{x_{1}},\co{x_{2}}) \\
    S_{3}(\co{x_{2}},\co{x_{3}}) \\
    S_{4}(\co{x_{3}},\co{x_{4}})
\end{bmatrix*}=\begin{bmatrix*}[l]
    \cg{z_{1}} \\
    \cg{z_{2}} \\
    \cg{z_{3}} \\
    \cg{z_{4}}
\end{bmatrix*} = \cg{\z_{1:4}}.
\label{eq:decomposed_S_sparse}
\end{equation}

Equivalently, its inverse map would be:

\begin{equation}
\smap^{-1}(\cg{\z_{1:4}})=\begin{bmatrix*}[l]
    S_{1}^{-1}(\cg{z_{1}}) \\
    S_{2}^{-1}(\cg{z_{2}};\co{x_{1}}) \\
    S_{3}^{-1}(\cg{z_{3}};\cancel{\co{x_{1}}},\co{x_{2}}) \\
    S_{4}^{-1}(\cg{z_{4}};\cancel{\co{x_{1}}},\cancel{\co{x_{2}}},\co{x_{3}})
\end{bmatrix*}=\begin{bmatrix*}[l]
    S_{1}^{-1}(\cg{z_{1}}) \\
    S_{2}^{-1}(\cg{z_{2}};\co{x_{1}}) \\
    S_{3}^{-1}(\cg{z_{3}};\co{x_{2}}) \\
    S_{4}^{-1}(\cg{z_{4}};\co{x_{3}})
\end{bmatrix*}=\begin{bmatrix*}[l]
    \co{x_{1}} \\
    \co{x_{2}} \\
    \co{x_{3}} \\
    \co{x_{4}}
\end{bmatrix*} = \co{\x_{1:4}}.
\label{eq:inverse_decomposed_S_sparse}
\end{equation}

Making use of conditional independence properties in this way is useful for two reasons:

\begin{enumerate}
    \item \textbf{Robustness}: Any conditional independence we can enforce by construction is statistical information the map does not have to pry from the samples or the model, improving the overall fidelity and robustness of the approximation to $\pi$ in settings with finite ensemble size.
    \item \textbf{Efficiency}: The removal of superfluous dependencies reduces the number of input arguments to many of the map components $S_k$, decreasing the evaluation and inversion complexity (e.g., see \cref{subsec:monotonicity} for evaluation complexity). This property is often called sparsification as it turns the Jacobian $\nabla \smap$ into a sparse matrix.
\end{enumerate}

This second property is the key to applying transport methods in high-dimensional systems. As each map component function $S_{k}$ generically depends on all previous arguments $\co{\x_{1:k-1}}$, the computational demand explodes with the dimension of the target $\co{\pi}$ when optimizing or evaluating the map. With sufficient conditional independence, however, sparse maps can overcome this dramatic increase in complexity. For instance, the Markov structure in \crefrange{eq:decomposed_S_sparse}{eq:inverse_decomposed_S_sparse} results in a sparse map $\smap$ with numerical complexity scaling linearly in the target dimension $K$.

A comprehensive account of the link between conditional independence and the sparsity of triangular maps is given in \citet{Spantini2018InferenceCouplings}, through two main lines of results. First, given a sparse undirected probabilistic graphical model that encodes Markov properties of the target distribution $\co{\pi}$, it is shown how to predict the sparsity pattern of the triangular map $\smap$. This process relies on an ordered graph elimination algorithm, and can thus be performed before learning the map itself. But the resulting sparsity pattern depends on the chosen ordering of the random variables, which underscores the fact that triangular maps are intrinsically \textit{anisotropic} objects: a good ordering is necessary in order to maximize sparsity. We comment further on methods for finding such orderings in \cref{sec:outlook}. Second, \citet{Spantini2018InferenceCouplings} show that a property somehow dual to sparsity is \textit{decomposability}: given the Markov structure of some distribution $\co{\pi}$ on $\mathbb{R}^K$,
the inverse of the map $\smap$ defined by $\cg{\smap_\sharp \pi} = \cg{\eta}$ can be represented as
the composition of finitely many low-dimensional triangular transport maps, where low dimensionality here means that each component map is a function only of a small number of inputs. The exact structure of this decomposition follows from, again, an ordered decomposition of the original graph. These results allow very high-dimensional problems to be broken into many smaller and more manageable parts, given some conditional independence. %

\subsubsection{Block-triangular maps}\label{subsubsec:block_triangular}

Recalling the block structure in \cref{subsubsec:conditioning}, we have so far assumed that the map component ``blocks'' in \cref{eq:inverse_decomposed_S_blocks} are internally triangular; i.e., $\smap_{1:k}$ has output $[S_1(\co{x_1}),\ldots,S_k(\co{\x_{1:k}})]$ and similar for $\smap_{k+1:K}$. We may instead use \textit{any} invertible functions $\smap_{1:k}:\real{k}\to\real{k}$ and $\smap_{k+1:K}:\real{K} \to \real{K-k}$, for example certain neural networks \citep[e.g., ][]{doi:10.1137/23M1581546}. This still permits sampling conditionals and can be numerically advantageous in high-dimensional systems, but can compromise the ability to exploit conditional independence within the map blocks. In the following, we will assume that the transport map is \textit{fully triangular}, even where we adopt the block-triangular structure for ease of notation.

%% file: sections/implementation.tex
In the preceding section, we have discussed the theoretical foundations of triangular transport, but stopped short of defining the precise method to evaluate component $S_k$ or the optimization problem which identifies the specific map $\S$ from $\co{\pi}$ to $\cg{\eta}$. In this section, we will fill these gaps, and provide guidance on the practical implementation of triangular transport methods in code.

\tmlr{First, though, a slight interlude about the ``work'' a map performs to transform $\co{\pi}$ and $\cg{\eta}$ into one another. If $\cg\eta$ is standard Gaussian and $\co\pi$ is also standard Gaussian distribution, then the transport map $\S$ is trivially the identity.
If $\co\pi$ is instead a general Gaussian distribution, an affine $\smap$ will suffice to transform between $\co\pi$ and $\cg{\eta}$.
Once $\co\pi$ becomes non-Gaussian, however, the transport map must be strictly nonlinear, as an affine $\smap$ cannot create non-Gaussian features from a Gaussian reference. The nonlinearity of a map is therefore related to how ``different'' our reference and target distributions are in various senses.
This intuitive idea underpins many practical considerations; for example, it is expedient to choose a reference distribution $\cg\eta$ that shares approximately the same tail behavior as the target pdf $\co{\pi}$, i.e., to ensure that the quotient $\co{\pi}(\x)/\cg{\eta}(\x)$ is approximately constant far from the origin. This allows us to use map component functions $S_k$ that are linear in their inputs outside a certain radius; learning a map that instead changes the tail behavior would require learning nonlinear behavior in the regime $\| \x \| \to \infty$, which is typically outside the support of the training data and thus intrinsically challenging.}

\subsection{Structuring the map components \texorpdfstring{$S_{k}$}{Sk}}\label{subsec:map_components}

A core component of any triangular transport map $\smap$ is the definition of the constituent map components $S_{k}$. In this subsection, we will introduce and discuss important features in the definition of these map components $S_{k}$. \tmlr{We will begin by discussing ways to satisfy the montonicity requirements of triangular maps (Section~\ref{subsec:monotonicity}), then discuss different parametrization structures for the map component functions (Section~\ref{subsubsec:parameterizations}), and conclude with a discussion of examples of suitable basis function families (Section~\ref{subsubsec:basis_functions}).}

\subsubsection{Monotonicity}\label{subsec:monotonicity}

Let us recall from \cref{subsec:change_of_variables} that each map component function must be monotone in its last argument $\co{x_{k}}$, that is to say, $\partial_{\co{x_{k}}}S_{k} > 0$ on the \textit{entire} domain (i.e., for any choice $\co{\x_{1:k-1}}$). Together with the triangular structure and linearity in the tails, this property guarantees that the resulting map $\S$ is bijective. 

This may not be immediately intuitive, so an illustration of this effect is provided in \cref{fig:monotone_functions_2D}, which shows an example in $\mathbb{R}^2$. Here, invertibility ensures that every tuple $(\co{\mathsf{X}_{1}},\co{\mathsf{X}_{2}})$ maps to a unique tuple $(\cg{\mathsf{Z}_{1}},\cg{\mathsf{Z}_{2}})$. Graphically, this means that if we overlay \cref{fig:monotone_functions_2D}C and \cref{fig:monotone_functions_2D}D on top of each other, any pair of contour lines between the subplots must not intersect more than once. This is achieved due to triangularity and monotonicity: %
\begin{figure}[t!]
  \centering
  \includegraphics[width=\textwidth]{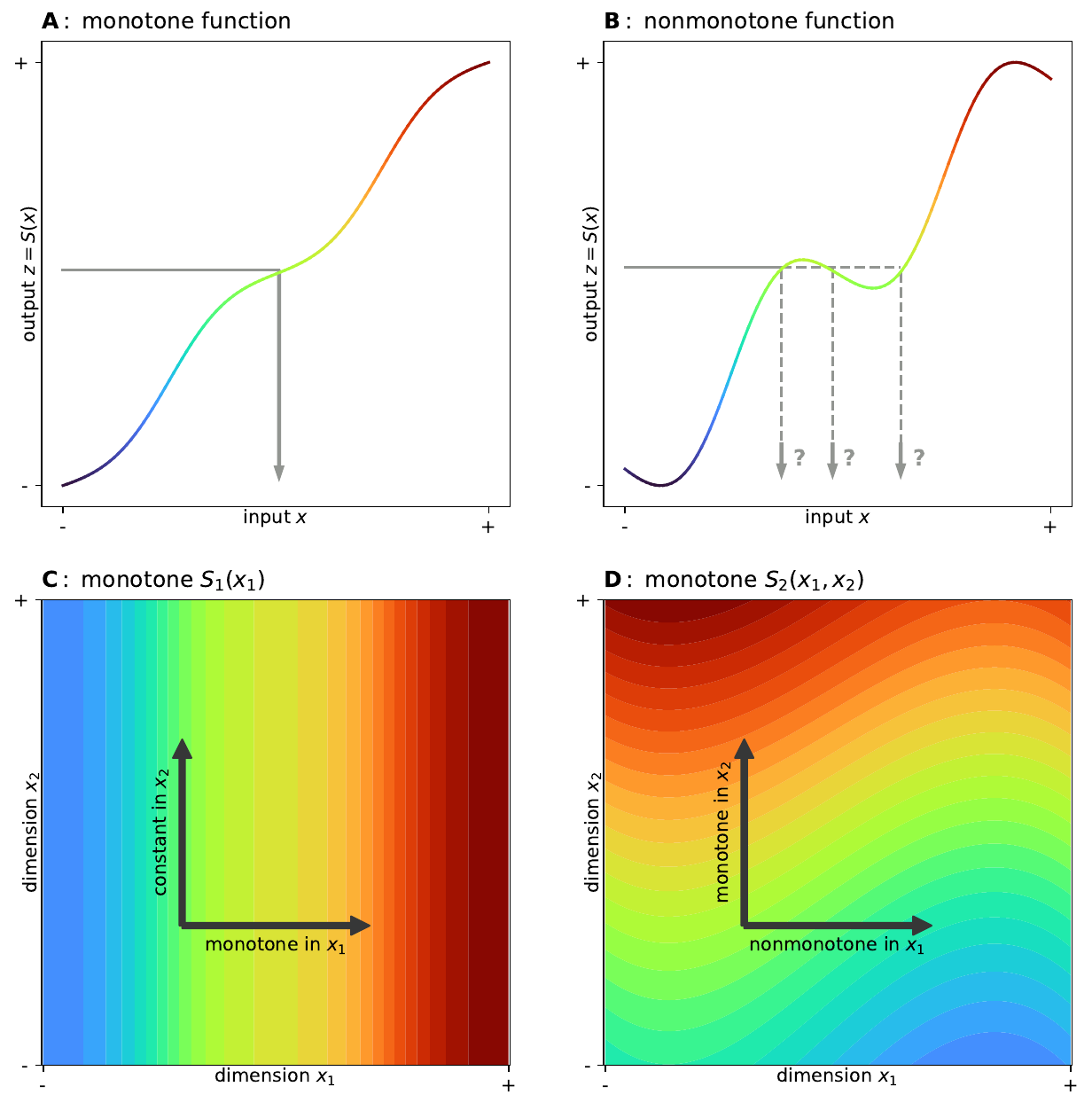}
  \caption{(A) Monotonicity guarantees that a function remains invertible. (B) Non-monotone map component functions have non-unique inverses. How do multivariate maps $\smap$ remain monotone? (C) The map component function $S_1(\co{x_1})=\cg{z_{1}}$ is monotone in $\co{x_{1}}$ but constant in $\co{x_{2}}$, as it does not depend on this input. (D) The map component function $S_2(\co{x_1},\co{x_2})=\cg{z_{2}}$ is monotone in $\co{x_{2}}$ but can be nonmonotone in $\co{x_{1}}$. Color indicates the magnitude of the map component's output in all subplots.}
  \label{fig:monotone_functions_2D}
\end{figure}
\begin{itemize}
    \item The \textbf{triangular structure} ensures that the contours of every map component $\smap_{1:k-1}(\co{\x_{1:k-1}})$ are independent of $\co{x_{k}}$ and are thus constant along $\co{x_{k}}$. In consequence, the contours of $S_{1}$ in \cref{fig:monotone_functions_2D}C are constant along $\co{x_{2}}$, which aligns them parallel to $\co{x_{2}}$ (see \cref{fig:monotone_functions_2D}C).
    \item The \textbf{monotonicity} of $S_1(\co{x_{1}})$ relates each $\co{\mathsf{X}_{1}}$ to a unique $\cg{\mathsf{Z}_{1}}$. We then obtain a unique tuple $(\cg{\mathsf{Z}_{1}},\cg{\mathsf{Z}_{2}})$ when we ensure that the second set of contours from $S_{2}(\co{\mathsf{X_1}},\co{x_{2}})$ increases monotonously along $\co{x_{2}}$ (see \cref{fig:monotone_functions_2D}D). This is achieved through the \textbf{monotonicity requirement} $\partial_{\co{x_{k}}}S_{k}>0$.
\end{itemize}

The same principle extends to higher dimensions: the triangular structure means that the $(k-1)$-dimensional contours of $\smap_{1:k-1}(\co{\x_{1:k-1}})$ are aligned along $\co{x_{k}}$. Monotonicity of $S_{k}$ along $\co{x_{k}}$ then ensures that for $(\co{\mathsf{X}_{1}},\dots,\co{\mathsf{X}_{k-1}})$ and any $\co{x_{k}}$, we obtain a unique tuple $(\cg{\mathsf{Z}_{1}},\dots,\cg{\mathsf{Z}_{k}})$. There are several strategies that can ensure each map component satisfies these monotonicity requirements. We will explore these strategies in the following.

\paragraph{Monotonicity through integration}

A general, powerful, but also computationally demanding method to enforce monotonicity makes use of a \textit{rectifier} and \textit{integration} \cite[e.g., ][]{Baptista2023OnMaps}. In this formulation, our starting point is a smooth, differentiable, but nonmonotone function $S_{k}^{\text{non}}:\real{k}\to\mathbb{R}$. An example of such a nonmonotone function is:

\begin{equation}
    S_{3}^{\text{non}}\left(\co{x_{1}},\co{x_{2}},\co{x_{3}}\right) = \underbrace{c_{0} + c_{1}\co{x_{1}} + c_{2}\co{x_{2}} + c_{3}\co{x_{1}x_{2}} + c_{4}\co{x_{1}}^{2}}_{\text{nonmonotone part }g\left(\co{x_{1}},\co{x_{2}}\right)} + \underbrace{c_{5}\co{x_{3}} + c_{6}\co{x_{1}}\co{x_{3}}^{2} + c_{7}\co{x_{2}}\co{x_{3}}}_{\text{pre-monotone part }\hat{g}\left(\co{x_{1}},\co{x_{2}},\co{x_{3}}\right)},
    \label{eq:premonotone_function}
\end{equation}

where the coefficients $\{ c_{i} \}$ parameterize $S_{k}^{\text{non}}$. In \cref{eq:premonotone_function},we have conceptually separated the terms of $S_{3}^{\text{non}}$ into a nonmonotone part $g$, which does not depend on $\co{x_{3}}$, and a ``pre-monotone'' part $\hat{g}$, which does. In general, the function $\hat{g}$ will not be monotone in $\co{x_{3}}$, and consequently neither will be $S_{3}^{\text{non}}$. We can make it monotone, however, by first applying a rectifier $r:\mathbb{R}\to\real{+}$ to $\hat{g}$, then integrating the rectified output over $\co{x_{k}}$:

\begin{equation}
\begin{aligned}
    & S_k(\co{x_1},\ldots,\co{x_k}) = \underbrace{g(\co{x_1},\ldots,\co{x_{k-1}})}_{\text{nonmonotone part}} + \underbrace{f(\co{x_1},\ldots,\co{x_k})}_{\text{monotone part}}, \\
    & \text{where } 
    \underbrace{f(\co{x_1},\ldots,\co{x_k})}_{\text{monotone function}} = \int_0^{\co{x_k}}\overbrace{r(\underbrace{\hat{g}(\co{x_1},\ldots,\co{x_{k-1}},t)}_{\text{nonmonotone function}})}^{\text{positive function}}\,\mathrm{d}t.
\end{aligned}
\label{eq:integrated_rectifier}
\end{equation}

The two steps combined in \cref{eq:integrated_rectifier} are as follows:

\begin{enumerate}
    \item \textbf{Rectification}: First, $\hat{g}$ is sent through the rectifier $r$, a function which maps its output to strictly positive values. Examples of useful rectifiers include exponential functions, the softplus function (i.e., $\log(1+\exp(x))$) or Exponential Linear Units \citep{Clevert2016FastELUs}.
    \item \textbf{Integration}: Numerically or analytically integrating the output of the resulting strictly positive function over $\co{x_{k}}$ yields a function $f$ monotone in $\co{x_{k}}$, which turn also guarantees the monotonicity of $S_{k}$ in $\co{x_{k}}$.
\end{enumerate}

\begin{figure}
  \centering
  \includegraphics[width=\textwidth]{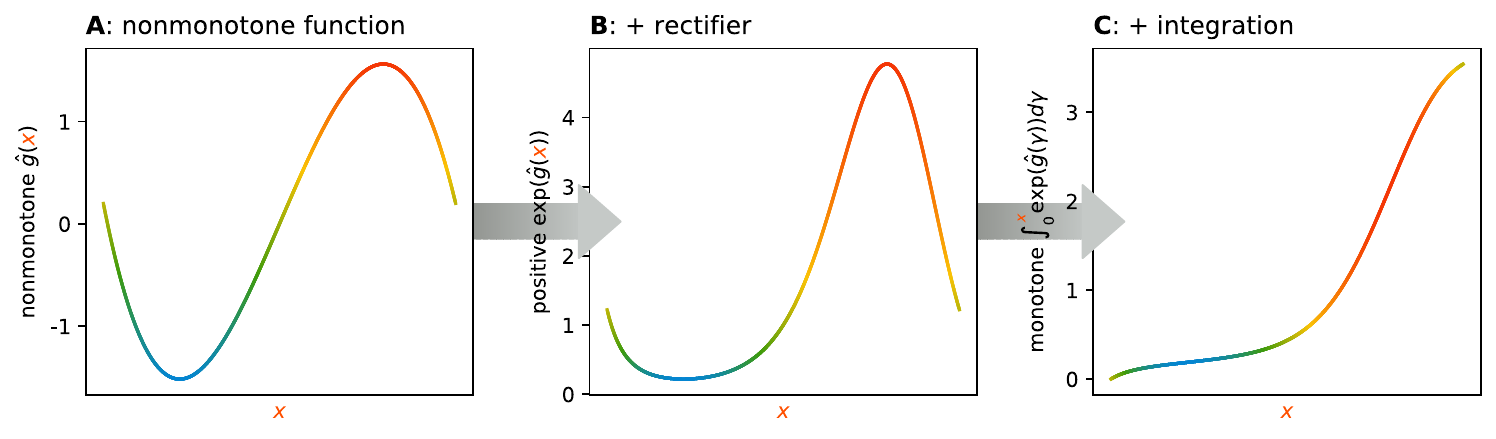}
  \caption{Integrated rectifiers create monotonicity. (A) We start with an arbitrary smooth, non-monotone function $\hat{g}(\co{x})$. (B) Applying a monotone rectifier, for example $\exp\left(\hat{g}(\co{x})\right)$, yields strictly positive function output. (C) Integrating a strictly positive function yields a monotone function $f$. Color indicates the magnitude of the nonmonotone function in (A).}
  \label{fig:integrated_rectifier}
\end{figure}

This process is visualized in \cref{fig:integrated_rectifier}, and has a number of important advantages. First, the use of an integrated rectifier places very few limitations on the structure of $S_{k}^{\text{non}}$. Furthermore, it also permits the introduction of \textit{cross-terms} such as $\co{x_{1}}\co{x_{k}}$ between the last argument $\co{x_{k}}$ and previous dimensions $\co{x_{1:k-1}}$, which are required for many of the more complex mapping operations (see \cref{subsubsec:parameterizations}). The drawback of this approach is that practically evaluating \cref{eq:integrated_rectifier} often demands one-dimensional numerical integration via quadrature, increasing the computational demand of the map's evaluation, optimization, and inversion.

An important practical consideration when using the integrated map formulation is to ensure that the pre-monotone term $\hat{g}$ reverts to a \textit{constant} in the tails of $\co{x_{k}}$. This can be ensured by defining $\hat{g}$ as a combination of a constant term and Hermite functions (see \cref{subsubsec:basis_functions}), which revert to zero in the tails. To understand why this is important, consider the process illustrated in \cref{fig:integrated_rectifier}. Through rectification and integration, increasingly negative values of $\hat{g}$ become flat stretches in $S_{k}$. If one or both tails of the monotone term $f$ become near-flat, it is possible for the root finding during the map's inversion (see \cref{subsubsec:inversion}) to fail for outlying values. If, however, $\hat{g}$ instead reverts to a positive constant in the tails, the resulting integrated monotone term $f$ will extrapolate linearly and thus generally remain more robust to outliers during the inversion. This is an artifact from the discussion above where, while $\pi$ and $\eta$ may ``look'' very different for the bulk of the pdf, we expect the tails of both of them to behave similarly \citep{Baptista2023OnMaps}.

\paragraph{Monotonicity through variable separation}

A more computationally efficient way to ensure monotonicity is to formulate a map component function $S_{k}$ which is both \textit{separable in} $\co{x_{k}}$ and \textit{linear in the coefficients}, though not generally linear in the inputs $\x$. An example for such a map component function is:

\begin{equation}
    S_{3}\left(\co{x_{1}},\co{x_{2}},\co{x_{3}}\right) = \underbrace{c_{0} + c_{1}\co{x_{1}} + c_{2}\co{x_{2}} + c_{3}\co{x_{1}x_{2}} + c_{4}\co{x_{1}}^{2}}_{\text{nonmonotone part }g\left(\co{x_{1}},\co{x_{2}}\right)} + \underbrace{c_{5}\co{x_{3}} + c_{6}\operatorname{erf}(\co{x_{3}})}_{\text{monotone part }f\left(\co{x_{3}}\right)},
    \label{eq:separable_map}
\end{equation}

where $c_{i}$ once more are the map component's coefficients, which are optimized to find the map from $\co{\pi}$ to $\cg{\eta}$, and $c_5,c_6 > 0$. The key aspects in \cref{eq:separable_map} are a clear separation into a nonmonotone term $g$, which may depend on all arguments except the last $\co{\x_{1:k-1}}$, and a monotone term $f$, which may \textit{only} depend on the last argument $\co{x_{k}}$. To ensure that $S_{k}$ remains monotone in $\co{x_{k}}$, all terms in $f$ must likewise be monotone basis functions (e.g., $\co{x_{k}}$ and $\operatorname{erf}(\co{x_{k}})$).

The advantages of this separable formulation are two-fold: First, \cref{eq:separable_map} has no need for numerical integration, which reduces computational demand substantially. Second, as we shall see in \cref{subsec:optimization}, this separable formulation allows for extremely efficient map optimization. The price for this efficiency is that variable separation does not allow for cross-terms between the last argument and previous arguments (e.g., $\co{x_{1}}\co{x_{k}}$), which limits the map's expressivity. We will explore the consequences of this limitation in the next section.

\subsubsection{Parameterization}\label{subsubsec:parameterizations}

Closely related to the concern of monotonicity is how each map component $S_k$ relates $\co{x_k}$, its last argument, to $\co{x_{1}},\ldots,\co{x_{k-1}}$.
This decision leads to three different kinds of map parameterizations, which permit maps of different complexity. In ascending order of complexity, these parameterizations are:

\paragraph{Marginal maps}
If both the nonmonotone part $g$ and the monotone part $f$ depend exclusively on input $\co{x_{k}}$, such map component functions $S_{k}$ result in a \textit{marginal} (or \textit{diagonal}) map. As the name implies, such maps only transform the marginals of $\co{\x}$ without capturing any dependencies between its dimensions. An example of a marginal map would be:

\begin{equation}
    S_{3}\left(\co{x_{3}}\right) = \underbrace{c_{0}}_{\text{nonmonotone part }g\left(-\right)} + \underbrace{c_{1}\co{x_{3}} + c_{2}\co{x_{3}}^{3}}_{\text{monotone part }f\left(\co{x_{3}}\right)}.
    \label{eq:parameterization_marginal}
\end{equation}

Recalling the implications of removing dependencies on earlier arguments $\co{\x_{1:k-1}}$ from the map component functions (see \cref{subsubsec:conditional_independence}), marginal maps are of no direct use for the conditioning operations in \cref{subsubsec:conditioning}. This is because marginal maps implicitly assume all entries of $\co{\x}$ are marginally independent, which in turn implies that $\co{\pi}(\co{\x_{k+1:K}}|\co{\x_{1:k}})=\co{\pi}(\co{\x_{k+1:K}})$, that is to say, nothing can be learned from $\co{\x_{1:k}}$ about $\co{\x_{k+1:K}}$. As marginal Gaussianization schemes, however, they can find limited use in Gaussian anamorphosis \citep{Schoniger2012ParameterTomography,Zhou2011AnFiltering} or as preconditioning tools in certain graph detection methods \citep{Liu2009TheGraphs}. Marginal maps are only included here for completeness' sake, and we use $g(-)$ to denote that the nonmonotone part is constant in $\co{\x}$, connecting better to the more complex parameterizations.

\paragraph{Separable maps} 

Closely related to the monotonicity scheme of the same name in the previous section, 
\textit{separable} maps separate the map component $S_k$ into the sum of a function $g:\mathbb{R}^{k-1}\to\mathbb{R}$ that takes in $\co{\x_{1:k-1}}$ and a univariate monotone function $f:\mathbb{R}\to\mathbb{R}$ that just takes in $\co{x_k}$:

\begin{equation}
    S_{3}\left(\co{x_{1}},\co{x_{2}},\co{x_{3}}\right) = \underbrace{c_{0} + c_{1}\co{x_{1}} + c_{2}\co{x_{1}}^{2} + c_{3}\co{x_{1}}\co{x_{2}}}_{\text{nonmonotone part }g\left(\co{x_{1}},\co{x_{2}}\right)} + \underbrace{c_{4}\co{x_{3}} + c_{5}\co{x_{3}}^{3}}_{\text{monotone part }f\left(\co{x_{3}}\right)}.
    \label{eq:parameterization_separable}
\end{equation}

Separable maps are one level of complexity above marginal maps. Such maps can consider nonlinear dependencies between the different dimensions, but do not permit cross-terms with the last argument $\co{x_{k}}$, which in turn limits the complexity of the target distributions $\co{\pi}$ they can recover.%
\footnote{\tmlr{Linearity in coefficients and the existence of cross terms are not mutually exclusive; we could choose a map with cross terms that is linear in its coefficients. One must be careful, however, to ensure monotonicity in the final argument $\co{x_k}$ of the map for \textit{any} choice of the other inputs $\co{\x_{1:k-1}}$.}}
This limitation is subtle, and will be discussed shortly. An important advantage of separable maps is that, when linear in the coefficients, their optimization can be partially solved in closed-form, which can improve computational efficiency substantially. More detail on this is provided in \cref{appendix:separable_optimization}.

\paragraph{Cross-term maps}

The most versatile map parameterization are \textit{cross-term} maps, which permit the greatest expressiveness at the cost of increased computational demand. As mentioned previously, cross-terms are basis functions which depend both on the last argument $\co{x_{k}}$ and preceding arguments $\co{\x_{1:k-1}}$. The presence of these terms requires that such maps must use integrated rectifiers (\cref{subsec:monotonicity}) to ensure monotone $S_{k}$:

\begin{equation}
    S_{2}\left(\co{x_{1}},\co{x_{2}}\right) = \underbrace{c_{0} + c_{1}\co{x_{1}} + c_{2}\co{x_{1}}^{2}}_{\text{nonmonotone part }g\left(\co{x_{1}}\right)} + \underbrace{\int_{0}^{\co{x_{2}}} \exp \bigg(\overbrace{c_{3}\co{x_{1}}^{2}\co{t}}^{\text{cross-term}} + c_{4}\co{t} + \overbrace{c_{5}\co{x_{1}}\co{t}^{2}}^{\text{cross-term}}\bigg) d\co{t}}_{\text{monotone part }f\left(\co{x_{1}},\co{x_{2}}\right)}.
    \label{eq:parameterization_crossterm}
\end{equation}

The presence of cross-terms permits increased control over the local details in the transformations from $\co{\pi}$ to $\cg{\eta}$. 
In principle, it is also possible to make \cref{eq:parameterization_crossterm} separable in $\co{x_{k}}$ by removing the dependencies of $f$ on $\co{\x_{1:k-1}}$, yielding a separable map with monotonicity through an integrated rectifier. Since the resulting $S_{k}$ would not be linear in the coefficients $c$, however, we cannot make use of a more efficient optimization scheme (\cref{appendix:separable_optimization}). In the following subsection, we will develop some intuition about the strengths and limitations of different map parameterizations.

\paragraph{Choosing a parameterization}

To provide more insight into the effects and limitations of each parameterization choice, 
we have illustrated the pullback $\co{\smap^\sharp\eta}$ of a transport map approximated using these parameterizations for each of three different target pdfs $\co{\pi}$ in \cref{fig:cross_terms}.
In every case, we begin by sampling the target distributions, proceed to optimize the different maps, then apply the inverse map (\cref{subsubsec:inversion}) to transform reference samples $\cg{\z}\sim\cg{\eta}$ into samples from the pullback $\co{\x}\sim\co{\S^{{\sharp}}\eta}$. We may make the following observations:

\begin{figure}[ht!]
  \centering
  \includegraphics[width=\textwidth]{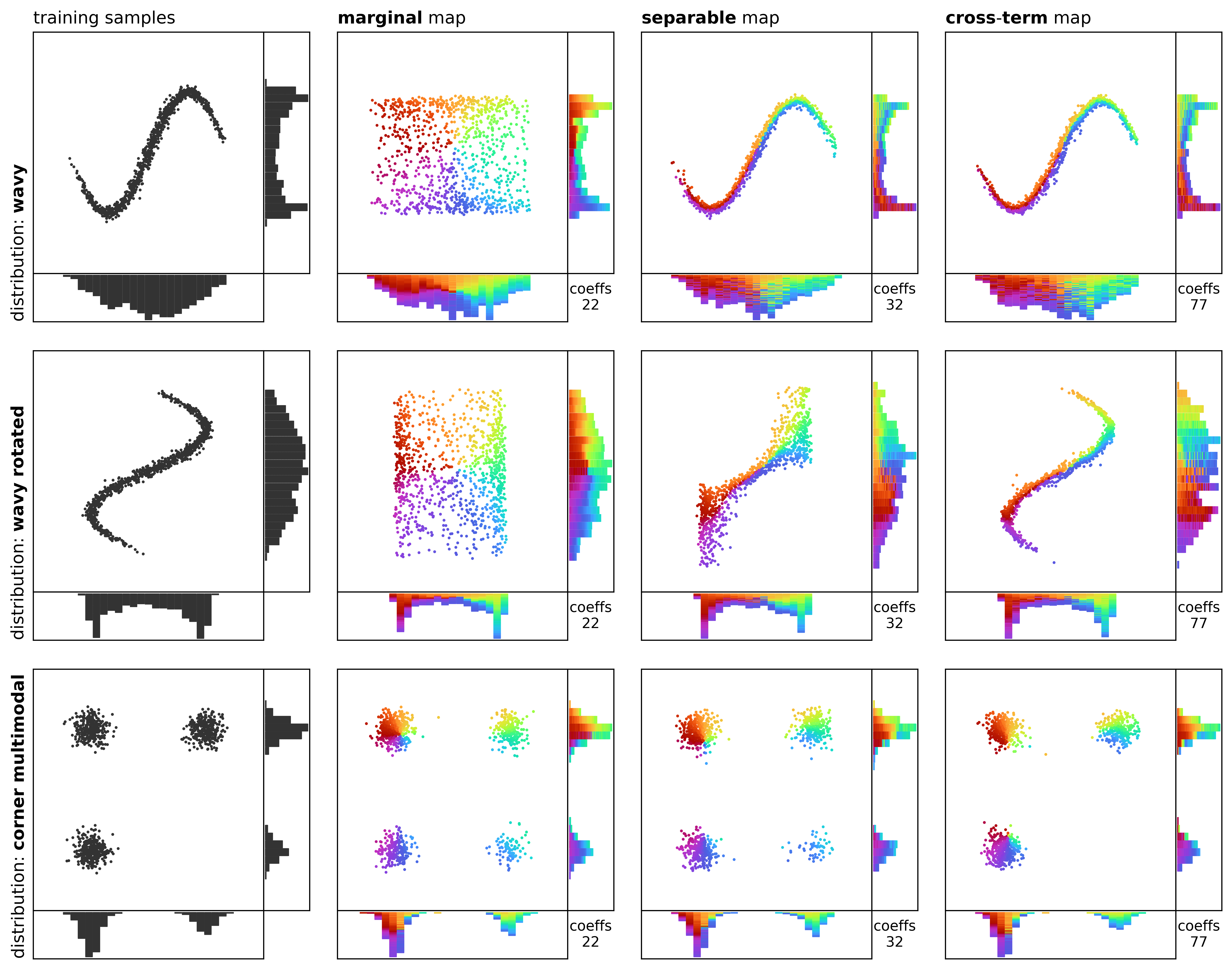}
  \caption{Pullback approximations $\co{\S^{{\sharp}}\eta}$ to different target distributions $\co{\pi}$ using nonlinear map components $S_{k}$ for marginal maps, separable maps, and cross-term maps. Some distributions require cross-term maps, for others simpler parameterizations may suffice. The variable ordering is $[\co{x_{1}},\co{x_{2}}]$, with $\co{x_{1}}$ plotted on the horizontal axis and $\co{x_{2}}$ on the vertical axis. Color represents the position of the reference samples $\cg{\z}$ relative to the mean of $\cg{\eta}$. The number of coefficients (i.e., parameters) for each map is in the bottom-right corner.}
  \label{fig:cross_terms}
\end{figure}

\begin{itemize}
    \item \textbf{Marginal} maps only reproduce the marginal densities of all three target distributions. As their structure implies independence between the marginals (\cref{subsubsec:conditional_independence}), however, the pullback approximation does not approximate the true target pdfs well. This effect can be subtle: in the corner-multimodal distribution, it is only noticeable through the emergence of a fourth phantom mode.
    \item \textbf{Separable} maps provide much better approximations to the target pdfs, but have some interesting caveats: while they successfully recover the wavy target distribution, they struggle with the distribution once it is rotated by $90$°, and they yield a slight---if less pronounced---fourth phantom mode for the corner-multimodal target. We will discuss the reason for this below.
    \item \textbf{Cross-term} maps prove most versatile, recovering all three target distributions well at the cost of a larger parameter space.
\end{itemize}

In practice, cross-terms provide greater control over local features. To understand why, let us take a closer look at a single map component inversion $S_{k}^{-1}$, which forms the basis of the map inversion (see \cref{subsubsec:inversion}) that ultimately yields the pushforward samples in \cref{fig:cross_terms}. Consider a generic map component function comprised of a nonmonotone part $g$ and a monotone part $f$:

\begin{equation}
    \cg{z_{k}} = S_{k}(\co{\x_{1:k-1}},\co{x_{k}}) = g(\co{\x_{1:k-1}}) + f(\co{\x_{1:k-1}},\co{x_{k}}).
\end{equation}

As discussed in \cref{subsubsec:inversion}, the inversion of $S_{k}$ relies on one-dimensional root finding. Reformulating this expression yields the root finding objective for $S_{k}^{-1}$, conditioned on the outcomes $\co{\x_{1:k-1}}$ of previous map component inversions:

\begin{equation}
    \cg{z_{k}} - g(\co{\x_{1:k-1}}) = f(\co{\x_{1:k-1}},\co{x_{k}}).
\end{equation}
If we now apply the simplifications of the three map parameterizations above, we obtain three different objectives for the root finding problem:

\begin{equation}
\begin{aligned}
    \cg{z_{k}} - g(-) &= f(\co{x_{k}}) && \text{(marginal maps)} \\
    \cg{z_{k}} - g(\co{\x_{1:k-1}}) &= f(\co{x_{k}}) && \text{(separable maps)} \\
    \cg{z_{k}} - g(\co{\x_{1:k-1}}) &= f(\co{\x_{1:k-1}},\co{x_{k}}) && \text{(cross-term maps)}
    \label{eq:conditional_inversion_types}
\end{aligned}
\end{equation} %

\begin{figure}[ht!]
  \centering
  \includegraphics[width=\textwidth]{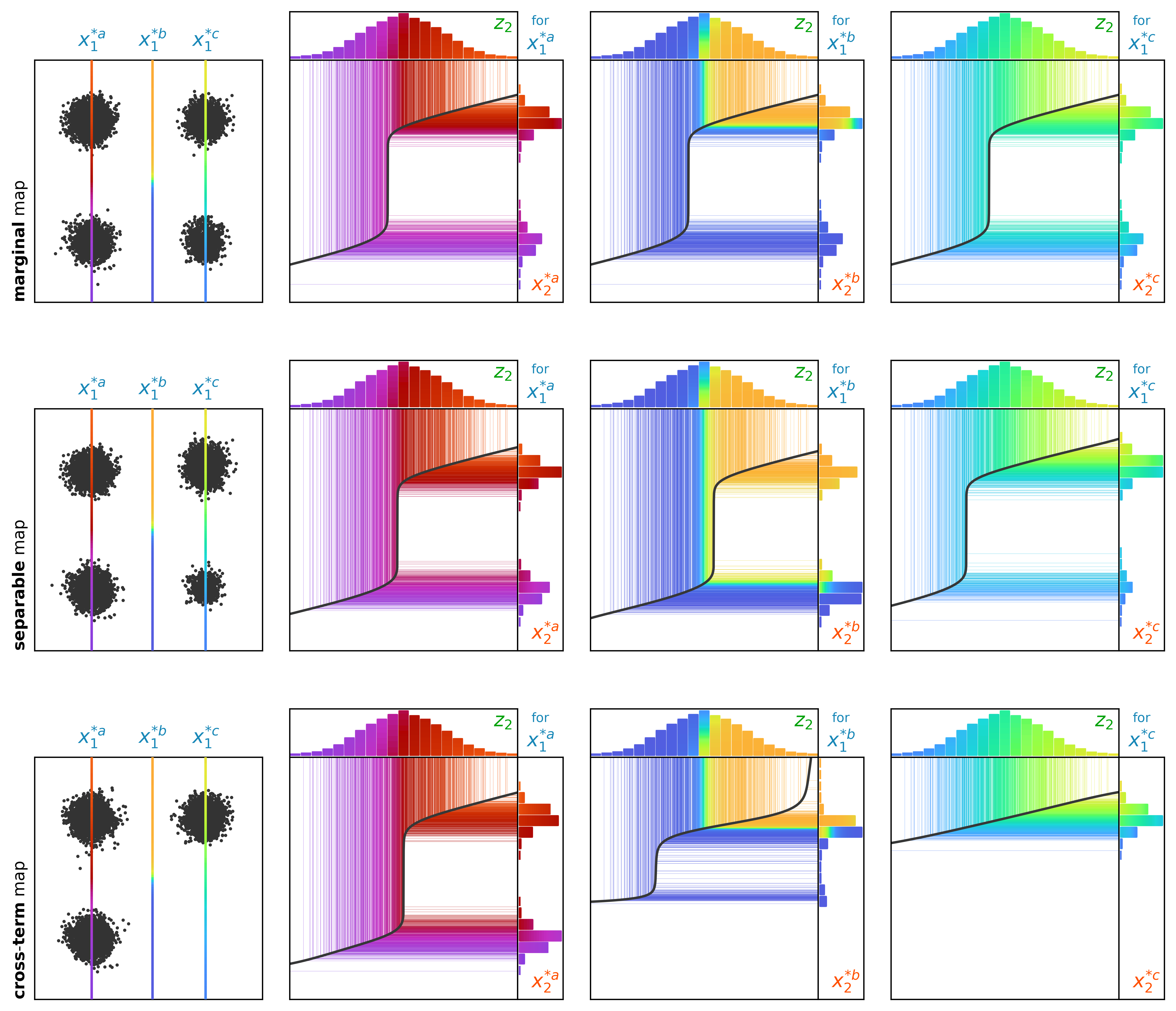}
  \caption{Conditional inversion $\co{x_{2}}=S_{2}^{-1}(\cg{z_{2}};\co{x_{1}})$ for marginal (top row), separable (center row), and cross-term (bottom row) maps. Each column corresponds to a conditional distribution for a different fixed $\cb{x_1^*}$. Each term in \cref{eq:conditional_inversion_types} has a different effect: $\cg{z_{2}}$ defines the horizontal starting position for the inversion, $f(\co{x_{1}},\co{x_{2}})$ defines the shape of coupling, and $g(\co{x_{1}})$ defines its horizontal offset. \textbf{Marginal} maps keep $f$ and $g$ fixed, and thus yield the same inverse $\co{x_{2}}$ for any previous $\co{x_{1}}$. \textbf{Separable} maps also keep $f$ constant, but can adjust the offset $g$. Finally, \textbf{cross-term} maps can adjust both $f$ and $g$ for different $\co{x_{1}}$, providing the greatest degree of flexibility.}
  \label{fig:cross_terms_analysis}
\end{figure}

These equations provide some insight into the differences between the pullback densities $\co{\S^{{\sharp}}\eta}$ we have observed in \cref{fig:cross_terms}. Each term above takes on a different role during the conditional inversion: the reference samples $\cg{z_{k}}$ are an independent input, unaffected by the map parameterization choice, and define a \textit{random initial offset} for the root finding. The nonmonotone term $g$ acts as an additional \textit{dynamic offset} for the inversion based on previous values, and the monotone term $f$ defines the \textit{shape of the inverse function} for the one-dimensional root finding over the unknown $\co{x_{k}}$. From this perspective, the differences between the map parameterizations emerge from how each handles the dependence on previous values $\co{\x_{1:k-1}}$:

\begin{enumerate}
    \item \textbf{Marginal} maps (\cref{fig:cross_terms_analysis}, top row) permit only a constant nonmonotone term $g(-)$, which results in no dynamic offset. Their monotone term $f(\co{x_{k}})$ likewise only depends on $\co{x_{k}}$. In consequence, marginal maps extract the same $\co{x_{k}}$ from a specific $\cg{z_{k}}$ for any given value of $\co{\x_{1:k-1}}$.
    \item \textbf{Separable} maps (\cref{fig:cross_terms_analysis}, center row) also feature monotone term $f(\co{x_{k}})$ that depends only on $\co{x_{k}}$, which keeps the inverse function's shape constant for all values of $\co{\x_{1:k-1}}$. Their off-diagonal term $g(\co{\x_{1:k-1}})$ can induce varying offsets for different $\co{\x_{1:k-1}}$, introducing a simple dependence on previous values.
    \item \textbf{Cross-term} maps (\cref{fig:cross_terms_analysis}, bottom row) can vary both the dynamic offset (nonmonotone term $g(\co{\x_{1:k-1}})$) and the inverse function's shape (monotone term $f(\co{\x_{1:k-1}},\co{x_{k}})$) with different $\co{\x_{1:k-1}}$, and thus provide the most expressive maps, but often at higher computational cost.
\end{enumerate}

These insights reveal why separable maps succeeded to recover the wavy distribution, but failed to capture its features once it is rotated (\cref{fig:cross_terms}). For the original wavy distribution, its (vertical) conditionals at different horizontal positions are mostly of the same shape, just shifted vertically, which makes them a perfect fit for recovery via separable maps. Once the distribution is rotated, however, the true vertical conditionals become more challenging to recover. Note that if we were to change the order of the target vector to $[\co{x_{2}},\co{x_{1}}]$ instead of $[\co{x_{1}},\co{x_{2}}]$, separable maps would succeed in recovering the rotated wavy distribution and instead fail for the standard one. This illustrates the subtle influence of input variable ordering on a map's approximation ability.

\subsubsection{Basis functions}\label{subsubsec:basis_functions}
A very important part of the parameterization of the map component function $S_{k}$ is its construction: %
what basis functions should be used to represent $S_k$ and how much nonlinearity should they permit? In a nutshell, simpler---perhaps separable---parameterizations are more computationally efficient, but can struggle to recover features of the target pdf $\co{\pi}$ that do not resemble the reference $\cg{\eta}$, as discussed at the beginning of this section. By contrast, more complex parameterizations add the flexibility required to capture ``localized'' features of $\co{\pi}$ (e.g., skewness, multi-modality, \ldots), but risk unfavourable bias-variance trade-offs. 
While there are alternative choices~\citep{lunde2026adaptive}, we discuss two common basis function families that trade off different advantages:

\paragraph{Polynomial basis functions}
In function approximation, polynomials are a canonical choice for forming an approximation; in particular, harkening to polynomial chaos expansion and other traditional stochastic function approximation literature \citep{LeMaitre2010SpectralUQ,Ernst2012OnExpansions}, we can choose orthogonal polynomials for our approximation function class. Since this work focuses on unbounded $\co{\x}$, we will consider the probabilists' Hermite polynomial family $\{\mathrm{He}_j\}_{j=1}^\infty$, which has the orthogonality property:
\begin{equation}\label{eq:orthogonality}
\int_{-\infty}^\infty \mathrm{He}_i(x)\mathrm{He}_j(x)\;\;\underbrace{\frac{1}{\sqrt{\tau}}\exp(-x^2/2)}_{\text{Gaussian weight}}\;\;\mathrm{d}x = j!\delta_{ij} = \begin{cases} j! & j=i\\0 & j\neq i\end{cases},
\end{equation}
where $\tau=6.283185\ldots$ is twice Archimedes' constant. Broadly, \cref{eq:orthogonality} shows what it means for polynomials of different orders (e.g., linear, quadratic, cubic, etc.) to be orthogonal under the Gaussian weight. Similar to orthogonal vectors in linear algebra, orthogonal polynomials (more generally, orthogonal functions) are often chosen as a basis for function approximation, as their elements contain no ``overlapping information'' with respect to a particular weight (which may or may not match $\cg{\eta}$). %
As these polynomials are dependent on the pdf, a few other orthogonal polynomial families with their respective weights and support include Legendre polynomials (respecting a constant weight on $(-1,1)$) and Laguerre polynomials (respecting weight $e^{-x}$ on $(0,\infty)$)~\citep{Szego1939OrthogonalPolynomials}.
These polynomial families may be suited for different problems depending on characteristics of $\cg{\eta}$ and $\co{\pi}$~\citep{wang2022minimax,sharp2026samplingiteratedapproximationgradientfree}.

Generally, if a polynomial family $\{\psi_j\}_{j=1}^\infty$ is orthogonal with respect to a weight $\rho$ (e.g., the probabilist Hermite polynomials take $\rho$ as the Gaussian weight),
one can approximate a given well-behaved function $f(x)$ with $\hat{f}(x) = c_1\psi_1(x)+c_2\psi_2(x)+\cdots+c_P\psi_P(x)$ using a finite number $P$ of these polynomials. Guarantees for this approximation are often given when measuring the error according to our weight $\rho$ via $\int (f(x) - \hat{f}(x))^2\ \mathrm{d}\rho(x)$~\citep{Ernst2012OnExpansions}. If we believe that $\cg{\eta}$ and $\co{\pi}$ are very similar, choosing polynomials orthogonal with respect to our reference $\cg{\eta}$ (which we know the orthogonal polynomial family of) will also perform well when the input is distributed according to $\co{\pi}$ (which we generally will not know the orthogonal polynomial family of). On the other hand, if $\cg{\eta}$ and $\co{\pi}$ are not very similar, such guarantees are not particularly helpful; practitioners see this manifest as a need for more complex map parameterizations (see \cref{subsubsec:parameterizations}).

As they have unbounded support, Hermite polynomials exert global influence and often require high-order terms to capture fine distributional features. While versatile, the use of polynomial basis functions has a few pitfalls which demand caution:

\begin{enumerate}

    \item \textbf{Importance of standardization}: In response to the above, we would like to make $\co\pi$ resemble $\cg\eta$ more closely by pre-processing the data we have from $\co\pi$. 
    When working with a Gaussian reference $\cg\eta$, it is thus generally advised to standardize $\co\x$ by trying to match the mean and variance of $\cg\eta$. We do this by (i) subtracting the empirical mean $\co{\bar{\boldsymbol{\mathsf{X}}}}=1/N\sum_{i=1}^{N}\co{\boldsymbol{\mathsf{X}}}^{i}$ and (ii) scaling each marginal to unit variance by dividing it through the unbiased estimator of the empirical standard deviation $\co{\sigma_{k}}$ where $\co{\sigma}_{\co{k}}^2=(N-1)^{-1}\sum_{i=1}^{N}(\co{\mathsf{X}}_{\co{k}}^{i}-\co{\bar{\mathsf{X}}_{k}})^2$ before formulating and applying a transport map%
    \footnote{%
    Alternatively, one may standardize the samples by forming the empirical cdf $\tilde{F}_{\co{\pi_{k}}}$ along the marginal for $\co{x_{k}}$, then standardize the samples by composing $\tilde{F}_{\co{\pi_{k}}}$ with the inverse standard Gaussian cdf $F_{\cg{\eta_{k}}}^{-1}$, that is to say, $F_{\cg{\eta_{k}}}^{-1} \circ \tilde{F}_{\co{\pi_{k}}}(\co{x_{k}})$, ensuring each marginal of the standardized samples has a Gaussian distribution.
    } %
    $\S$. All subsequent map operations are then implemented in this standardized space. Finally, the map's outputs are transformed back into target space by reversing the standardization. In effect, standardization wraps a separate linear transformation around the transport map, and thereby does not affect the validity of the operations inside this wrapper. In the subsequent sections, we will assume by default that the samples have been standardized. As pointed out in \citet{Morrison2022DiagonalMatrices}, invertible diagonal transformations retain the conditional independence structure of the target distribution. In other words, there is no loss of information while working in the marginally rescaled space.
    
    \item \textbf{Edge control}: A practical challenge of polynomials is their growth in the tails for higher-order terms. This often leads to volatility if the map is evaluated or inverted far from zero. To address this issue, one useful option is the use of \textit{edge-controlled terms} such as \textit{Hermite functions} $\mathcal{H}_{j}$. This variant of basis functions are defined as probabilist's Hermite polynomials $\mathrm{He}_{j}$ of order $j$ multiplied with a Gaussian weight, which reverts the polynomial's output to zero far from zero:
\end{enumerate}
\begin{equation}
    \mathcal{H}_{j}\left(x\right) = \mathrm{He}_{j}\left(x\right)\exp\left(-\frac{x^2}{4}\right).
    \label{eq:Hermite_function}
\end{equation}
A nice feature of this form is that $\mathcal{H}_i(x)\mathcal{H}_j(x) = \mathrm{He}_i(x)\mathrm{He}_j(x)\exp(-x^2/2)$, and so $\{\mathcal{H}_j\}_{j=1}^\infty$ will inherit the same error properties as $\{\mathrm{He}_{j}\}$ when measuring error without any weight $\rho$, i.e., $\int (f(x) - \hat{f}(x))^2\ \mathrm{d}x$.

\begin{figure}[ht!]
  \centering
  \includegraphics[width=\textwidth]{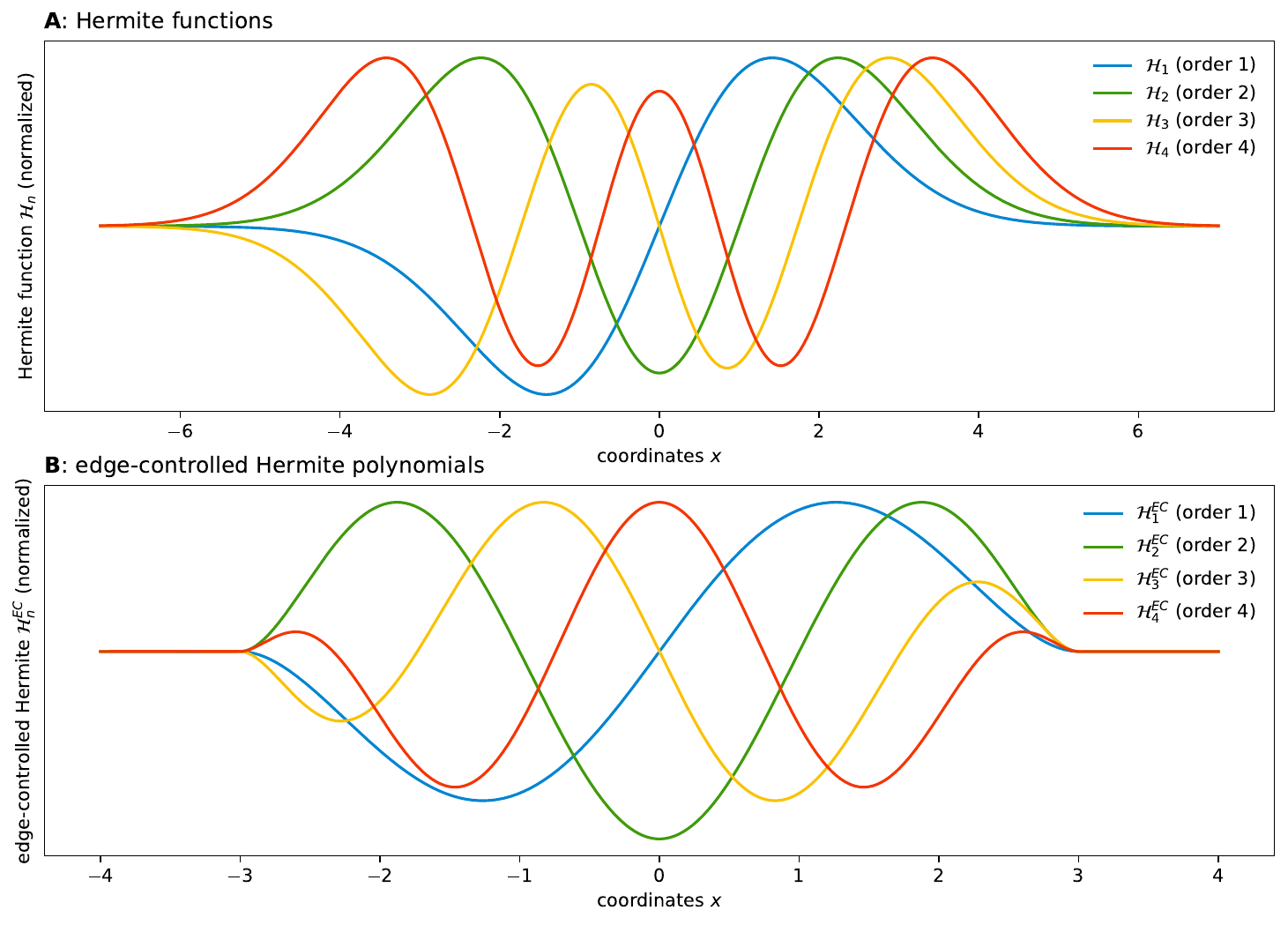}
  \caption{Two types of edge-controlled basis functions based on Hermite polynomials. (A) Hermite functions are multiplied with a Gaussian weight, causing them to approach zero in the tails. (B) Replacing the Gaussian weight with a scaled cubic spline leads to a different edge-controlled basis function. This enforces limited support, ensuring the function reverts to zero at a finite distance $r$.}
  \label{fig:edge_controlled_functions}
\end{figure}

Illustrations of the first few orders of Hermite functions are provided in \cref{fig:edge_controlled_functions}A. Note that expressing the map exclusively in terms of Hermite functions causes the map component $S_{k}$ to revert to zero in the tails, which can be undesirable whenever extrapolation may be required. In practice, 
limiting the Gaussian weight term to polynomials of order two or larger is a good compromise, thus we recommend retaining the linear terms without a weight.

A practical issue with Hermite functions can emerge for target distributions $\co{\pi}$ which include sharp features. From \citep{Szego1939OrthogonalPolynomials}, the largest maximizer $x^*_j$ of $\mathrm{He}_j$ grows at a rate of $x^*_j = \sqrt{j} + \mathcal{O}(j^{-1/6})$, which means that if $\mathcal{H}_j$ has nontrivial values on $(-r,r)$, we will have that $\mathcal{H}_{4j}$ has nontrivial values on $(-2r,2r)$; e.g., it is reasonable to estimate from \cref{fig:edge_controlled_functions}(A) that $r>4$ for $j=2$ and so $\mathcal{H}_8$ has nontrivial values on $(-8,8)$. Due to this ``global'' nature of polynomials influencing the edges of the Hermite functions, maps for such distributions often involve high-order terms with high-magnitude coefficients of opposing signs partially cancelling each other. While this is often unproblematic within the support of the training samples, the resulting high-magnitude coefficients can extend the influence of some Hermite functions farther into the tails, resulting in undesirable tail effects. As an alternative, we might prefer to employ a weight term that reverts the weights to zero at a finite distance. For instance, using a cubic spline weight yields a different \textit{edge-controlled} Hermite polynomial:

\begin{equation}
    \mathcal{H}_{j}^{\text{EC}}\left(x\right) = \mathrm{He}_{j}\left(x\right)\left(2\min\left(1,|x|/r\right)^3 - 3\min\left(1,|x|/r\right)^2 + 1\right),
    \label{eq:edge_controlled_Hermite}
\end{equation}

where $r$ is a finite edge-control radius specified by the user. Examples of such basis functions are illustrated in \cref{fig:edge_controlled_functions}B. When using these functions in the integrated-rectified formulation, this will create $\smap$ that is linear outside of the hypercube $(-r,r)^d$.

\paragraph{Radial basis functions}

A second useful class of basis functions is \textit{radial basis functions} (RBFs). RBFs exert more local influence than combinations of polynomial basis functions, and generally require two additional parameters: a position parameter $\mu$ and a scale (or bandwidth) parameter $\sigma$. To avoid the need to optimize these parameters along with the map's coefficients $c_{i}$, one can estimate $\mu$ and $\sigma$ from the empirical quantiles along the marginals of the training samples $\{\co{\boldsymbol{\mathsf{X}}}^{i}\}_{i=1}^N$~\citep{Spantini2022CouplingFiltering}. For the positions $\mu_{i}$, they propose to place each RBF at the empirical quantiles $q_{i/(j+1)}\left(\co{x_{k}}\right)$ for $i=1,\dots,j$, where $j$ is the total number of RBFs along the $k$-th marginal. Based on the resulting $\mu_{i}$, the corresponding scales $\sigma_{i}$ are determined by averaging the distances to neighbouring RBFs. To simplify notation, it can be useful to define local coordinates for each RBF:

\begin{equation}
    \co{x_{k}^{\text{loc},i}} = \frac{\co{x_{k}} - \mu_{i}}{\sqrt{\tau}\sigma_{i}},
    \label{eq:RBF_local_coordinates}
\end{equation}

where the $\tau$ is again twice the Archimedes' constant. Radial basis functions are particularly useful for multimodal distributions because they render the map selectively expressive wherever the target distribution's samples are located. Where a superposition of many high-order polynomials may be required to recover isolated modes, often only a few RBFs may suffice. Besides conventional RBFs, three useful related basis functions are \textit{left edge terms} (LET), \textit{integrated RBFs} (iRBF), and \textit{right edge terms} (RET):

\begin{equation}
    \begin{aligned}
        \operatorname{RBF}\left(\co{x_{k}^{\text{loc},i}}\right) & = \frac{1}{\sqrt{\tau}}\exp\left(-{\co{x_{k}^{\text{loc},i}}}^2\right),\\        \operatorname{iRBF}\left(\co{x_{k}^{\text{loc},i}}\right) & = \frac{1}{2}\left(1 + \operatorname{erf}\left(\co{x_{k}^{\text{loc},i}}\right)\right),\\        \operatorname{LET}\left(\co{x_{k}^{\text{loc},i}};\sigma_{i}\right) & = \frac{1}{2}\left(\sigma_{i}\sqrt{\tau}\co{x_{k}^{\text{loc},i}}\left(1 - \operatorname{erf}\left(\co{x_{k}^{\text{loc},i}}\right)\right) - \frac{4\sigma_{i}}{\sqrt{\tau}}\exp\left(-{\co{x_{k}^{\text{loc},i}}}^{2}\right)\right), \\
        \operatorname{RET}\left(\co{x_{k}^{\text{loc},i}};\sigma_{i}\right) & = \frac{1}{2}\left(\sigma_{i}\sqrt{\tau}\co{x_{k}^{\text{loc},i}}\left(1 + \operatorname{erf}\left(\co{x_{k}^{\text{loc},i}}\right)\right) + \frac{4\sigma_{i}}{\sqrt{\tau}}\exp\left(-{\co{x_{k}^{\text{loc},i}}}^{2}\right)\right).
    \end{aligned}
    \label{eq:RBF_terms}
\end{equation}

\begin{figure}
  \centering
  \includegraphics[width=\textwidth]{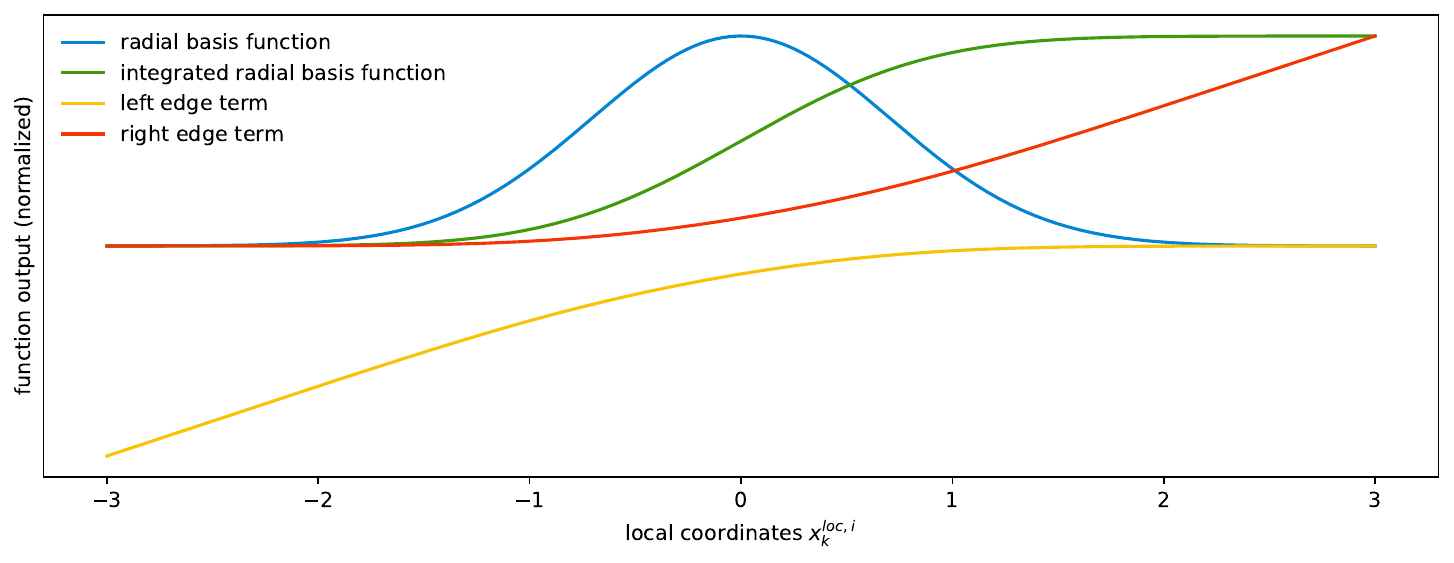}
  \caption{Illustration of different useful RBF basis functions.}
  \label{fig:radial_basis_functions}
\end{figure}

Illustrations of these functions are provided in \cref{fig:radial_basis_functions}. Note that iRBFs, LETs, and RETs are monotone functions, and thus an excellent choice for a \textit{linear separable} map (see \cref{subsubsec:parameterizations}). Other sigmoid-like functions and monotone ``edge terms'' that revert to linear in the tails include logistic and softplus functions, which may be computationally faster.

\subsection{Optimization}\label{subsec:optimization}

With the structure and parameterization of the map component functions $S_{k}$ defined, we may now address the question of how to identify the specific map $\S$ that relates the target distribution $\co{\pi}$ to the reference distribution $\cg{\eta}$. In general, we have two different ways to estimate the map, depending on the type of information available: \textit{maps from samples}, and \textit{maps from densities}. Both follow a very similar approach, but differ fundamentally in the direction in which they define the transport map:

\noindent
\begin{minipage}{.5\textwidth}
  \textbf{Maps from samples}
  \begin{itemize}
      \item require target samples $\co{\x}\sim\co{\pi}$
      \item require evaluations of the reference $\cg{\eta}$
      \item seek a %
      map $\S$ such that $\cg{\S_\sharp\pi} = \cg{\eta}$
  \end{itemize}
\end{minipage}
\begin{minipage}{.5\textwidth}
  \textbf{Maps from densities}
  \begin{itemize}
      \item require reference samples $\cg{\z}\sim\cg{\eta}$
      \item require evaluations of unnormalized target $\co{\tilde{\pi}}$
      \item seek a %
      map $\R$ such that $\co{\R_{\sharp} \eta} = \co{\pi}$
  \end{itemize}
\end{minipage}

\subsubsection{Maps from samples}\label{subsubsec:maps_from_samples}

In the preceding sections, we have often implicitly assumed a \textit{maps from samples} scenario. In other words, the target pdf $\co{\pi}$ is only known through samples
, i.e., we have a collection of samples $\co{\boldsymbol{\mathsf{X}}}^{i} \sim \co{\pi}$. In this case, we seek to minimize the ``distance'' between the target pdf $\co{\pi}$ and the pullback pdf $\co{\smap^{\sharp}\eta}$. While several notions of distance between pdfs are possible, the Kullback--Leibler divergence is commonly used for its computational tractability. For two arbitrary pdfs $p, q$ and a generic RV $\ba$, we define their Kullback--Leibler divergence $D(p\|q)$ as:
\begin{equation}
    D(p\|q) = \mathbb{E}_{p} \left[\log\frac{p(\ba)}{q(\ba)}\right] = \int p(\ba) \log\frac{p(\ba)}{q(\ba)} d\ba.
\end{equation}

Thus, we can formulate our optimization problem as finding the map $\smap_{\text{opt}}$ that minimizes the  difference, measured in terms of the Kullback--Leibler divergence, between the target $\co{\pi}$ and its approximation $\co{\smap^{\sharp}\eta}\approx\co{\pi}$:
\begin{equation}
    \S_{\text{opt}} = \arg \min_{\S\in\mathcal{F}} D(\co{\pi}\|\co{\S^{\sharp}\eta}),
    \label{eq:KLD}
\end{equation}
where $\mathcal{F}$ is some appropriate class of functions  (see \cref{subsec:map_components}). This is often referred to as the minimization of the \textit{forward} Kullback--Leibler divergence. We note that $D(\co{\pi}\|\co{\S^{\sharp}\eta})$ can be expanded as:
\begin{equation}
    D(\co{\pi}\|\co{\S^{\sharp}\eta})  
    = \co{\mathbb{E}_{\pi}}\left[\log\frac{\co{\pi}(\co{\x})}{\co{\S^{\sharp}\eta}(\co{\x})}\right]
     = \co{\mathbb{E}_{\pi}} \left[\log \co{\pi}(\co{\x}) \right] - \co{\mathbb{E}_{\pi}} \left[\log \co{\S^{\sharp}\eta}(\co{\x}) \right], 
     \label{eq:expansion_KL_from_samples}
\end{equation}
where the first term $\co{\mathbb{E}_{\pi}} \left[\log \co{\pi}(\co{\x}) \right]$ does not depend on the map $\smap$. Thus, we can alternatively view \cref{eq:KLD} as an attempt to maximize the log-likelihood (or minimize the negative log-likelihood) of the map's pullback density $\co{\S^{\sharp}\eta}\approx\co{\pi}$ over the target density $\co{\pi}$:
\begin{equation}
    \S_{\text{opt}} = \arg \min_{\S\in\mathcal{F}}\co{\mathbb{E}_{\pi}}\left[-\log \co{\S^{\sharp}\eta}(\co{\x})\right].
    \label{eq:argmax_pullback_samples}
\end{equation}

\begin{figure}
  \centering
  \includegraphics[width=\textwidth]{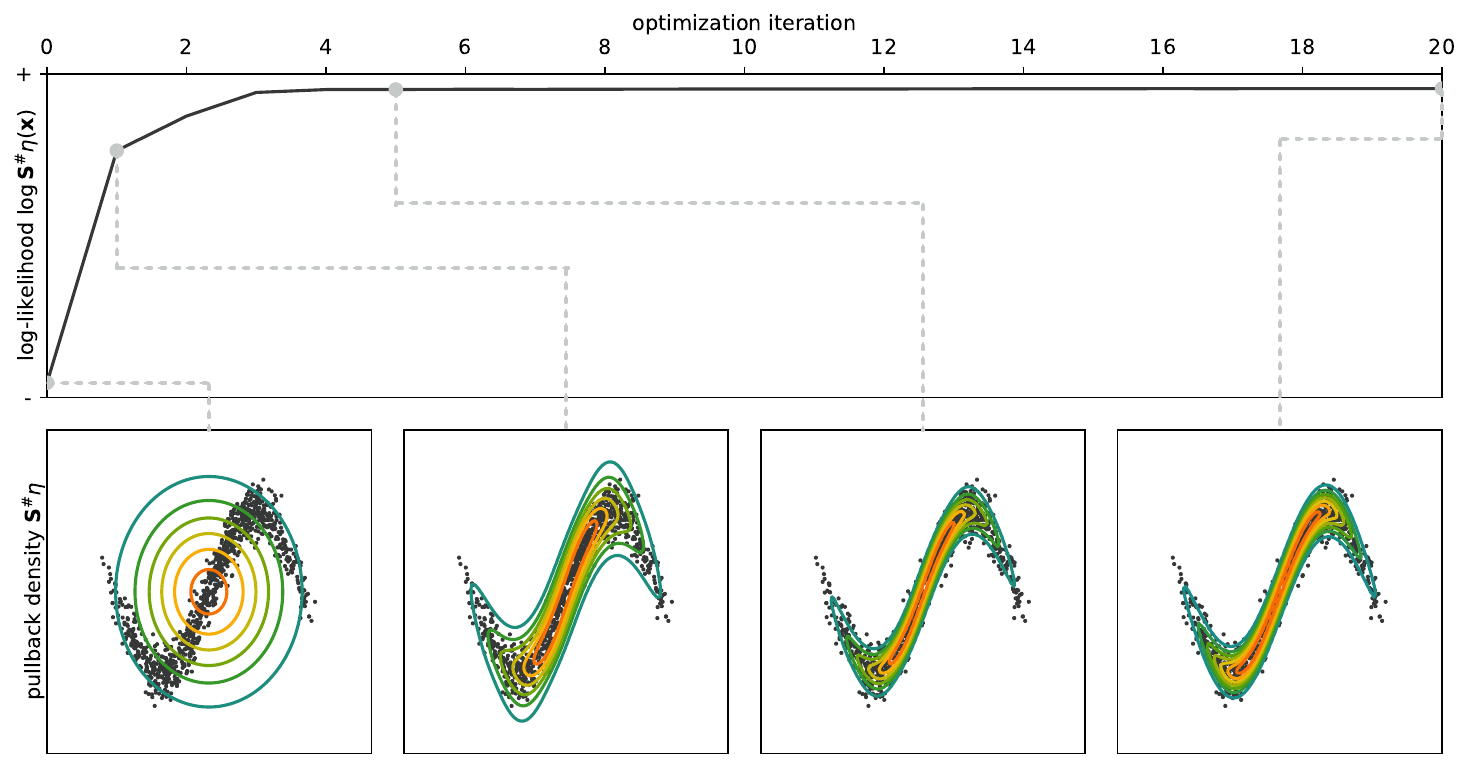}
  \caption{The optimization objective of \textit{maps from samples} maximizes the log-likelihood of a set of fixed samples $\co{\x}\sim\co{\pi}$ over the map's pullback density $\co{\S^{{\sharp}}\eta}\approx\co{\pi}$. In practical terms, the optimization seeks maps $\S$ which mold the reference pdf $\cg{\eta}$ to the target samples $\co{\x}$.}
  \label{fig:optimize_from_samples}
\end{figure}

An illustration of the optimization process is provided in \cref{fig:optimize_from_samples}. Substituting in \cref{eq:change_of_variables_multidimensional} for $\co{\S^{\sharp}\eta}(\co{\x})$ into the cost function \cref{eq:argmax_pullback_samples} and expanding the logarithms yields:
\begin{equation}
    \co{\mathbb{E}_{\pi}}\left[-\log \co{\S^{\sharp}\eta}(\co{\x})\right] = \co{\mathbb{E}_{\pi}}\left[-\log \cg{\eta}(\S(\co{\x}))  -  \log |\det \mathbf{\nabla}_{\co{\x}} \S(\co{\x}) |\right].
\end{equation}

Due to the lower triangular structure of $\S$, we know that $\mathbf{\nabla}_{\co{\x}} \S$ is a lower triangular matrix. As established in \cref{eq:triangular_determinant}, the determinant of a lower triangular matrix is given by the product of its diagonal entries:
\begin{equation}
    \log \det \mathbf{\nabla}_{\co{\x}} \S(\co{\x})=\log\left[\prod_{k=1}^{K}\frac{\partial S_{k}(\co{\x})}{\partial \co{x_{k}}}\right]=\sum_{k=1}^{K}\log\frac{\partial S_{k}(\co{\x})}{\partial \co{x_{k}}}.
    \label{eq:triangular_log_det}
\end{equation}

Mind that, due to the monotonicity of $S_{k}$ in $\co{x_{k}}$ (see \cref{subsec:monotonicity}), the derivative $\partial_{\co{x_k}}S_{k}$ will always be positive, and the logarithm in the right-hand side sum of \cref{eq:triangular_log_det} will be defined.
Plugging this identity into the cost function, we only now enforce that our reference $\cg{\eta}$ is multivariate Gaussian by substituting its pdf into the expectation:
\begin{equation}
    \co{\mathbb{E}_{\pi}} \left[-\log \co{\S^{\sharp}\eta}(\co{\x})\right]  = \co{\mathbb{E}_{\pi}} \left[- \log\frac{1}{\tau^{D/2}} + \frac{1}{2}\sum_{k=1}^{K}S_{k}(\co{\x})^{2} - \sum_{k=1}^{K}\log\frac{\partial S_{k}(\co{\x})}{\partial \co{x_{k}}}\right].
\end{equation}
where $\tau$ is again twice the Archimedes' constant. Recognizing that the first term is constant with respect to $\S$, it can be discarded from the objective function. Using $N$ samples from the target distribution $\co{\boldsymbol{\mathsf{X}}}^{i}\sim \co{\pi}$ and merging the two sums over $K$, we obtain a Monte Carlo approximation of the loss function:
\begin{equation}
    \mathcal{J}(\S) = \sum_{i=1}^{N}\sum_{k=1}^{K}\left(\frac{1}{2}S_{k}(\co{\boldsymbol{\mathsf{X}}}^{i})^{2} - \log\frac{\partial S_{k}(\co{\boldsymbol{\mathsf{X}}}^{i})}{\partial \co{x_{k}}} \right).
\end{equation}

From this expression, we can recognize that the summands of the full objective function $\mathcal{J}(\S)$ are independent of each other. As a consequence, we define separate objective functions $\mathcal{J}_{k}(S_{k})$ for each map component $S_{k}$:
\begin{equation}
\mathcal{J}_{k}(S_{k}) = \sum_{i=1}^{N}\left(\frac{1}{2}S_{k}(\co{\boldsymbol{\mathsf{X}}}^{i})^{2} - \log\frac{\partial S_{k}(\co{\boldsymbol{\mathsf{X}}}^{i})}{\partial \co{x_{k}}} \right).
\label{eq:objective_function_from_samples}
\end{equation}

\cref{eq:objective_function_from_samples} may then be minimized with off-the-shelf software developed for other optimization and machine learning tasks. If the linear separable formulation in \cref{subsec:monotonicity} is chosen, it is only necessary to optimize the monotone part $f$ of each map component $S_{k}$ numerically, as the coefficients for the nonmonotone terms $g$ can be derived in closed form. More detail on this is provided in \cref{appendix:separable_optimization}. Note that since the objective functions in \cref{eq:objective_function_from_samples} are independent of each other, the map component functions $S_{k}$ can be optimized in arbitrary order, even in parallel. Each objective function $\mathcal{J}_k$ balances a first quadratic term that drives samples $S_{k}(\co{\boldsymbol{\mathsf{X}}}^{i})^{2}$ to zero with a second term that acts as a log-barrier function to prevent null diagonal entries in the Jacobian of the transport map \citep{LeProvost2021AFlows}.

\subsubsection{Maps from densities}\label{subsubsec:maps_from_densities}

\begin{figure}
  \centering
  \includegraphics[width=\textwidth]{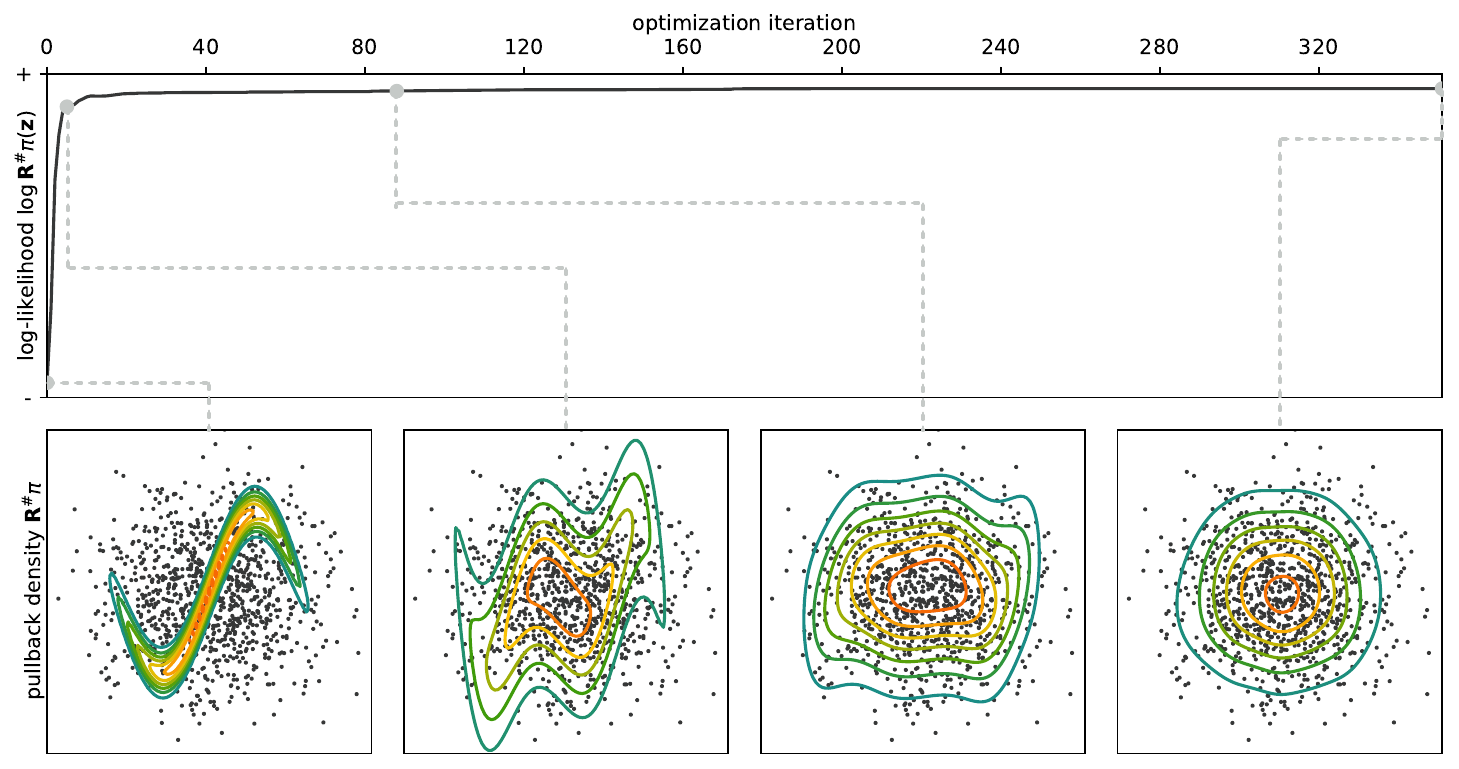}
  \caption{The optimization objective of \textit{maps from densities} maximizes the log-likelihood of a set of fixed samples $\cg{\boldsymbol{\mathsf{Z}}}\sim\cg{\eta}$ over the map's pullback density $\cg{\R^{{\sharp}}\tilde{\pi}}\approx\cg{\eta}$. In practical terms, the optimization seeks maps $\R$ which mold the unnormalized target pdf $\co{\tilde{\pi}}$ to the reference samples $\cg{\boldsymbol{\mathsf{Z}}}$.}
  \label{fig:optimize_from_densities}
\end{figure}

Alternatively, we choose a different optimization strategy in the case that we can evaluate the target pdf $\co{\pi}$, but only up to a constant of proportionality; i.e., we have access to evaluating density $\co{\tilde{\pi}} = \co{m} \co{\pi}$ for some unknown $\co{m}>0$. A key difference when learning a \textit{map from density}
is that we choose to define the triangular map in the opposite direction as before: where we previously considered a forward map $\S$ that maps target samples $\co{\boldsymbol{\mathsf{X}}}$ to reference samples $\cg{\boldsymbol{\mathsf{Z}}}$, we now define a forward map $\R$ mapping reference samples $\cg{\boldsymbol{\mathsf{Z}}}$ to target samples $\co{\boldsymbol{\mathsf{X}}}$. %
Correspondingly, the pushforward distribution $\co{\R_{{\sharp}}\eta}$ now approximates the target distribution $\co{\pi}$ via the forward map $\R$, and the pullback distribution $\cg{\R^{{\sharp}}\pi}$ now approximates the reference distribution $\cg{\eta}$ via the inverse map $\R^{-1}$. To find the optimal $\R_{\text{opt}}$, we seek to minimize the Kullback--Leibler divergence between the reference distribution $\cg{\eta}$ and the pullback distribution $\cg{\R^{\sharp}\pi}$ \citep{Marzouk2017SamplingIntroductionPUB}:
\begin{equation}
    \R_{\text{opt}}\in \arg \min_{\R\in\mathcal{F}}D(\cg{\eta}\|\cg{\R^{\sharp}\pi}),
    \label{eq:argmax_pullback_density}
\end{equation}

where $\mathcal{F}$ is some appropriate class of functions (see \cref{subsubsec:basis_functions}) and $\in$ is used to suggest multiple minimizers (see \citet{Baptista2023OnMaps} for a discussion of why this could happen). This is often referred to as the minimization of the \textit{reverse} Kullback--Leibler divergence. We visualize minimization of \cref{eq:argmax_pullback_density} in \cref{fig:optimize_from_densities}. Similar to \cref{eq:expansion_KL_from_samples} in the maps from samples scenario, we expand $D(\cg{\eta}\|\cg{\R^{\sharp}\pi})$ as:
\begin{equation}
    D(\cg{\eta}\|\cg{\R^{\sharp}\pi})  
    = \cg{\mathbb{E}_{\eta}}\left[\log\frac{\cg{\eta}(\cg{\z})}{\cg{\R^{\sharp}\pi}(\cg{\z})}\right]
     = \cg{\mathbb{E}_{\eta}}\left[\log \cg{\eta}(\cg{\z}) \right] - \cg{\mathbb{E}_{\eta}}\left[ \log \cg{\R^{\sharp}\pi}(\cg{\z}) \right], 
\end{equation}

where the first term $\cg{\mathbb{E}_{\eta}}\left[\log \cg{\eta}(\cg{\z}) \right]$ does not depend on the map $\R$. Thus, \cref{eq:argmax_pullback_density} can be viewed as an attempt to maximize the log-likelihood (or minimize the negative log-likelihood) of the map's pullback density $\cg{\R^{\sharp}\pi}\approx\cg{\eta}$ over the reference density $\eta$:
\begin{equation}
    \R_{\text{opt}}\in \arg \min_{\R\in\mathcal{F}}\cg{\mathbb{E}_{\eta}}\left[-\log \cg{\R^{\sharp}\pi}(\cg{\z})\right].
    \label{eq:objective_raw_maps_from_densities}
\end{equation}

We may now substitute the pullback density from the change-of-variables formula (\cref{eq:change_of_variables_multidimensional}):
\begin{equation}
    \cg{\eta}\left(\cg{\z}\right)\approx\cg{\R^{{\sharp}}\pi}\left(\cg{\z}\right) = \co{\pi}\left(\R\left(\cg{\z}\right)\right)\left|\det \mathbf{\nabla}_{\cg{\z}}\R\left(\cg{\z}\right)\right|.
\end{equation}

Equivalent to \cref{eq:triangular_log_det}, we express the log determinant of the forward map $\R$ as a product of its diagonal entries:
\begin{equation}
    \log \det \mathbf{\nabla}_{\cg{\z}} \R(\cg{\z})=\log\left[\prod_{k=1}^{K}\frac{\partial R_{k}(\cg{\z})}{\partial \cg{z_{k}}}\right]=\sum_{k=1}^{K}\log\frac{\partial R_{k}(\cg{\z})}{\partial \cg{z_{k}}}.
\end{equation}

Substituting the two equations above into the cost function of \cref{eq:objective_raw_maps_from_densities}, we obtain:
\begin{equation}
    \cg{\mathbb{E}_{\eta}}\left[-\log \cg{\R^{\sharp}\pi}(\cg{\z})\right] =  - \cg{\mathbb{E}_{\eta}} \left[ \log \co{\pi}(\R(\cg{\z})) \right] - \cg{\mathbb{E}_{\eta}} \left[ \sum_{k=1}^{K}\log\frac{\partial R_{k}(\cg{\z})}{\partial \cg{z_{k}}} \right].
\end{equation}

Since we assume we only have access to an unnormalized target density $\co{\tilde{\pi}}$, we can expand the first term $\cg{\mathbb{E}_{\eta}} \left[ \log \co{\pi}(\R(\z)) \right]$ using the identity $\co{\pi}=\co{\tilde{\pi}}/\co{m}$, where $\co{m}$ is the unknown normalization factor:
\begin{equation}
    \cg{\mathbb{E}_{\eta}}\left[-\log \cg{\R^{\sharp}\pi}(\cg{\z})\right] = - \cg{\mathbb{E}_{\eta}} \left[ \log \co{\tilde{\pi}}(\R(\z)) \right]  + \cg{\mathbb{E}_{\eta}} \left[ \log \co{m} \right] - \cg{\mathbb{E}_{\eta}} \left[ \sum_{k=1}^{K}\log\frac{\partial R_{k}(\cg{\z})}{\partial \cg{z_{k}}} \right].
\end{equation}

Recognizing that the normalization factor $\co{m}$ does not depend on the map $\R$, we can discard it from the objective function:
\begin{equation}
    \R_{\text{opt}}\in \arg \min_{\R\in\mathcal{F}}- \cg{\mathbb{E}_{\eta}} \left[ \log \co{\tilde{\pi}}(\R(\z)) \right] - \cg{\mathbb{E}_{\eta}} \left[ \sum_{k=1}^{K}\log\frac{\partial R_{k}(\cg{\z})}{\partial \cg{z_{k}}} \right].
    \label{eqn:argmax_pullback_density_unnorm}
\end{equation}

Using $N$ samples from the reference distribution $\cg{\boldsymbol{\mathsf{Z}}}^i \sim \cg{\eta}$, we obtain a Monte Carlo approximation of the loss function:
\begin{equation}
    \mathcal{J}(\R) = \sum_{i=1}^{N}\left(-\log\co{\tilde{\pi}}\left(\R\left(\cg{\boldsymbol{\mathsf{Z}}}^{i}\right)\right)-\sum_{k=1}^{K}\log\frac{\partial R_{k}(\cg{\boldsymbol{\mathsf{Z}}}^{i})}{\partial \cg{z_{k}}}\right).
    \label{eq:Final_Objective_Map_from Densities}
\end{equation}

This yields the objective function for optimizing maps from densities. Two comments are in order regarding this optimization problem. First, observe that opposed to maps from samples, we have to first generate an ensemble of $N$ i.i.d.\ samples $\cg{\boldsymbol{\mathsf{Z}}}^i \sim \cg{\eta}$. 
Second, since we now find a map $\rmap$ to pulls the reference back to the target, we still avoid evaluating the inverse map $\rmap^{-1}$ when calculating the loss $\mathcal{J}(\rmap)$ for some candidate map $\rmap\in\mathcal{F}$.
Opposed to optimizing a map from samples (\cref{subsubsec:maps_from_samples}), however, this objective function cannot generally be subdivided into individual objectives for each constituent map component $R_{k}$. As such, all map components are usually found via a single optimization problem.

We note that there are other loss functions that we could consider beyond the Kullback--Leibler divergence, for instance the variance diagnostic \citep{ElMoselhy2012BayesianMaps,richter2020vargrad}.%
We do not consider these options here in greater detail, %
as the Kullback--Leibler objective is widely used and quite tractable. Moreover, in the maps-from-samples setting, a minimizer of the forward Kullback--Leibler divergence corresponds to a \textit{maximum likelihood estimator} of the transport map within the chosen function class; this link is useful for both interpretation and theory~\citep{wang2022minimax}.

\subsubsection{Regularization}
When the number of samples $N$ is small relative to either the target's dimensionality $K$ or the number of parameters for the map components, using more complex maps (i.e., increasing the number of the parameters) risks overfitting.
In turn, this might lead to numerical instabilities. To prevent these issues, we can introduce an $L^1$ or $L^2$ regularization penalty to the objective functions introduced in the preceding section. 
In the maps from samples (\cref{subsubsec:maps_from_samples}) case, we take map component $S_k$ for some fixed $k$, and let $c_i$ parameterize this component via, e.g., polynomial coefficients. The regularized objective function would be:

\begin{equation}
\begin{aligned}
    &\mathcal{J}_{k}^{\text{L1}}(S_{k}) = \sum_{i=1}^{N}\left(\frac{1}{2}S_{k}(\co{\boldsymbol{\mathsf{X}}}^{i})^{2} - \log\frac{\partial S_{k}(\co{\boldsymbol{\mathsf{X}}}^{i})}{\partial \co{x_{k}}} \right) + \sum_{i=1}^{j} \lambda_{i}|c_{i}| && L^1 \ \text{regularization}, \\
    &\mathcal{J}_{k}^{\text{L2}}(S_{k}) = \sum_{i=1}^{N}\left(\frac{1}{2}S_{k}(\co{\boldsymbol{\mathsf{X}}}^{i})^{2} - \log\frac{\partial S_{k}(\co{\boldsymbol{\mathsf{X}}}^{i})}{\partial \co{x_{k}}} \right) + \sum_{i=1}^{j} \lambda_{i} c_{i}^{2} && L^2 \ \text{regularization}. \\
\end{aligned}
\label{eq:objective_function_reg}
\end{equation}

where we add a $\lambda$-weighted penalty term to the objective functions, which penalizes non-zero parameters $c_{i}$ for each of the $i=1,\ldots,P_k$ basis functions of map component $S_k$ (see \cref{subsubsec:basis_functions}). This creates an optimum closer to where the parameters are zero, which can be interpreted as penalizing map complexity.
The regularization factors $\lambda_{i}$ are user-specified hyperparameters that should be tuned to the problem at hand, ensuring that the penalization is large enough to have an effect and small enough to avoid excessive biasing. For maps from densities, the optimization objective follows an equivalent structure to \cref{eq:objective_function_reg}. Similar to other applications with regularization parameters, a bit of experimentation is required to find a suitable balance between regularization and bias.

In practice, the use of $L^1$ regularization is a common strategy to discover sparsity in the triangular map, which we recall corresponds to conditional independence (see \cref{subsubsec:conditional_independence}). In many cases, it is more efficient to impose sparsity and parsimony by construction. We will discuss practical strategies to learn a parsimonious degree of map complexity in \cref{subsec:adaptation}. 

\section{Practical heuristics}\label{sec:practice}

With the theoretical foundations and implementation-related details established, we now discuss a few important features and tricks that help to further enhance the potential of triangular transport maps. 

\subsection{Composite maps for conditional sampling}\label{subsec:composite_maps}

\tmlr{Composite maps are used to improve conditional samples from imperfect maps.}
For complex target distributions $\co{\pi}$, it is often infeasible to define and optimize a map of sufficient complexity to capture all features of $\co{\pi}$. %
In such cases, the map $\smap$ will fail to completely normalize the target $\co{\pi}$, making certain features persist in the pushforward $\cg{\smap_\sharp\pi}$. At the same time, the pullback $\co{\smap^\sharp\eta}$ may not be a good approximation of $\co{\pi}$. 
\Cref{fig:imperfect_maps} illustrates this for two different levels of map complexity.

This imperfection has important consequences for Bayesian inference with triangular transport maps. Since the (conventional) map's conditional inverse only samples a conditional of the pullback distribution $\co{\S^{{\sharp}}\eta}$, not of the real target $\co{\pi}$, the quality of the resulting posterior samples will depend on the mismatch between $\co{\S^{{\sharp}}\eta}$ and $\co{\pi}$. 
Yet, we are often given a sample $\co{\boldsymbol{\mathsf{X}}} = (\co{\boldsymbol{\mathsf{X}}_{1:k}}, \co{\boldsymbol{\mathsf{X}}_{k+1:K}})$ of a joint distribution $\pi(\co{\x_{1:k}},\co{\x_{k+1:K}})$ and would like to create a high-quality sample $\co{\boldsymbol{\mathsf{X}}_{k+1:K}^{*}}$ conditioned on some fixed $\cb{\x_{1:k}^*}$. To do this, we apply a \textit{composite map} to each sample $\co{\x}$, defined as:
\begin{equation}
    \co{\x^*_{k+1:K}} = 
    \S_{k+1:K}^{-1}(\cdot\ ;\,\cb{\x_{1:k}^{*}})\circ\S_{k+1:K}(\co{\x_{k+1:K}};\,\co{\x_{1:k}}).
    \label{eq:composite_map}
\end{equation}
Note that $\cb{\x_{1:k}^{*}}$ and $\co{\x_{1:k}}$ are two different objects: The \cb{former} comes from the real world (i.e., it is a fixed value, independent from the random sample $\co{\x}$). For instance, $\cb{\x_{1:k}^{*}}$ may be specific measurement values on which we want to condition $\co{\pi}$. By contrast, the \co{latter} are jointly sampled with $\co{\x_{k+1:K}}$ from $\co{\pi}$.
While this may initially seem complicated, the key idea of composite maps is as follows:

\begin{figure}[ht!]
  \centering
  \includegraphics[width=\textwidth]{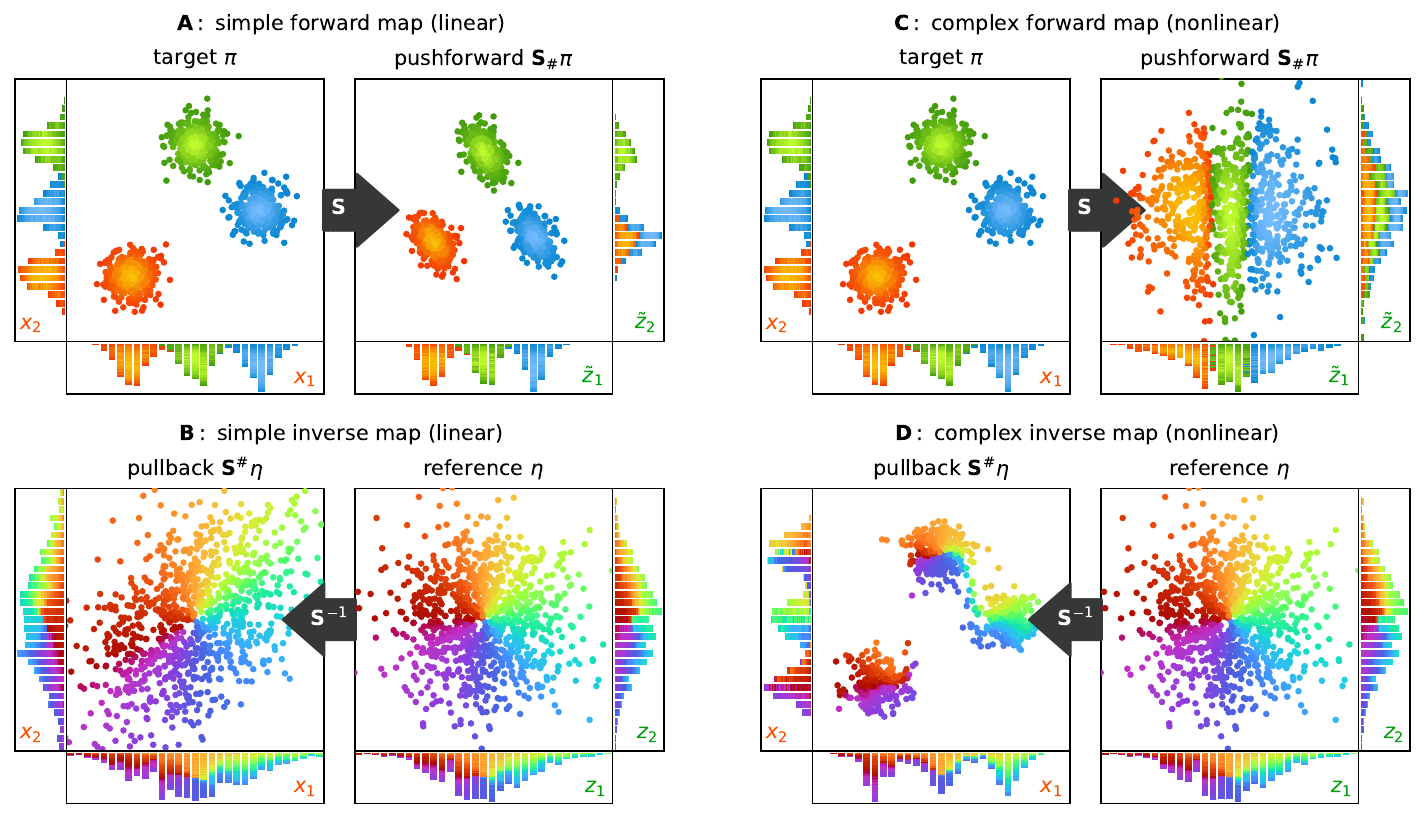}
  \caption{Simple maps capture only simple features. (A) For a non-Gaussian target $\co{\pi}$, a simple linear map does not yield a Gaussian pushforward $\cg{\smap_{\sharp}\pi}\neq\cg{\eta}$. (B) Likewise, its pullback $\co{\smap^{\sharp}\eta}\neq\co{\pi}$ does not provide a good approximation to the target. (C \& D) More complex nonlinear maps yield better approximations. Color in (A) and (C) indicates the sample's original cluster membership, whereas in (B) and (D) color indicates a sample position's angle w.r.t.\ $\cg{\eta}$'s mean.}
  \label{fig:imperfect_maps}
\end{figure}

\begin{figure}[ht!]
  \centering
  \includegraphics[width=\textwidth]{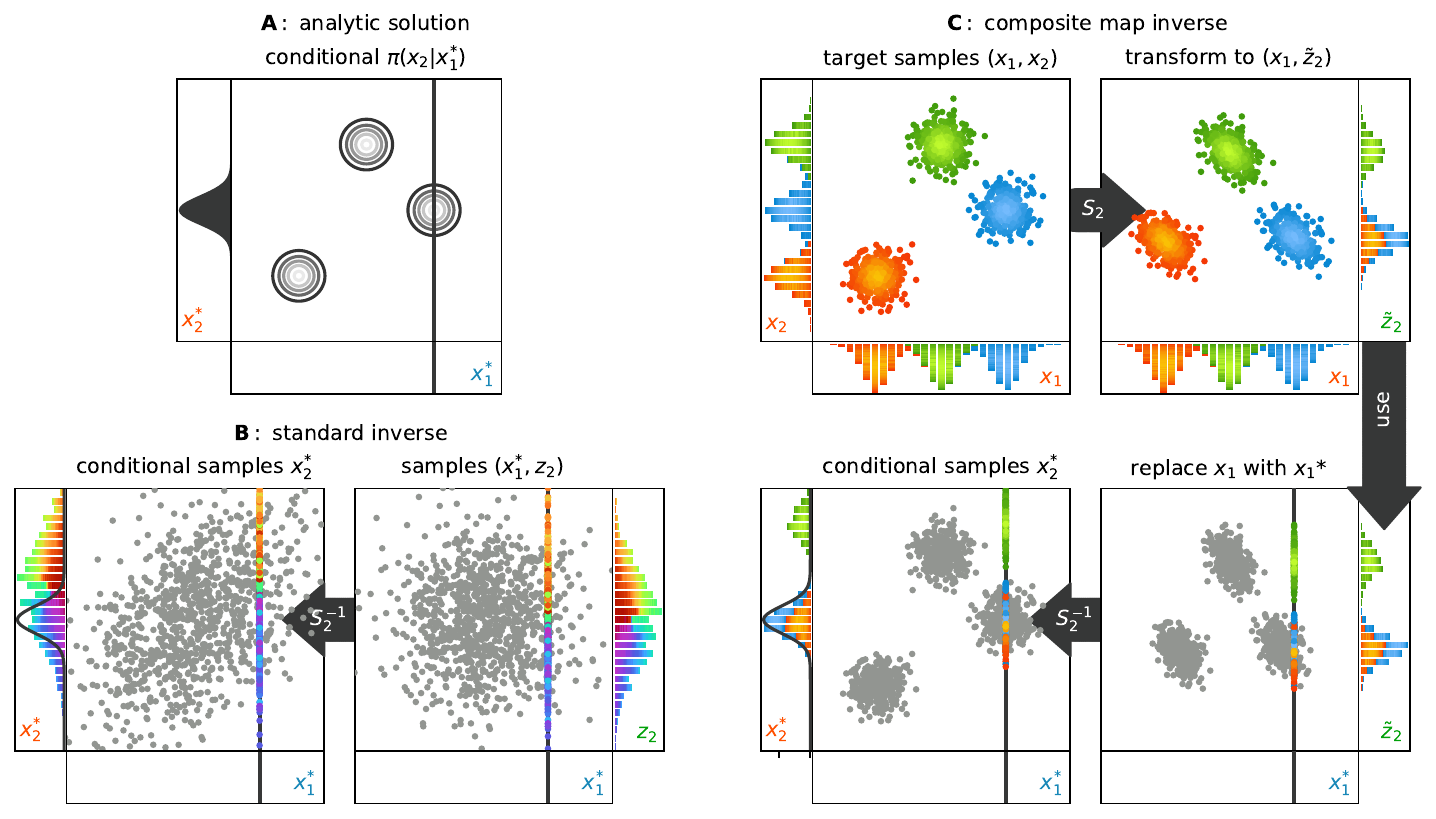}
  \caption{Composite maps yield better conditionals for imperfect maps. (A) The true solution for the conditioning operation. For the simple linear map, 
  (B) using reference samples $\cg{z}\sim\cg{\eta}$ directly samples a conditional of the pullback $\co{\S^{{\sharp}}\eta}\neq\co{\pi}$, which here is unrepresentative of the true conditional. (C) Using pushforward samples $\cg{\tilde{z}}\sim\cg{\S_{{\sharp}}\pi}$ instead as part of a composite inversion yields a better approximation to the true conditional. The black line in the $\co{x_{2}^{*}}$ histograms in B, C, and D shows the true conditional distribution from subplot A for comparison.}
  \label{fig:composite_maps}
\end{figure}

\begin{enumerate}
    \item We have an approximate map $\smap$. Its \textbf{forward evaluation} thus does not transform samples from the true target $\co{\x}\sim\co{\pi}$ into samples from the reference $\cg{\z}\sim\cg{\eta}$, but instead yields samples from the pushforward $\cg{\tilde{\z}}\sim\cg{\S_{{\sharp}}\pi} \neq \cg{\eta}$. These pushforward samples $\cg{\tilde{\z}}$ preserve features of the target distribution $\co{\pi}$ the map did not capture. For instance, a linear map $\smap$ can only shift, scale, and rotate the target distribution $\co{\pi}$, which is insufficient if $\co{\pi}$ is non-Gaussian (\cref{fig:imperfect_maps}A).
    \item Likewise, an approximate \textbf{inverse map} does not transform samples from the reference $\cg{\z}\sim\cg{\eta}$ into samples from the true target $\co{\x}\sim\co{\pi}$, instead yielding samples from the pullback $\co{\tilde{\x}}\sim\co{\S^{{\sharp}}\eta}\neq\co{\pi}$ which lack any features of $\co{\pi}$ that the map has not captured. For instance, the pullback $\co{\S^{{\sharp}}\eta}$ of a linear map $\smap$ only samples a Gaussian approximation of the target distribution (\cref{fig:imperfect_maps}B).
    \item Applying an invertible approximate forward map, directly followed by its inverse, cancels the map's approximation error and always restores the original target samples. By construction, we know that the composition of $\smap$ with its inverse $\smap^{-1}$ maps a sample to itself, i.e., if $\cg{\tilde{\z}} = \smap(\co{\x})\sim\cg{\smap_\sharp\pi}$, then $\co{\x} = \smap^{-1}(\cg{\tilde{\z}})$.
    \item This restoration of unresolved features persists in part during the conditioning operation. Instead of reference samples $\cg{\z}\sim\cg{\eta}$ (\cref{fig:composite_maps}B), we might thus use pushforward samples $\cg{\tilde{\z}}\sim\cg{\smap_{{\sharp}}\pi}$ to preserve some features of $\co{\pi}$. This generally reduces the map's approximation error (\cref{fig:composite_maps}C); see Proposition~11 in \citet{Baptista2022ProbabilisticTransport}.
\end{enumerate}

\cref{fig:composite_maps} illustrates the composite map's preservation of uncaptured features during conditional inversion for a Gaussian mixture target $\co{\pi}$ (\cref{fig:composite_maps}A). The pullback approximation $\co{\S^{{\sharp}}\eta}$ from a linear transport map is Gaussian, and thus clearly inadequate for a multimodal Gaussian mixture target $\co{\pi}$. Inverting the map with reference samples $\cg{z}\sim\cg{\eta}$ only samples conditionals of the (multivariate Gaussian) pullback (\cref{fig:composite_maps}B). However, using reference samples from the pushforward $\cg{z}\sim\cg{\S_{{\sharp}}\pi}$ preserves complex features of the target $\co{\pi}$, yielding better conditional samples $\co{x_{2}^{*}}$ (\cref{fig:composite_maps}C).

Keep in mind that if we instead use a sufficiently complex, nonlinear map formulation (e.g., \cref{fig:imperfect_maps}D), the map approximation retrieves the target distribution sufficiently well. In this case, there is little practical difference between the use of reference $\cg{\z}\sim\cg{\eta}$ or pushforward samples $\cg{\tilde{\z}}\sim\cg{\S_{{\sharp}}\pi}$, as both distributions will be quite similar ($\cg{\S_{{\sharp}}\pi} \approx \cg{\eta}$). In the general case, however, we may not be certain about the quality of our map approximation without explicit validation. In consequence, we recommend the use of composite maps as a safer strategy for conditional sampling.

\paragraph{\tmlr{Implementation pseudocode}}

\tmlr{In Appendix~\ref{appendix:pseudocodes} we provide pseudocodes with both the linear combination of basis functions and the rectified integration formulations for map evaluation. We also provide pseudocode for evaluating inverse and composite maps.}

\subsection{Map adaptation}\label{subsec:adaptation}

Identifying a suitable degree of map complexity can be a challenging task. For one, conditional independence structures (see \cref{subsubsec:conditional_independence}) are not always known a-priori. Similarly, finding a suitable degree of complexity for each map component to achieve an optimal bias-variance trade-off is not an easy task. To address these issues, we may draw on map adaptation algorithms. \tmlr{Examples of and pseudocode for such algorithms are provided by~\citet{Baptista2023OnMaps,lunde2026adaptive}.}

\begin{figure}
  \centering
  \includegraphics[width=\textwidth]{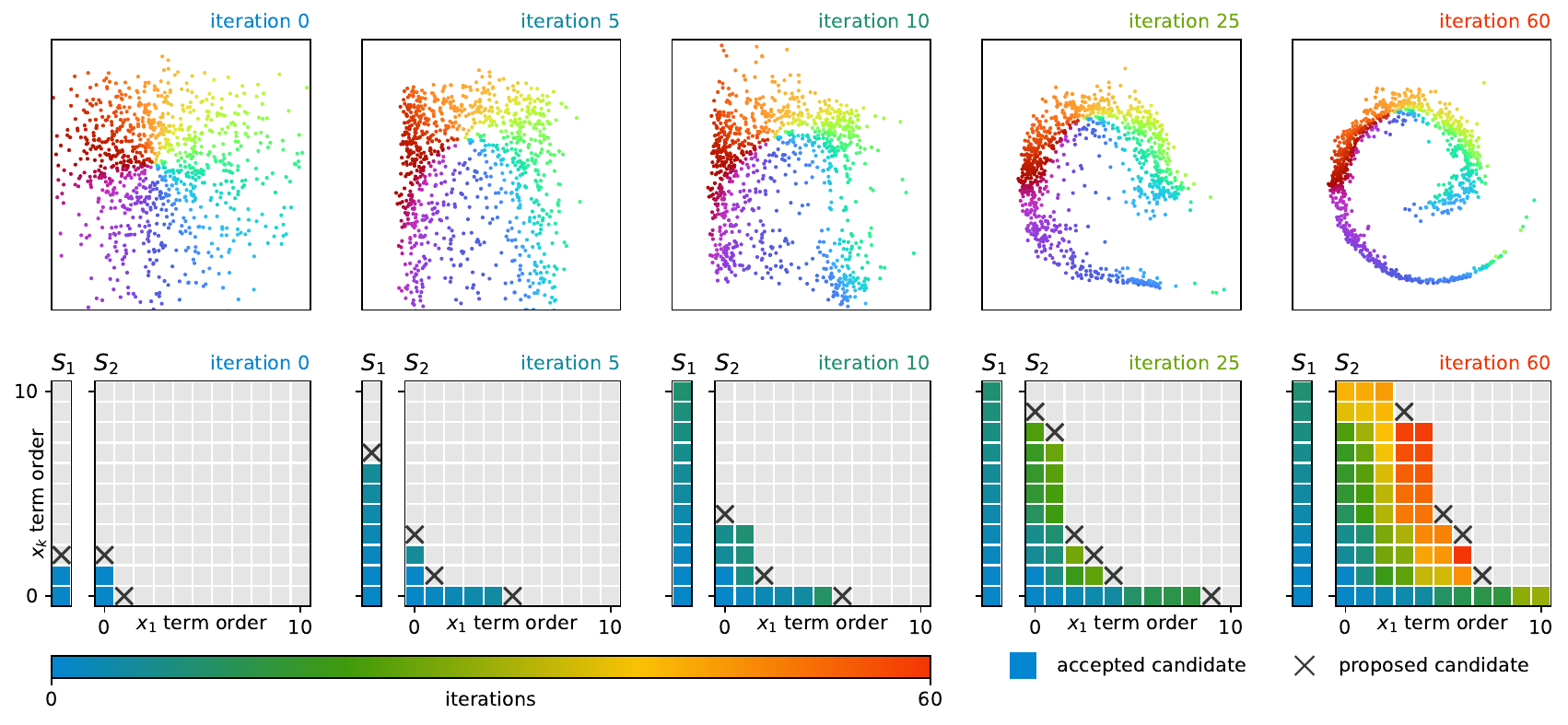}
  \caption{Adaptive transport maps gradually expand the complexity and expressiveness of the map components, adding terms corresponding to the largest absolute gradient of the objective function (\cref{eq:objective_function_from_samples}) among a set of candidate terms. This identifies the necessary degree of map complexity gradually. The lower-row grids illustrate the order and composition of the map components at each iteration. Each filled cell is an accepted candidate, with coordinates $(3,5)$ in the $S_2$ block corresponding to a term $\mathcal{H}_{3}(x_{1})\mathcal{H}_{5}(x_{2})$.}
  \label{fig:map_adaptation}
\end{figure}

In brief, 
the method starts with a triangular map that is viable and particularly simple; for example, one may use the identity map $S_{k}(\co{x_k})=\co{x_k}$.
The algorithm then proposes a list of candidate basis functions for each $S_{k}$ (\cref{fig:map_adaptation}). These candidate terms either (i) extend the dependence of $S_{k}$ to a previous state dimension $\co{x_{1:k-1}}$, or (ii) incrementally increase the complexity of an existing dependence by adding higher degrees of nonlinearity.

To determine which candidate basis functions are added to the map, the algorithm in~\citet{Baptista2023OnMaps} calculates the gradient of the optimization objective function (see \cref{subsec:optimization}) with respect to the candidates' coefficients. It then adds the candidate corresponding to the steepest derivative. The idea behind this approach is that candidate terms with steep optimization objective gradients promise to improve the map's approximation significantly, whereas those with flat gradients add little beyond complexity. The algorithm then proposes new candidates to the expanded map further and repeats the procedure until a user defined stopping criterion is met.

This process gradually expands the complexity and expressiveness of each map component. Since this process is iterative, this adaptation algorithm can be expensive if the number of candidate basis functions is large, as is the case for $S_{k}$ with many variable dependencies, particularly if the map parameterization permits cross-terms (\cref{fig:map_adaptation}). The use of maps without cross-terms can drastically reduce this computational demand, 
as it would restrict the expansion and exploration of candidates to basis functions corresponding to the outer edges in \cref{fig:map_adaptation}, making a two-dimensional result look like the letter ``L.''

%% file: sections/outlook.tex
\subsection{State of the art}
Measure transport methods comprise an active and quickly evolving field in statistics and machine learning. Within this broader field, triangular maps occupy an important niche: transparent and versatile, they provide a powerful set of tools for conditional sampling from limited information. This, in turn, makes them highly useful for Bayesian inference.

To summarize, triangular transport methods have a number of important \textbf{advantages}: %
\begin{enumerate}
    \item \textbf{Parsimony}: The ability to fine-tune the parameterization of a triangular map, coupled with the sparse variable dependence that a triangular map inherits from conditional independence, allows triangular transport to be adapted to the computational demands, data availability, and distributional complexity of a given problem---increasing efficiency and accuracy.
    
    \item \textbf{Numerical convenience}: Triangular maps are simple to learn and invert, and many common parameterizations provide optimization guarantees for the problem of learning maps from samples (see, e.g., \cref{appendix:separable_optimization} and \citet{Baptista2023OnMaps}). 
    
    \item \textbf{Transparency}: Triangular maps provide a close correspondence between the map component functions $S_{k}$ and a specific factorization of the target distribution $\co{\pi}$. This link makes it easy to predict the impact of changes to the map on the resulting statistical model, 
    and often allows us to decompose an inference or sampling task into smaller problems that are themselves interpretable. %

\end{enumerate}

At the same time, triangular maps face some important challenges. Ensuring the scalability of triangular transport with the dimension of the target distribution demands exploiting conditional independence to produce a sparse map, which in turn requires an appropriate variable ordering. 
We reiterate that, under mild assumptions, a triangular map exists between any target and reference~\citep[][Section 2.3]{Santambrogio2015OptimalMathematicians}, which includes any permutation of our target random variables. In practice, however, these maps may vary in sparsity and other notions of complexity: 
see \cref{subsubsec:conditional_independence} and \cref{fig:cross_terms}, respectively, and note that the rotated ``wavy'' distribution in \cref{fig:cross_terms} actually corresponds to a different variable ordering. %
There are many approaches to finding  orderings that maximize sparsity: for example, minimum degree algorithms \citep[e.g., ][]{Amestoy1996AnAlgorithm,Cuthill1969ReducingMatrices} from graph theory, or other algorithms inspired by sparse Cholesky factorizations~\citep{baptista2024learning,schaefer2021sparse}. Setting aside sparsity, however, the impact of variable ordering on the difficulty of \textit{approximating} a given map component function $S_k$ is generally more difficult to predict, as it depends on finer properties of the target distribution at hand. %

\subsection{Broader context within transport}
There are many ways to parameterize a triangular transport map that are not discussed here---for example, performing a closed-form integration over a parameterization of the target pdf. This can be done using squared polynomials~\citep{zanger2024sequential}, parameterizing the square-root of the pdf~\citep{dolgov2020approximation}, or even a composition of polynomials parameterizing this square-root pdf~\citep{cui2022deep}, often in the context of numerical tensor methods that permit efficient integration. Regarding the \textit{approximation} of triangular transport maps, there are several works elucidating approximation rates by, e.g., polynomials or neural networks \citep{zech2022sparseI,zech2022sparseII,baptista2024approximation,westermann2023measure}; other works analyse the \textit{statistical} consistency and convergence of triangular maps learned from finite samples~\citep{wang2022minimax,pmlr-v151-irons22a}.

More generally, triangular maps sit within the broader field of measure transport, which has seen remarkable computational advances in recent years. Many other approaches to constructing transport maps exchange triangular structure for different assumptions or desiderata.
For example, \textit{optimal} transport methods seek a mapping between two distributions that is as ``close'' to the identity function as possible, where closeness is encoded by a particular integrated \textit{transport cost} (for instance, the distance between the input and output of the map)~\citep{peyre2019computational}. %
Many works seek to approximate optimal transport maps by searching over an appropriate function class---e.g., by parameterizing classes of convex functions (rather than our monotone triangular formulation) and writing the map as the gradient of such a function~\citep{makkuva2020optimal,wang2023efficient}. %
There are some connections between triangular maps and optimal transport maps. First, each component $S_k$ of a monotone triangular map is optimal in its last (scalar) input $x_k$, conditioned on the first $k-1$ inputs $x_{1:k-1}$. \textit{Conditional} optimal transport maps \citep{tabak2021data,carlierVectorQuantileRegression2016,doi:10.1137/23M1581546,pooladian2025conditional} generalize this idea from strictly triangular to block triangular structure. %
As described in \citet{Knothe_to_Brenier_2010}, triangular maps also arise as the limit of optimal transport maps obtained with increasingly anisotropic quadratic transport cost.

Normalizing flows~\citep{papamakarios2021normalizing} are another widely used class of transport methods, which typically do not seek to approximate some canonical (i.e., optimal or triangular) map, but rather use a composition of many simpler invertible transformations to construct a transport map, increasing the number of functions (layers) in the composition until the desired expressivity is reached. Autoregressive normalizing flows in fact use triangular maps as a building block, with parameterizations ranging 
in complexity from relatively simple to the complex rectified formulations discussed in \cref{subsec:monotonicity}~\citep{wehenkel2019unconstrained,jaini2019sum}. %
These flows interleave such triangular layers with permutations of the variables, which mitigates ordering issues described above at the cost of sacrificing the conditional independence and sparsity properties of triangular maps. %
Such constructions are partly inspired by earlier autoregressive models that directly parameterize marginal conditional densities~\citep{bond2021deep}. %
Finally, many contemporary flow-based generative models \citep{song2020score,albergo2023stochastic,lipman2023flow,liu2023flow} can be understood as dynamical or continuous-time representations of transport maps; here the transport is represented by the flow map of a system of ordinary differential equations (ODEs) or by the corresponding evolution of densities under a system of stochastic differential equations. Some models of this kind explicitly employ triangular structure~\citep{heng2021gibbs}, but triangular maps have also proven useful in the analysis of more general neural ODE models~\citep{marzouk2024distribution}. %

Within this broader landscape, what is the \textbf{role of triangular transport}? We believe that the greatest strength of triangular transport lies in its extraordinary capacity for \textit{parsimony}: Triangular maps may not only naturally exploit conditional independence---a special %
property among transport-based %
samplers---but they allow us to fine-tune the level of nonlinearity with which we resolve statistical dependencies. In consequence, these methods allow us to tailor the maps precisely to the system's demands by building some structure of the target distribution into the map itself. In the small sample size regime, we believe this property holds the key to challenging the 
prevalence of linear methods, which %
are still the state of the art %
for practical settings from meteorological data assimilation to subsurface parameter inference.

\subsection{Where to go from here} %

As a relatively new method, much of the potential of triangular maps has yet to be realized. This includes a number of promising research directions in applications, methodology, and theory.
To unlock the full potential of their parsimony, a key effort for the operationalization of triangular transport is the development of efficient \textbf{adaptation strategies}. One of the greatest challenges of the parameterizations discussed in this paper is their scaling with the dimension of the target distribution. One way to mitigate this ``curse of dimensionality'' is to only include terms one knows will help, which is the problem that the adaptive algorithms above attempt to tackle. As briefly discussed in \cref{subsec:adaptation}, these algorithms automatically seek to identify parsimonious maps for general inference problems. Increasing the efficiency of these methods will ease their applicability to higher-dimensional systems. Promising research avenues include exploring connections to graph structure learning %
algorithms that 
estimate conditional independence structure~\citep{baptista2024learning,liaw2025learning,drton2017structure,zheng2020learning} %
or developing information criteria~\citep{lunde2025ensemble,konishi1996generalised} %
to help identify suitable levels of map complexity.

A closely-related subject is research into the properties, advantages, and drawbacks of different \textbf{parameterizations} of triangular transport. While we have explored many practical details in this tutorial, we have still only scratched the surface of this topic. A promising avenue is to focus on known computational bottlenecks of existing parameterizations. 
As a simple example, it is known the $\mathrm{erf}$ function (see \cref{eq:RBF_terms}) has unstable evaluations near the extremes of its domain and is often computationally expensive; exploring more efficient alternatives is of cross-cutting utility. Similarly, some parameterization heuristics such as centering RBFs at marginal quantiles of the target distribution remain largely ad hoc, especially given strong dependencies in $\co{\pi}$; what can we do to improve these heuristics? 
What kinds of parameterizations should be employed when the distributions at hand are heavy-tailed: do the ``edge terms'' of maps need to increase sub- or super-linearly (depending on the direction of the map), and does the use of composite maps relieve the need to model tails when performing conditional sampling?
More broadly, is there a practical role for \textit{nonparametric} representations of transport maps \citep{pooladian2021entropic} in high-dimensional problems? And whether in parametric or nonparametric settings, how can \textit{regularization} or penalization schemes for learning maps from limited information be more rigorously designed and scaled?
Relatedly, it is common nowadays to have multi-fidelity data sources (e.g., models that can be resolved at different discretization levels); how can we efficiently learn %
maps via samples and model evaluations of differing fidelities?
\tmlr{How can we learn maps from partially observed samples or censored data (i.e., when some samples have missing entries)? Can triangular transport be used in this and other settings for data imputation?}

Furthermore, there are many open questions at the intersection of triangular transport with numerical analysis, relevant to probabilistic modelling and uncertainty quantification. For example, there has been recent work marrying deterministic quadrature or quasi-Monte Carlo methods with transport~\citep{Cui2025Quasi-MonteInterpolation,klebanov2023quadrature,liu2024transport}, promising higher-order convergence rates for the approximation of expectations weighted by complex distributions. It is interesting to consider how triangular structure could be leveraged in this setting and to understand broader links to cubature~\citep{cools1997constructing} or other high-dimensional integration schemes. 

Finally, as we have argued that sparsity provides a route to scalability, it is important to understand the impact of \textit{approximate} sparsity in transport, and hence \textit{approximate} conditional independence in continuous non-Gaussian distributions---e.g., the error incurred by discarding weak conditional dependencies. (See~\citet{johnson2007recursive,jog2015model} for analyses in the Gaussian case.) This line of work has direct links to \textit{localization} methods widely used in data assimilation, whose analysis is a topic of much current research \citep{al2024non,tong2023localized,gottwald2024localized}. Moreover, sparsity is only one form of low dimensionality. Other work seeks scalable inference and probabilistic modelling via explicit dimension reduction---for instance, by searching for low-dimensional subspaces that capture important interactions between parameters and data, or by identifying low-dimensional ``updates'' from prior to posterior in the Bayesian setting~\citep{baptista2022gradient,zahm2022certified,constantine2015active,fukumizu2007statistical,hardle1989investigating,Cui2014Likelihood-informedProblems}. %
These notions correspond naturally to specific transport representations~\citep{brennan2020greedy,cao2024lazydino,cui2025subspace,chen2025rotated} 
and enable the solution of problems that would otherwise---without the exploitation of structure---be intractable. %

\subsection{Toolboxes}

We hope that this tutorial provided an accessible introduction to the theory and implementation of triangular transport. As we conclude, we want to leave you, the reader, with some concrete numerical tools to explore your own applications of triangular transport---if you do not want to code your own. The code to reproduce the figures and examples in this tutorial is available on 
\href{https://github.com/MaxRamgraber/Triangular-Transport-Tutorial}{GitHub} and has been implemented using the \href{https://github.com/MaxRamgraber/Triangular-Transport-Toolbox}{Triangular Transport Toolbox}, written in Python. 
Another implementation is available via the \href{https://measuretransport.github.io/MParT/}{\textbf{M}onotone \textbf{Par}ameterization \textbf{T}oolbox (MParT)}~\citep{parno2022mpartJOSS}, which is written in C++ but features bindings to Python, Julia, and MATLAB, with a focus on core functionality and efficiency. A package often used is \href{https://transportmaps.mit.edu/docs/}{TransportMaps}, a Python toolbox with many features appearing in modern papers.
Regarding triangular transport parameterizations beyond those presented here, there is code to reproduce sum-of-squares polynomial transport in Julia using \href{https://github.com/benjione/SequentialMeasureTransport.jl}{SequentialMeasureTransport.jl}, square-root pdf tensor-based methods using the \href{https://github.com/DeepTransport/deep-tensor}{Deep Tensor Toolbox} in MATLAB, and \href{https://github.com/DeepTransport/deep-tensor-py/tree/main}{\texttt{deep\_tensor}} gives a PyTorch implementation of similar tensor-based pdf approximations.

%% file: sections/supplementary1.tex
\newcommand{\bfPsi}{\boldsymbol{\Psi}}
\newcommand{\bfP}{\co{\mathbf{P}}}
\newcommand{\bfPnonk}{\co{\mathbf{P}_{k}^{\text{non}}}}
\newcommand{\bfPmonk}{\co{\mathbf{P}_{k}^{\text{mon}}}}
\newcommand{\bfQ}{\co{\mathbf{Q}_k}}
\newcommand{\bfR}{\co{\mathbf{R}_k}}
\newcommand{\bfb}{\co{\mathbf{b}}^{i}_{\co{k}}}
\newcommand{\bfc}{\mathbf{c}}
\newcommand{\nonk}{_k^{\text{non}}}
\newcommand{\monk}{_k^{\text{mon}}}
\newcommand{\xkm}{\co{\boldsymbol{\mathsf{X}}_{1:k-1}}}
\newcommand{\xkmi}{\co{\boldsymbol{\mathsf{X}}}_{\co{1:k-1}}^i}
\newcommand{\xk}{\co{\boldsymbol{\mathsf{X}}_k}}
\newcommand{\xki}{\co{\boldsymbol{\mathsf{X}}}_{\co{k}}^{i}}
\newcommand{\xeli}{\co{\boldsymbol{\mathsf{X}}}^i}

\section{Optimizing linear separable maps}\label{appendix:separable_optimization}

As established in Section~\ref{subsubsec:parameterizations}, choosing a \textit{linear separable} map parameterization permits more efficient map optimization. First, for the purpose of this material we denote gradient of some scalar-valued function $r(\mathbf{z})$ with respect to $\mathbf{z}$ evaluated at $\mathbf{z}^*$ as $\nabla_\mathbf{z} r\big|_{\mathbf{z}^*}$ and the derivative of a univariate function $s(z)$ with respect to scalar $z$ evaluated at $z^*$ as $\partial_z s\big|_{z^*}$. Then, recall for a separable map that is \textit{linear in the coefficients}, each map component function $S_{k}$ is defined as:

\begin{equation}
\begin{aligned}
    S_{k}\left(\co{x_{1}},\dots,\co{x_{k-1},\co{x_{k}}}\right) &= g(\co{\x_{1:k-1}}) + f(\co{x_{k}})\\
    &= \co{c_{k,1}^{\text{non}}}\psi_{k,1}^{\text{non}}(\co{\x_{1:k-1}}) + \dots + \co{c_{k,m}^{\text{non}}}\psi_{k,m}^{\text{non}}(\co{\x_{1:k-1}}) + \co{c_{k,1}^{\text{mon}}}\psi_{k,1}^{\text{mon}}(\co{x_{k}}) + \dots + \co{c_{k,n}^{\text{mon}}}\psi_{k,n}^{\text{mon}}(\co{x_{k}}) \\
    &= \boldsymbol{\Psi}_{k}^{\text{non}}(\co{\x_{1:k-1}}) \co{\mathbf{c}_{k}^{\text{non}}} + \boldsymbol{\Psi}_{k}^{\text{mon}}(\co{x_{k}})\co{\mathbf{c}_{k}^{\text{mon}}},
\end{aligned}
\label{apeq:separable_map_discrete}
\end{equation}

where $\psi_{k,j}^{\text{non}}:\mathbb{R}^{k-1}\to\mathbb{R}$ and $\psi_{k,j}^{\text{mon}}:\mathbb{R}\to\mathbb{R}$ are the $j$-th basis functions of map component $S_{k}$, associated with the nonmonotone and monotone terms, respectively. Then, $\boldsymbol{\Psi}_{k}^{\text{non}}:\mathbb{R}^{k-1}\to \mathbb{R}^{1\times m}$ and $\boldsymbol{\Psi}_{k}^{\text{mon}}:\mathbb{R}\to \mathbb{R}^{1\times n}$ are vectors of basis function evaluations for $g$ and $f$, and $\co{\mathbf{c}_{k}^{\text{non}}}\in\mathbb{R}^{ m \times 1}$ and $\co{\mathbf{c}_{k}^{\text{mon}}}\in\mathbb{R}^{ n \times 1}$ are the corresponding column vectors of coefficients. Therefore, the expressions $\boldsymbol{\Psi}_k^{\text{non}}(\co{\x_{1:k-1}})\co{\mathbf{c}^{\text{non}}_k}$ and $\boldsymbol{\Psi}_k^{\text{mon}}(\co{x_k})\co{\mathbf{c}_k^{\text{mon}}}$ are both inner product functions of $\co{\x}$, i.e. scalar-valued. For \textit{maps from samples} (Section~\ref{subsubsec:maps_from_samples}), we consider samples $\co{\boldsymbol{\mathsf{X}}}^{1},\ldots,\co{\boldsymbol{\mathsf{X}}}^{N}\sim\co{\pi}$ and recall the following optimization objective for $S_{k}$:

\begin{equation}
    \mathcal{J}_{k}(S_{k}) = \sum_{i=1}^{N}\left(\frac{1}{2}S_{k}(\xeli)^{2} - \log\partial_{\co{x_k}}S_k\big|_{\xeli} \right).
\end{equation}

Plugging in Equation~\ref{apeq:separable_map_discrete}, we obtain:
\begin{equation}
\begin{aligned}
    \mathcal{J}_k(S_k) &= \sum_{i=1}^N \frac{1}{2}\left[\bfPsi\nonk(\xkmi)\co{\bfc\nonk} + \bfPsi\monk(\xki)\co{\bfc\monk} \right]^2 - \log\partial_{\co{x_k}}\left[\bfPsi\monk(\xkmi)\co{\bfc\nonk} + \bfPsi\monk(\xk)\co{\bfc\monk}\right]_{\xki} \\
    &= \sum_{i=1}^N \frac{1}{2}\left[\bfPsi\nonk(\xkmi)\co{\bfc\nonk} + \bfPsi\monk(\xki)\co{\bfc\monk} \right]^2 - \log\partial_{\co{x_k}}\bfPsi\monk\big|_{\xki}\co{\bfc\monk}.
\end{aligned}\label{apeq:plugin}
\end{equation}

Note that the samples $\xeli$ are defined by the target distribution $\co{\pi}$, and are thus fixed during optimization. In consequence, the basis function evaluation vectors $\bfPsi\nonk$ and $\bfPsi\monk$ are also fixed for a given map parameterization (see Section~\ref{subsubsec:parameterizations}), and are thus independent of the coefficients $\co{\bfc\nonk}$ and $\co{\bfc\monk}$. We can simplify Equation~\ref{apeq:plugin} further by absorbing the sum over the first term, and defining a new variable for the partial derivative in the second term. To this end, we form matrices $\bfPnonk\in\mathbb{R}^{N\times m}$ and $\bfPmonk\in\mathbb{R}^{N\times n}$, and vectors $\bfb\in\mathbb{R}^{1\times n}$, which are each defined element-wise as:
\[[\bfPnonk]_{ij}=\psi^{\text{non}}_{k,j}(\xkmi), \quad [\bfPmonk]_{ij}=\psi^{\text{mon}}_{k,j}(\xki),\quad [\bfb]_j = \partial_{\co{x_k}}\psi^{\text{mon}}_{k,j}\big|_{\co\xki},\]
where the $ij$ entry is the $j$th indexed basis function evaluated at sample index $i$ for both matrices $\bfPnonk$ and $\bfPmonk$, and the log of the derivative of $\bfPsi\monk$ with respect to $\xk$ is evaluated at $\xki$ for the vectors $\bfb$. As above, these matrices and vectors can be pre-computed. The optimization objective now simplifies further to:
\[\mathcal{J}_k(S_k) = \frac{1}{2}\left\|\bfPnonk\co{\bfc\nonk} + \bfPmonk\co{\bfc\monk}\right\|^2 - \sum_{i=1}^N \log\bfb\co{\bfc\monk}.\]
This function is similar to that of an objective for an interior point method and, remarkably, becomes quadratic in $\bfc\nonk$. At this point, we add L2 (i.e., Tikhonov) regularization on both $\co{\bfc\monk}$ and $\co{\bfc\nonk}$ according to the guidance in Section~\ref{sec:implementation}.
\begin{equation}\label{apeq:reg_loss}
\mathcal{J}_k(S_k;\lambda) = \frac{1}{2}\left\|\bfPnonk\co{\bfc\nonk} + \bfPmonk\co{\bfc\monk}\right\|^2 - \sum_{i=1}^N \log\bfb\co{\bfc\monk} + \frac{\lambda}{2}(\|\co{\bfc\nonk}\|^2 + \|\co{\bfc\monk}\|^2).
\end{equation}
In practice, this means the optimal coefficients $\co{\widehat{\bfc}\nonk}$ for the nonmonotone basis function evaluations minimizing \eqref{apeq:reg_loss} must satisfy:
\begin{align}
0 &\equiv \nabla_{\co{\bfc\nonk}} \mathcal{J}_k(S_k;\lambda) \big|_{\co{\widehat{\bfc}\nonk}}\nonumber\\
&= \nabla_{\co{\bfc\nonk}}\frac{1}{2}\left\|\bfPnonk\co{\bfc\nonk} + \bfPmonk\co{\bfc\monk}\right\|^2\bigg|_{\co{\widehat{\bfc}\nonk}}+\lambda\co{\bfc\nonk}\nonumber\\
&= ({\bfPnonk}^{\top}\bfPnonk+\lambda\mathbf{I})\co{\widehat{\bfc}\nonk} + {\bfPnonk}^\top\bfPmonk\co{\bfc\monk}.\label{apeq:nonk_loss}
\end{align}
In consequence, for a given choice of coefficients parameterizing the monotone functions, $\co{\bfc\monk}$, we can find the optimal choice of $\co{\widehat{\bfc}\nonk}$ by solving the \textit{normal equations}; for this scenario, we assume that $m < N$, i.e. the number of samples surpasses the number of basis functions (and thus ${\bfPnonk}^\top\bfPnonk$ is full rank).
The normal equations are a well-studied class of problems in numerical linear algebra~\citep{Trefethen2022NumericalEdition}.
Assuming linearly independent basis functions, this affords a solution for $\co{\widehat{\bfc}\nonk}$ as a function of $\co{\bfc\monk}$ given as:
\begin{equation}\label{apeq:cnon_hat}
\begin{aligned}
({\bfPnonk}^\top\bfPnonk+\lambda\mathbf{I})\co{\widehat{\bfc}\nonk} &= -{\bfPnonk}^\top\bfPmonk\co{\bfc\monk} \\ \co{\widehat{\bfc}\nonk} &= -\underbrace{({\bfPnonk}^\top\bfPnonk+\lambda\mathbf{I})^{-1}{\bfPnonk}^\top}_{\co{\mathbf{M}_{k,\lambda}}}\bfPmonk\co{\bfc\monk}\\
&\stackrel{\text{def}}{=} -\co{\mathbf{M}_{k,\lambda}}\bfPmonk\co{\bfc\monk}.
\end{aligned}
\end{equation}
Following best practices from numerical linear algebra, this inversion should not be computed explicitly; rather, one should use a numerical solver for the systems induced. Then, substituting Expression~\eqref{apeq:cnon_hat} into the optimization objective, we obtain a new objective for $\co{\bfc\monk}$ (i.e., entirely independent from the nonmonotone coefficients),
\begin{align}
    \mathcal{J}\monk(\co{\bfc\monk};\lambda) &= \frac{1}{2}\|-\bfPnonk\co{\mathbf{M}_{k,\lambda}}\bfPmonk\co{\bfc\monk} + \bfPmonk\co{\bfc\monk}\|^2 - \sum_{i=1}^N \log\bfb\co{\bfc\monk} + \frac{\lambda}{2}(\|\co{\widehat{\bfc}\nonk}\|^2+\|\co{\bfc\monk}\|^2)\nonumber\\
    &= \frac{1}{2}\|\underbrace{(\mathbf{I}-\bfPnonk\co{\mathbf{M}_{k,\lambda}})\bfPmonk}_{\co{\mathbf{A}_{k,\lambda}}}\co{\bfc\monk}\|^2  - \sum_{i=1}^N\log\bfb\co{\bfc\monk} + \frac{\lambda}{2}(\|\underbrace{\co{\mathbf{M}_{k,\lambda}}\bfPmonk}_{\co{\mathbf{D}_{k,\lambda}}}\co{\bfc\monk}\|^2 + \|\co{\bfc\monk}\|^2)\nonumber\\
    &\stackrel{\text{def}}{=} \frac{1}{2}\|\co{\mathbf{A}_{k,\lambda}\bfc\monk}\|^2 - \sum_{i=1}^N\log\bfb\co{\bfc\monk} + \frac{\lambda}{2}(\|\co{\mathbf{D}_{k,\lambda}}\co{\bfc\monk}\|^2 + \|\co{\bfc\monk}\|^2),
\end{align}
where $\co{\mathbf{A}_{k,\lambda}}$, $\co{\mathbf{D}_{k,\lambda}}$ and $\bfb$ can be precomputed prior to the optimization routine via evaluation of the basis functions. Remarkably, this method translates the original loss function into a very simple constrained convex optimization problem:
\begin{equation}\label{apeq:overdet_soln}
\boxed{
\co{\widehat{\bfc}\monk} = \argmin_{\co{\bfc\monk}\geq \mathbf{0}} \frac{1}{2}\|\co{\mathbf{A}_{k,\lambda}\bfc\monk}\|^2 - \sum_{i=1}^N \log\bfb\co{\bfc\monk} + \frac{\lambda}{2}(\|\co{\mathbf{D}_{k,\lambda}}\co{\bfc\monk}\|^2 + \|\co{\bfc\monk}\|^2),\quad \co{\widehat{\bfc}\nonk} = -\co{\mathbf{M}_{k,\lambda}}\bfPmonk\co{\widehat{\bfc}\monk}.}
\end{equation}
Note that the (element-wise) constraint of $\co{\bfc\monk}\geq \mathbf{0}$ is vital to maintain monotonicity; this can be enforced explicitly during optimization using particular optimization algorithms, e.g., L-BFGS-B (i.e. Low-memory BFGS with box constraints), implicitly by constructing an objective with a log-barrier term, or employing a convex transformation of the optimization objective (e.g. optimize over $\mathbf{p}\monk := \log\co{\bfc\monk}$). Since we often will have coefficients with zero values, the first methodology might be preferable (though no empirical results appear here). This optimization objective, notably, requires no evaluations of the map during optimization, that is to say, the optimization is as fast as the combination of the implementation of linear algebra algorithms called and the optimization routine used. %

\newcommand{\bfU}{\co{\mathbf{U}_k}}
\newcommand{\bfV}{\co{\mathbf{V}_k}}
\newcommand{\bfSig}{\co{\boldsymbol{\Sigma}_k}}

In the case where we have more parameters than samples, it is worth noting that using this formulation is remarkably sensitive to $\lambda$ because ${\bfPnonk}^\top\bfPnonk$ is no longer full rank. Thus the calculation of $\co{\mathbf{M}_{k,\lambda}}$ solves a system with possibly poor numerical properties (the industry of methods for \textit{ill-conditioned} systems is dedicated to such problems). For vanishingly small $\lambda$, this corresponds to the problem of overfitting and the fact that we have infinite choices of $\co{\widehat{\bfc}\nonk}$ for any given choice of $\co{\bfc\monk}$.

\section{Pseudo-codes}\label{appendix:pseudocodes}
\tmlr{Below we introduce several short pseudo-codes for some methods discussed in the paper. Excepting \cref{alg:forward_map}, all the algorithms are defined for any parameterization of a fully-triangular map. On the other hand, map evaluation is shown for the two types of parameterizations shared in this paper; other approximation classes (i.e., beyond basis functions) can be substituted for the nonmonotone function $g$, premonotone function $\hat{g}$, or monotone function $f$; further, other references shared in this work consider alternative parameterizations altogether.}

\SetKwInOut{Input}{Input}
\SetKwInOut{Output}{Output}
\SetKwInput{Assert}{assert}
\begin{algorithm}[!ht]
\SetAlgoLined
\DontPrintSemicolon
\setstretch{1.15}
\Input{Map $\S$, target sample $\co{\boldsymbol{\mathsf{X}}_{1:K}}$, monotone and nonmonotone coefficients $\{(\mathbf{c}_k^{\mathrm{mon}},\,\mathbf{c}_k^{\mathrm{non}})\}_{k=1}^K$.}
\Output{Pushforward sample $\cg{\widetilde{\boldsymbol{\mathsf{Z}}}_{1:K}}$}
\For{$k \leftarrow 1$ \KwTo $K$}{
    Evaluate nonmonotone function $g(x_{1:k-1};\,\mathbf{c}^{\mathrm{non}}_k) = \sum_{i=1}^{M} c_{ik}^{\mathrm{non}} \psi_i(\co{x_{1:k-1}})$\;
    Define function $G_k:\mathbb{R}\to\mathbb{R}$ as the mapping $t \mapsto\hat{g}(\co{x_{1:k-1}}, t;\,\mathbf{c}^{\mathrm{mon}}_k) = \sum_{i=1}^M c^{\mathrm{mon}}_{ik} \psi_i(\co{x_{1:k-1}},t)$\;
    \Switch{Parameterization type}{
        \uCase{Linear combination of monotone functions}{
            \Assert{$G_k(t)$ is monotone for all $t$}
            Define $f(\co{x_{1:k}};\,\mathbf{c}^{\mathrm{mon}}) = G_k (\co{x_k})$\;
        }
        \uCase{Monotonicity through integration}{
            Calculate or estimate $f(\co{x_{1:k}};\,\mathbf{c}^{\mathrm{mon}}) = \int_0^{\co{x_k}} r(G_k (t))\,\mathrm{d}t$\;
        }
    }
    Create pushforward sample $\cg{\widetilde{\boldsymbol{\mathsf{Z}}}_{k}} = S_k(\co{{\boldsymbol{\mathsf{X}}}_{1:k}}) = g(\co{x_{1:k-1}};\,\mathbf{c}^{\mathrm{non}}_k) + f(\co{x_{1:k}};\,\mathbf{c}^{\mathrm{mon}}_k)$\;
}

\caption{Forward map evaluation for implementations provided in \cref{sec:implementation}}
\label{alg:forward_map}
\end{algorithm}

\begin{algorithm}[!ht]
\SetAlgoLined
\DontPrintSemicolon
\setstretch{1.15}

\Input{Trained transport map $\S$, reference sample $\cg{\boldsymbol{\mathsf{Z}}_{1:K}}$}
\Output{Pullback sample $\co{\widetilde{\boldsymbol{\mathsf{X}}}_{1:K}}$}
\For{$k \leftarrow 1$ \KwTo $K$}{
    Define residual function $H_k:\mathbb{R}\to\mathbb{R}$ as the map $t\mapsto S_k(\co{\widetilde{\boldsymbol{\mathsf{X}}}_{1:k-1}},t) - \cg{\boldsymbol{\mathsf{Z}}_{k}}$\;
    One-dimensional root finding: Find $\co{\widetilde{\boldsymbol{\mathsf{X}}}_k}$ such that $H_{k}(\co{\widetilde{\boldsymbol{\mathsf{X}}}_k}) = 0$\;
}

\caption{Inverse map evaluation}
\label{alg:inverse_map}
\end{algorithm}

\begin{algorithm}[!ht]
\SetAlgoLined
\DontPrintSemicolon
\setstretch{1.15}

\Input{Trained lower block $\S_{k+1:K}$ of map $\S$, conditioning sample $\cb{\boldsymbol{\mathsf{X}}_{1:k}^{*}}$, reference sample $\cg{\boldsymbol{\mathsf{Z}}_{k+1:K}}$}
\Output{Conditional pullback sample $\co{\widetilde{\boldsymbol{\mathsf{X}}}_{k+1:K}^{*}}$}

\For{$j \leftarrow k+1$ \KwTo $K$}{
    Define residual function $H_k:\mathbb{R}\to\mathbb{R}$ as the map $t\mapsto S_k(\cb{\boldsymbol{\mathsf{X}}_{1:k}^{*}},\co{\widetilde{\boldsymbol{\mathsf{X}}}_{k+1:j-1}},t) - \cg{\boldsymbol{\mathsf{Z}}_{j}}$\;
    One-dimensional root finding: Find $\co{\widetilde{\boldsymbol{\mathsf{X}}}_k}$ such that $H_{k}(\co{\widetilde{\boldsymbol{\mathsf{X}}}_k}) = 0$\;
  
}

\caption{Conditional inverse map evaluation}
\label{alg:conditional_inverse_map}
\end{algorithm}

\begin{algorithm}[!ht]
\SetAlgoLined
\DontPrintSemicolon
\setstretch{1.15}
\Input{Trained lower block $\S_{k+1:K}$ of map $\S$, conditioning sample $\cb{\boldsymbol{\mathsf{X}}_{1:k}^{*}}$, target sample $\co{\boldsymbol{\mathsf{X}}_{1:K}}$}
\Output{Conditional sample $\co{\widetilde{\boldsymbol{\mathsf{X}}}_{k+1:K}^{*}}$}

\For{$j \leftarrow k+1$ \KwTo $K$}{
    Generate pushforward sample $\cg{\widetilde{\boldsymbol{\mathsf{Z}}}_{j}} = S_{j}(\co{{\boldsymbol{\mathsf{X}}}_{1:k}},\co{{\boldsymbol{\mathsf{X}}}_{k+1:j-1}}, \co{{\boldsymbol{\mathsf{X}}}_j})$\;

    Define residual function $H_k:\mathbb{R}\to\mathbb{R}$ as the map $t\mapsto S_k(\cb{\boldsymbol{\mathsf{X}}_{1:k}^{*}},\co{\widetilde{\boldsymbol{\mathsf{X}}}_{k+1:j-1}},t) - \cg{\widetilde{\boldsymbol{\mathsf{Z}}}_{j}}$\;

    One-dimensional root finding: Find $\co{\widetilde{\boldsymbol{\mathsf{X}}}_k}$ such that $H_{k}(\co{\widetilde{\boldsymbol{\mathsf{X}}}_k}) = 0$\;
}

\caption{Composite maps for conditional inverse map evaluation}
\label{alg:composite_maps}
\end{algorithm}

%% file: references.bib
@article{doi:10.1137/23M1581546,
    title = {Conditional Sampling with Monotone {GANs}: From Generative Models to Likelihood-Free Inference},
    year = {2024},
    journal = {SIAM/ASA Journal on Uncertainty Quantification},
    author = {Baptista, Ricardo and Hosseini, Bamdad and Kovachki, Nikola and Marzouk, Youssef},
    number = {3},
    pages = {868--900},
    volume = {12},
    doi = {10.1137/23M1581546}
}

@article{Knothe_to_Brenier_2010,
    title = {{From Knothe's transport to Brenier's map and a continuation method for optimal transport}},
    year = {2009},
    journal = {SIAM Journal on Mathematical Analysis},
    author = {Carlier, Guillaume and Galichon, Alfred and Santambrogio, Filippo},
    number = {6},
    pages = {2554--2576},
    volume = {41},
    doi = {10.1137/080740647},
    issn = {00361410},
    arxivId = {0810.4153}
}

@article{akbari2023learningcausalgraphsmonotone,
  title={Learning causal graphs via monotone triangular transport maps},
  author={Akbari, Sina and Ganassali, Luca and Kiyavash, Negar},
  journal={arXiv preprint arXiv:2305.18210},
  year={2023}
}

@inproceedings{pmlr-v151-irons22a,
    title = {Triangular Flows for Generative Modeling: Statistical Consistency, Smoothness Classes, and Fast Rates},
    year = {2022},
    booktitle = {Proceedings of The 25th International Conference on Artificial Intelligence and Statistics},
    author = {Irons, Nicholas J. and Scetbon, Meyer and Pal, Soumik and Harchaoui, Zaid},
    pages = {10161--10195},
    series = {Proceedings of Machine Learning Research},
    volume = {151},
}

@inproceedings{LeProvost2021AFlows,
    title = {{A low-rank nonlinear ensemble filter for vortex models of aerodynamic flows}},
    year = {2021},
    booktitle = {AIAA Scitech 2021 Forum},
    author = {Le Provost, Mathieu and Baptista, Ricardo and Marzouk, Youssef and Eldredge, Jeff D.},
    isbn = {9781624106095},
    doi = {10.2514/6.2021-1937}
}

@article{Zhou2011AnFiltering,
    title = {{An approach to handling non-Gaussianity of parameters and state variables in ensemble Kalman filtering}},
    year = {2011},
    journal = {Advances in Water Resources},
    author = {Zhou, Haiyan and G{\'{o}}mez-Hern{\'{a}}ndez, J. Jaime and Hendricks Franssen, Harrie Jan and Li, Liangping},
    number = {7},
    pages = {844--864},
    volume = {34},
    isbn = {0309-1708},
    doi = {10.1016/j.advwatres.2011.04.014},
    issn = {03091708}
}

@article{Amestoy1996AnAlgorithm,
    title = {{An approximate minimum degree ordering algorithm}},
    year = {1996},
    journal = {SIAM Journal on Matrix Analysis and Applications},
    author = {Amestoy, Patrick R. and Davis, Timothy A. and Duff, Iain S.},
    doi = {10.1137/S0895479894278952},
    issn = {08954798}
}

@book{Gelman2013BayesianAnalysis,
    title = {{{Bayesian} Data Analysis}},
    year = {2013},
    author = {Gelman, Andrew and Carlin, John B. and Stern, Hal S. and Dunson, David B. and Vehtari, Aki and Rubin, Donald B.},
    pages = {677},
    publisher = {CRC Press},
    isbn = {978-1-4398-4095-5}
}

@article{ElMoselhy2012BayesianMaps,
    title = {{{Bayesian} inference with optimal maps}},
    year = {2012},
    journal = {Journal of Computational Physics},
    author = {El Moselhy, Tarek A. and Marzouk, Youssef},
    doi = {10.1016/j.jcp.2012.07.022},
    issn = {10902716},
}

@inproceedings{VanDenOord2016ConditionalDecoders,
    title = {{Conditional image generation with PixelCNN decoders}},
    year = {2016},
    booktitle = {Advances in Neural Information Processing Systems},
    author = {van den Oord, Aäron and Kalchbrenner, Nal and Vinyals, Oriol and Espeholt, Lasse and Graves, Alex and Kavukcuoglu, Koray},
    issn = {10495258},
    arxivId = {1606.05328}
}

@article{Knothe1957ContributionsBodies,
    title = {{Contributions to the theory of convex bodies}},
    year = {1957},
    journal = {Michigan Mathematical Journal},
    author = {Knothe, Herbert},
    number = {1},
    pages = {39--52},
    volume = {4}
}

@article{Spantini2022CouplingFiltering,
    title = {Coupling techniques for nonlinear ensemble filtering},
    year = {2022},
    journal = {SIAM Review},
    author = {Spantini, Alessio and Baptista, Ricardo and Marzouk, Youssef},
    number = {4},
    volume = {64},
    doi = {10.1137/20M1312204},
    issn = {23318422},
    arxivId = {1907.00389v2},
}

@article{Martinez-Sanchez2024DecomposingComponents,
    title = {{Decomposing causality into its synergistic, unique, and redundant components}},
    year = {2024},
    journal = {Nature Communications},
    author = {Martinez-Sanchez, Alvaro and Arranz, Gonzalo and Lozano-Duran, Adrian},
    number = {1},
    pages = {9296},
    volume = {15},
    doi = {10.1038/s41467-024-53373-4},
    issn = {2041-1723}
}

@article{Lopez-Marrero2024DensitySciences,
    title = {{Density estimation via measure transport: Outlook for applications in the biological sciences}},
    year = {2024},
    journal = {Statistical Analysis and Data Mining: An ASA Data Science Journal},
    author = {L{\'{o}}pez-Marrero, Vanessa and Johnstone, Patrick R. and Park, Gilchan and Luo, Xihaier},
    number = {3},
    month = {6},
    pages = {e11687},
    volume = {17},
    publisher = {John Wiley {\&} Sons, Ltd},
    doi = {10.1002/sam.11687},
    issn = {1932-1864}
}

@article{Morrison2022DiagonalMatrices,
    title = {{Diagonal nonlinear transformations preserve structure in covariance and precision matrices}},
    year = {2022},
    journal = {Journal of Multivariate Analysis},
    author = {Morrison, Rebecca and Baptista, Ricardo and Basor, Estelle},
    doi = {10.1016/j.jmva.2022.104983},
    issn = {10957243}
}

@article{Ramgraber2023EnsembleFramework,
    title = {{Ensemble transport smoothing. Part I: Unified framework}},
    year = {2023},
    journal = {Journal of Computational Physics: X},
    author = {Ramgraber, Maximilian and Baptista, Ricardo and McLaughlin, Dennis and Marzouk, Youssef},
    month = {11},
    pages = {100134},
    volume = {17},
    doi = {10.1016/j.jcpx.2023.100134},
    issn = {25900552}
}

@article{Ramgraber2023EnsembleUpdates,
    title = {{Ensemble transport smoothing. Part II: Nonlinear updates}},
    year = {2023},
    journal = {Journal of Computational Physics: X},
    author = {Ramgraber, Maximilian and Baptista, Ricardo and McLaughlin, Dennis and Marzouk, Youssef},
    month = {11},
    pages = {100133},
    volume = {17},
    doi = {10.1016/j.jcpx.2023.100133},
    issn = {25900552}
}

@inproceedings{Clevert2016FastELUs,
    title = {{Fast and accurate deep network learning by exponential linear units (ELUs)}},
    year = {2016},
    booktitle = {4th International Conference on Learning Representations, ICLR 2016 - Conference Track Proceedings},
    author = {Clevert, Djork Arné and Unterthiner, Thomas and Hochreiter, Sepp},
    arxivId = {1511.07289}
}

@article{Spantini2018InferenceCouplings,
  title={Inference via low-dimensional couplings},
  author={Spantini, Alessio and Bigoni, Daniele and Marzouk, Youssef},
  journal={Journal of Machine Learning Research},
  volume={19},
  number={66},
  pages={1--71},
  year={2018}
}

@article{Cui2014Likelihood-informedProblems,
    title = {{Likelihood-informed dimension reduction for nonlinear inverse problems}},
    year = {2014},
    journal = {Inverse Problems},
    author = {Cui, Tiangang and Martin, James and Marzouk, Youssef and Solonen, Antti and Spantini, Alessio},
    doi = {10.1088/0266-5611/30/11/114015},
    issn = {13616420}
}

@article{Kobyzev2021NormalizingMethods,
  title={Normalizing flows: An introduction and review of current methods},
  author={Kobyzev, Ivan and Prince, Simon JD and Brubaker, Marcus A.},
  journal={IEEE transactions on pattern analysis and machine intelligence},
  volume={43},
  number={11},
  pages={3964--3979},
  year={2020},
  publisher={IEEE}
}

@book{Trefethen2022NumericalEdition,
    title = {{Numerical Linear Algebra, Twenty-fifth Anniversary Edition}},
    year = {2022},
    author = {Trefethen, Lloyd N. and Bau, David},
    month = {1},
    publisher = {Society for Industrial and Applied Mathematics},
    address = {Philadelphia, PA},
    isbn = {978-1-61197-715-8},
    doi = {10.1137/1.9781611977165}
}

@article{Ernst2012OnExpansions,
    title = {{On the convergence of generalized polynomial chaos expansions}},
    year = {2012},
    journal = {ESAIM: Mathematical Modelling and Numerical Analysis},
    author = {Ernst, Oliver G. and Mugler, Antje and Starkloff, Hans Jörg and Ullmann, Elisabeth},
    doi = {10.1051/m2an/2011045},
    issn = {28047214}
}

@article{Baptista2023OnMaps,
    title = {On the Representation and Learning of Monotone Triangular Transport Maps},
    year = {2023},
    journal = {Foundations of Computational Mathematics},
    author = {Baptista, Ricardo and Marzouk, Youssef and Zahm, Olivier},
    month = {11},
    doi = {10.1007/s10208-023-09630-x},
    issn = {1615-3375}
}

@article{Huan2024OptimalComputations,
    title = {{Optimal experimental design: Formulations and computations}},
    year = {2024},
    journal = {Acta Numerica},
    author = {Huan, Xun and Jagalur, Jayanth and Marzouk, Youssef},
    edition = {2024/09/04},
    pages = {715--840},
    volume = {33},
    publisher = {Cambridge University Press},
    doi = {DOI: 10.1017/S0962492924000023},
    issn = {0962-4929}
}

@book{Santambrogio2015OptimalMathematicians,
    title = {Optimal Transport for Applied Mathematicians},
    year = {2015},
    author = {Santambrogio, Filippo},
    editor = {Bezis, Haim},
    pages = {376},
    publisher = {Birkh{\"{a}}user},
    address = {Orsay},
    isbn = {978-3-319-20827-5},
    doi = {10.1007/978-3-319-20828-2}
}

@book{Villani2007OptimalNew,
  title={Optimal transport: old and new},
  author={Villani, Cédric and others},
  volume={338},
  year={2009},
  publisher={Springer}
}

@book{Szego1939OrthogonalPolynomials,
    title = {{Orthogonal Polynomials}},
    year = {1939},
    author = {Szeg{\H{o}}, Gábor},
    publisher = {American Mathematical Soc.}
}

@article{Schoniger2012ParameterTomography,
    title = {{Parameter estimation by ensemble Kalman filters with transformed data: Approach and application to hydraulic tomography}},
    year = {2012},
    journal = {Water Resources Research},
    author = {Sch{\"{o}}niger, Anneli and Nowak, Wolfgang and Hendricks Franssen, Harrie Jan},
    number = {4},
    volume = {48},
    isbn = {1944-7973},
    doi = {10.1029/2011WR010462},
    issn = {00431397}
}

@phdthesis{Baptista2022ProbabilisticTransport,
    title = {{Probabilistic modeling and Bayesian inference via triangular transport}},
    year = {2022},
    author = {Baptista, Ricardo Miguel},
    school = {Massachusetts Institute of Technology}
}

@article{Cui2025Quasi-MonteInterpolation,
    title = {{Quasi-Monte Carlo methods for mixture distributions and approximated distributions via piecewise linear interpolation}},
    year = {2025},
    journal = {Advances in Computational Mathematics},
    author = {Cui, Tiangang and Dick, Josef and Pillichshammer, Friedrich},
    number = {1},
    pages = {10},
    volume = {51},
    doi = {10.1007/s10444-025-10223-1},
    issn = {1572-9044}
}

@inproceedings{Cuthill1969ReducingMatrices,
    title = {{Reducing the bandwidth of sparse symmetric matrices}},
    year = {1969},
    booktitle = {Proceedings of the 1969 24th National Conference, ACM 1969},
    author = {Cuthill, Elizabeth H. and McKee, James},
    doi = {10.1145/800195.805928}
}

@article{Rosenblatt1952RemarksTransformation,
    title = {Remarks on a Multivariate Transformation},
    year = {1952},
    journal = {The Annals of Mathematical Statistics},
    author = {Rosenblatt, Murray},
    doi = {10.1214/aoms/1177729394},
    issn = {0003-4851}
}

@article{Liu2009TheGraphs,
    title = {{The nonparanormal: Semiparametric estimation of high dimensional undirected graphs}},
    year = {2009},
    journal = {Journal of Machine Learning Research},
    author = {Liu, Han and Lafferty, John and Wasserman, Larry},
    doi = {10.1184/r1/6610712},
    issn = {15324435},
    arxivId = {0903.0649}
}


%% file: references_daniel.bib
@article{zanger2024sequential,
  title={Sequential transport maps using {SoS} density estimation and $\alpha$-divergences},
  author={Zanger, Benjamin and Zahm, Olivier and Cui, Tiangang and Schreiber, Martin},
  journal={arXiv preprint arXiv:2402.17943},
  year={2024}
  }

@article{dolgov2020approximation,
  title={Approximation and sampling of multivariate probability distributions in the tensor train decomposition},
  author={Dolgov, Sergey and Anaya-Izquierdo, Karim and Fox, Colin and Scheichl, Robert},
  journal={Statistics and Computing},
  volume={30},
  pages={603--625},
  year={2020},
  publisher={Springer}
  }

@article{cui2022deep,
  title={Deep composition of Tensor-Trains using squared inverse {Rosenblatt} transports},
  author={Cui, Tiangang and Dolgov, Sergey},
  journal={Foundations of Computational Mathematics},
  volume={22},
  number={6},
  pages={1863--1922},
  year={2022},
  publisher={Springer}
}

@article{peyre2019computational,
  title={Computational optimal transport: With applications to data science},
  author={Peyr{\'e}, Gabriel and Cuturi, Marco and others},
  journal={Foundations and Trends{\textregistered} in Machine Learning},
  volume={11},
  number={5-6},
  pages={355--607},
  year={2019},
  publisher={Now Publishers, Inc.}
}

@article{tabak2021data,
  title={Data driven conditional optimal transport},
  author={Tabak, Esteban G and Trigila, Giulio and Zhao, Wenjun},
  journal={Machine Learning},
  volume={110},
  pages={3135--3155},
  year={2021},
  publisher={Springer}
}

@article{wang2023efficient,
  title={Efficient neural network approaches for conditional optimal transport with applications in {Bayesian} inference},
  author={Wang, Zheyu Oliver and Baptista, Ricardo and Marzouk, Youssef and Ruthotto, Lars and Verma, Deepanshu},
  journal={SIAM Journal on Scientific Computing},
  volume={47},
  number={4},
  pages={C979--C1005},
  year={2025},
  publisher={SIAM}
}

@inproceedings{makkuva2020optimal,
  title={Optimal transport mapping via input convex neural networks},
  author={Makkuva, Ashok and Taghvaei, Amirhossein and Oh, Sewoong and Lee, Jason},
  booktitle={International Conference on Machine Learning},
  pages={6672--6681},
  year={2020},
  organization={PMLR}
}

@inproceedings{
  song2020score,
  title={Score-Based Generative Modeling through Stochastic Differential Equations},
  author={Yang Song and Jascha Sohl-Dickstein and Diederik P Kingma and Abhishek Kumar and Stefano Ermon and Ben Poole},
  booktitle={International Conference on Learning Representations},
  year={2021},
}

@article{albergo2023stochastic,
  title={Stochastic interpolants: A unifying framework for flows and diffusions},
  author={Albergo, Michael and Boffi, Nicholas M and Vanden-Eijnden, Eric},
  journal={Journal of Machine Learning Research},
  volume={26},
  number={209},
  pages={1--80},
  year={2025}
}

@article{papamakarios2021normalizing,
  title={Normalizing flows for probabilistic modeling and inference},
  author={Papamakarios, George and Nalisnick, Eric and Rezende, Danilo Jimenez and Mohamed, Shakir and Lakshminarayanan, Balaji},
  journal={Journal of Machine Learning Research},
  volume={22},
  number={57},
  pages={1--64},
  year={2021}
}

@article{wehenkel2019unconstrained,
  title={Unconstrained monotonic neural networks},
  author={Wehenkel, Antoine and Louppe, Gilles},
  journal={Advances in Neural Information Processing Systems},
  volume={32},
  year={2019}
}

@article{bond2021deep,
  title={Deep generative modelling: A comparative review of {VAEs}, {GANs}, normalizing flows, energy-based and autoregressive models},
  author={Bond-Taylor, Sam and Leach, Adam and Long, Yang and Willcocks, Chris G},
  journal={IEEE Transactions on Pattern Analysis and Machine Intelligence},
  volume={44},
  number={11},
  pages={7327--7347},
  year={2021},
  publisher={IEEE}
}

@article{zech2022sparseI,
  title={Sparse approximation of triangular transports, part {I}: The finite-dimensional case},
  author={Zech, Jakob and Marzouk, Youssef},
  journal={Constructive Approximation},
  volume={55},
  number={3},
  pages={919--986},
  year={2022},
  publisher={Springer}
}

@article{zech2022sparseII,
  title={Sparse approximation of triangular transports, part {II}: The infinite-dimensional case},
  author={Zech, Jakob and Marzouk, Youssef},
  journal={Constructive Approximation},
  volume={55},
  number={3},
  pages={987--1036},
  year={2022},
  publisher={Springer}
}

@article{wang2022minimax,
  title={On minimax density estimation via measure transport},
  author={Wang, Sven and Marzouk, Youssef},
  journal={arXiv preprint arXiv:2207.10231},
  year={2022}
}

@article{marzouk2024distribution,
  title={Distribution learning via neural differential equations: A nonparametric statistical perspective},
  author={Marzouk, Youssef and Ren, Zhi Robert and Wang, Sven and Zech, Jakob},
  journal={Journal of Machine Learning Research},
  volume={25},
  number={232},
  pages={1--61},
  year={2024}
}

@article{baptista2024learning,
  author  = {Ricardo Baptista and Rebecca Morrison and Olivier Zahm and Youssef Marzouk},
  title   = {Learning Non-{G}aussian Graphical Models via {H}essian Scores and Triangular Transport},
  journal = {Journal of Machine Learning Research},
  year    = {2024},
  volume  = {25},
  number  = {85},
  pages   = {1--46},
}

@article{schaefer2021sparse,
  title={Sparse {C}holesky Factorization by {Kullback--Leibler} Minimization},
  author={Sch\"{a}fer, Florian and Katzfuss, Matthias and Owhadi, Houman},
  journal={SIAM Journal on Scientific Computing},
  volume={43},
  number={3},
  pages={A2019--A2046},
  year={2021},
  publisher={SIAM}
}

@book{LeMaitre2010SpectralUQ,
  title={Spectral methods for uncertainty quantification: With applications to computational fluid dynamics},
  author={Le Ma{\^\i}tre, Olivier and Knio, Omar M},
  year={2010},
  publisher={Springer Science \& Business Media}
}

@article{rubio2023transport,
  title={A transport approach to sequential simulation-based inference},
  author={Rubio, Paul-Baptiste and Marzouk, Youssef and Parno, Matthew},
  journal={arXiv preprint arXiv:2308.13940},
  year={2023}
}

@article{brennan2020greedy,
  title={Greedy inference with structure-exploiting lazy maps},
  author={Brennan, Michael and Bigoni, Daniele and Zahm, Olivier and Spantini, Alessio and Marzouk, Youssef},
  journal={Advances in Neural Information Processing Systems},
  volume={33},
  pages={8330--8342},
  year={2020}
}

@incollection{Marzouk2017SamplingIntroductionPUB,
    title = {{Sampling via measure transport: An introduction}},
    year = {2017},
    booktitle = {Handbook of Uncertainty Quantification},
    author = {Marzouk, Youssef and Moselhy, Tarek and Parno, Matthew and Spantini, Alessio},
    publisher = {Springer, Cham},
    isbn = {9783319123851},
    doi = {10.1007/978-3-319-12385-1_23}
}

@article{zhao2024tensor,
  title={Tensor-train methods for sequential state and parameter learning in state-space models},
  author={Zhao, Yiran and Cui, Tiangang},
  journal={Journal of Machine Learning Research},
  volume={25},
  number={244},
  pages={1--51},
  year={2024}
}

@article{katzfuss2024scalable,
  title={Scalable {Bayesian} transport maps for high-dimensional non-{Gaussian} spatial fields},
  author={Katzfuss, Matthias and Sch{\"a}fer, Florian},
  journal={Journal of the American Statistical Association},
  volume={119},
  number={546},
  pages={1409--1423},
  year={2024},
  publisher={Taylor \& Francis}
}

@article{richter2020vargrad,
  title={{VarGrad}: a low-variance gradient estimator for variational inference},
  author={Richter, Lorenz and Boustati, Ayman and N{\"u}sken, Nikolas and Ruiz, Francisco and Akyildiz, Omer Deniz},
  journal={Advances in Neural Information Processing Systems},
  volume={33},
  pages={13481--13492},
  year={2020}
}

@inproceedings{rezende2015variational,
  title={Variational inference with normalizing flows},
  author={Rezende, Danilo and Mohamed, Shakir},
  booktitle={International Conference on Machine Learning},
  pages={1530--1538},
  year={2015},
  organization={PMLR}
}

@article{klebanov2023quadrature,
  title={Transporting higher-order quadrature rules – {Quasi}-{Monte} {Carlo} points and sparse grids for mixture distributions},
  author={Klebanov, Ilja and Sullivan, Timothy John},
  journal={Statistics and Computing},
  volume={36},
  number={1},
  pages={46},
  year={2026},
  publisher={Springer}
}

@inproceedings{
pooladian2025conditional,
title={Conditional simulation via entropic optimal transport: Toward non-parametric estimation of conditional {Brenier} maps},
author={Aram-Alexandre Pooladian and Ricardo Baptista and Michael Brennan and Youssef Marzouk and Jonathan Niles-Weed},
booktitle={The 28th International Conference on Artificial Intelligence and Statistics},
year={2025},
}

@article{cao2024lazydino,
  title={{LazyDINO}: Fast, scalable, and efficiently amortized {Bayesian} inversion via structure-exploiting and surrogate-driven measure transport},
  author={Cao, Lianghao and Chen, Joshua and Brennan, Michael and O'Leary-Roseberry, Thomas and Marzouk, Youssef and Ghattas, Omar},
  journal={Journal of Machine Learning Research},
  volume={27},
  number={24},
  pages={1--71},
  year={2026}
}

@article{cui2025subspace,
  title={Subspace accelerated measure transport methods for fast and scalable sequential experimental design, with application to photoacoustic imaging},
  author={Cui, Tiangang and Koval, Karina and Herzog, Roland and Scheichl, Robert},
  journal={arXiv preprint arXiv:2502.20086},
  year={2025}
}

@article{lunde2025ensemble,
  title={An Ensemble Information Filter: Retrieving {M}arkov-information from the {SPDE} discretisation},
  author={Lunde, Berent Å. S.},
  journal={arXiv preprint arXiv:2501.09016},
  year={2025}
}

@inproceedings{liaw2025learning,
  title={Learning local neighborhoods of non-{Gaussian} graphical models},
  author={Liaw, Sarah and Morrison, Rebecca and Marzouk, Youssef and Baptista, Ricardo},
  booktitle={Proceedings of the AAAI Conference on Artificial Intelligence},
  volume={39},
  number={18},
  pages={18711--18718},
  year={2025}
}

@article{baptista2024bayesian,
  title={Bayesian model calibration for block copolymer self-assembly: Likelihood-free inference and expected information gain computation via measure transport},
  author={Baptista, Ricardo and Cao, Lianghao and Chen, Joshua and Ghattas, Omar and Li, Fengyi and Marzouk, Youssef and Oden, J. Tinsley},
  journal={Journal of Computational Physics},
  volume={503},
  pages={112844},
  year={2024},
  publisher={Elsevier}
}

@article{carlierVectorQuantileRegression2016,
  title = {Vector Quantile Regression: {{An}} Optimal Transport Approach},
  shorttitle = {Vector Quantile Regression},
  author = {Carlier, Guillaume and Chernozhukov, Victor and Galichon, Alfred},
  year = {2016},
  month = jun,
  journal = {The Annals of Statistics},
  volume = {44},
  number = {3},
  issn = {0090-5364},
  doi = {10.1214/15-AOS1401},
}

@article{baptista2024approximation,
  title={An approximation theory framework for measure-transport sampling algorithms},
  author={Baptista, Ricardo and Hosseini, Bamdad and Kovachki, Nikola and Marzouk, Youssef and Sagiv, Amir},
  journal={Mathematics of Computation},
  year={2024}
}

@article{westermann2023measure,
  title = {Measure Transport via Polynomial Density Surrogates},
  author = {Westermann, Josephine and Zech, Jakob},
  year = 2025,
  journal = {Foundations of Data Science},
  volume = {7},
  number = {3},
  pages = {838--887},
  issn = {2639-8001},
  doi = {10.3934/fods.2025001},
  urldate = {2026-09-28}
}

@inproceedings{jaini2019sum,
  title={Sum-of-squares polynomial flow},
  author={Jaini, Priyank and Selby, Kira A. and Yu, Yaoliang},
  booktitle={International Conference on Machine Learning},
  pages={3009--3018},
  year={2019},
  organization={PMLR}
}

@inproceedings{
lipman2023flow,
title={Flow Matching for Generative Modeling},
author={Yaron Lipman and Ricky T. Q. Chen and Heli Ben-Hamu and Maximilian Nickel and Matthew Le},
booktitle={The Eleventh International Conference on Learning Representations },
year={2023},
}

@inproceedings{
liu2023flow,
title={Flow Straight and Fast: Learning to Generate and Transfer Data with Rectified Flow},
author={Xingchao Liu and Chengyue Gong and Qiang Liu},
booktitle={The Eleventh International Conference on Learning Representations },
year={2023},
}

@article{heng2021gibbs,
  title={{Gibbs} flow for approximate transport with applications to {Bayesian} computation},
  author={Heng, Jeremy and Doucet, Arnaud and Pokern, Yvo},
  journal={Journal of the Royal Statistical Society Series B: Statistical Methodology},
  volume={83},
  number={1},
  pages={156--187},
  year={2021},
  publisher={Oxford University Press}
}

@article{konishi1996generalised,
  title={Generalised information criteria in model selection},
  author={Konishi, Sadanori and Kitagawa, Genshiro},
  journal={Biometrika},
  volume={83},
  number={4},
  pages={875--890},
  year={1996},
  publisher={Oxford University Press}
}

@article{pooladian2021entropic,
  title={Entropic estimation of optimal transport maps},
  author={Pooladian, Aram-Alexandre and Niles-Weed, Jonathan},
  journal={arXiv preprint arXiv:2109.12004},
  year={2021}
}

@article{liu2024transport,
  title={Transport quasi-{Monte} {Carlo}},
  author={Liu, Sifan},
  journal={SIAM Journal on Scientific Computing},
  volume={48},
  number={4},
  pages={C658--C683},
  year={2026},
  publisher={SIAM}
}

@article{drton2017structure,
  title={Structure learning in graphical modeling},
  author={Drton, Mathias and Maathuis, Marloes H.},
  journal={Annual Review of Statistics and Its Application},
  volume={4},
  number={1},
  pages={365--393},
  year={2017},
  publisher={Annual Reviews}
}

@article{cools1997constructing,
  title={Constructing cubature formulae: {T}he science behind the art},
  author={Cools, Ronald},
  journal={Acta Numerica},
  volume={6},
  pages={1--54},
  year={1997},
  publisher={Cambridge University Press}
}

@article{baptista2022gradient,
  title={Gradient-based data and parameter dimension reduction for {Bayesian} models: an information theoretic perspective},
  author={Baptista, Ricardo and Marzouk, Youssef and Zahm, Olivier},
  journal={arXiv preprint arXiv:2207.08670},
  year={2022}
}

@book{constantine2015active,
  title={Active subspaces: Emerging ideas for dimension reduction in parameter studies},
  author={Constantine, Paul},
  year={2015},
  publisher={SIAM}
}

@article{fukumizu2007statistical,
  title={Statistical consistency of kernel canonical correlation analysis.},
  author={Fukumizu, Kenji and Bach, Francis R. and Gretton, Arthur},
  journal={Journal of Machine Learning Research},
  volume={8},
  number={2},
  year={2007}
}

@article{grashorn2024transport,
  title={Transport Map Coupling Filter for State-Parameter Estimation},
  author={Grashorn, Jan and Broggi, Matteo and Chamoin, Ludovic and Beer, Michael},
  journal={arXiv preprint arXiv:2407.02198},
  year={2024}
}

@INPROCEEDINGS{grange2024DistributedFiltering,
  author={Grange, Daniel and Baptista, Ricardo and Taghvaei, Amirhossein and Tannenbaum, Allen and Phillips, Sean},
  booktitle={2024 American Control Conference (ACC)}, 
  title={Distributed Nonlinear Filtering using Triangular Transport Maps}, 
  year={2024},
  volume={},
  number={},
  pages={3062-3067},
  doi={10.23919/ACC60939.2024.10645057}}

@article{parno2016multiscale,
  title={A multiscale strategy for {B}ayesian inference using transport maps},
  author={Parno, Matthew and Moselhy, Tarek and Marzouk, Youssef},
  journal={SIAM/ASA Journal on Uncertainty Quantification},
  volume={4},
  number={1},
  pages={1160--1190},
  year={2016},
  publisher={SIAM}
}

@article{hardle1989investigating,
  title={Investigating smooth multiple regression by the method of average derivatives},
  author={H{\"a}rdle, Wolfgang and Stoker, Thomas M.},
  journal={Journal of the American Statistical Association},
  volume={84},
  number={408},
  pages={986--995},
  year={1989},
  publisher={Taylor \& Francis}
}

@article{johnson2007recursive,
  title={A recursive model-reduction method for approximate inference in {G}aussian {M}arkov random fields},
  author={Johnson, Jason K. and Willsky, Alan S.},
  journal={IEEE Transactions on Image Processing},
  volume={17},
  number={1},
  pages={70--83},
  year={2007},
  publisher={IEEE}
}

@inproceedings{jog2015model,
  title={On model misspecification and {KL} separation for {G}aussian graphical models},
  author={Jog, Varun and Loh, Po-Ling},
  booktitle={2015 IEEE International Symposium on Information Theory (ISIT)},
  pages={1174--1178},
  year={2015},
  organization={IEEE}
}

@article{al2024non,
  title={Non-asymptotic analysis of ensemble {K}alman updates: effective dimension and localization},
  author={Al-Ghattas, Omar and Sanz-Alonso, Daniel},
  journal={Information and Inference: A Journal of the IMA},
  volume={13},
  number={1},
  pages={iaad043},
  year={2024},
  publisher={Oxford University Press}
}

@article{tong2023localized,
  title={Localized ensemble {K}alman inversion},
  author={Tong, Xin T. and Morzfeld, Matthias},
  journal={Inverse Problems},
  volume={39},
  number={6},
  pages={064002},
  year={2023},
  publisher={IOP Publishing}
}

@article{gottwald2024localized,
  title={Localized {Schr}{\"o}dinger bridge sampler},
  author={Gottwald, Georg A and Reich, Sebastian},
  journal={Journal of Computational Physics},
  pages={114583},
  year={2025},
  publisher={Elsevier}
}

@article{zeng2023bounded,
  title={Bounded {KRnet} and its applications to density estimation and approximation},
  author={Zeng, Li and Wan, Xiaoliang and Zhou, Tao},
  journal={SIAM Journal on Scientific Computing},
  volume={47},
  number={6},
  pages={C1294--C1318},
  year={2025},
  publisher={SIAM}
}

@inproceedings{
xi2023triangular,
title={Triangular Monotonic Generative Models Can Perform Causal Discovery},
author={Quanhan Xi and Sebastian Gonzalez and Benjamin Bloem-Reddy},
booktitle={Causal Representation Learning Workshop at NeurIPS 2023},
year={2023},
}

@article{backhoff2017causal,
  title={Causal transport in discrete time and applications},
  author={Backhoff, Julio and Beiglböck, Mathias and Lin, Yiqing and Zalashko, Anastasiia},
  journal={SIAM Journal on Optimization},
  volume={27},
  number={4},
  pages={2528--2562},
  year={2017},
  publisher={SIAM}
}

@article{parno2022mpartJOSS,
  title={{MParT}: Monotone parameterization toolkit},
  author={Parno, Matthew and Rubio, Paul-Baptiste and Sharp, Daniel and Brennan, Michael and Baptista, Ricardo and Bonart, Henning and Marzouk, Youssef},
  journal={Journal of Open Source Software},
  volume={7},
  number={80},
  pages={4843},
  year={2022}
}

@article{zahm2022certified,
  title={Certified dimension reduction in nonlinear {Bayesian} inverse problems},
  author={Zahm, Olivier and Cui, Tiangang and Law, Kody and Spantini, Alessio and Marzouk, Youssef},
  journal={Mathematics of Computation},
  volume={91},
  number={336},
  pages={1789--1835},
  year={2022}
}

@article{sharp2026samplingiteratedapproximationgradientfree,
  title={Sampling through iterated approximation: Gradient-free and multi-fidelity {Bayesian} inference via transport},
  author={Sharp, Daniel and van Bloemen Waanders, Bart and Marzouk, Youssef},
  journal={arXiv preprint arXiv:2603.12448},
  year={2026}
}

@article{lunde2026adaptive,
  title={Adaptive Nonlinear Data Assimilation through {P}-Spline Triangular Measure Transport},
  author={Lunde, Berent Å. S. and Ramgraber, Maximilian},
  journal={arXiv preprint arXiv:2603.19058},
  year={2026}
}

@InProceedings{zheng2020learning,
  title = 	 {Learning Sparse Nonparametric {DAGs}},
  author =       {Zheng, Xun and Dan, Chen and Aragam, Bryon and Ravikumar, Pradeep and Xing, Eric},
  booktitle = 	 {Proceedings of the Twenty Third International Conference on Artificial Intelligence and Statistics},
  pages = 	 {3414--3425},
  year = 	 {2020},
  editor = 	 {Chiappa, Silvia and Calandra, Roberto},
  volume = 	 {108},
  series = 	 {Proceedings of Machine Learning Research},
  month = 	 {26--28 Aug},
  publisher =    {PMLR},
}

@inproceedings{
bryutkinNeuralTriangularTransport2025,
title={Neural Triangular Transport Maps: A New Approach Towards Sampling in Lattice {QCD}},
author={Andrey Bryutkin and Youssef Marzouk},
booktitle={NeurIPS 2025 AI for Science Workshop},
year={2025},
}

@book{bogachevMeasureTheory2007,
  title = {Measure {{Theory}}},
  author = {Bogachev, Vladimir I.},
  year = 2007,
  publisher = {Springer Berlin Heidelberg},
  address = {Berlin, Heidelberg},
}

@article{kovalTractableOptimalExperimental2024,
  title = {Tractable Optimal Experimental Design Using Transport Maps},
  author = {Koval, Karina and Herzog, Roland and Scheichl, Robert},
  year = 2024,
  month = dec,
  journal = {Inverse Problems},
  volume = {40},
  number = {12},
  pages = {125002},
}

@book{figalliInvitationOptimalTransport2021,
    address = {Berlin, Germany},
    series = {{EMS} textbooks in mathematics},
    title = {An invitation to optimal transport, {Wasserstein} distances, and gradient flows},
    isbn = {978-3-98547-010-5},
    publisher = {EMS Press},
    author = {Figalli, Alessio and Glaudo, Federico},
    year = {2021},
}

@book{devroyeNonUniformRandomVariate1986,
    address = {New York, NY},
    title = {Non-Uniform Random Variate Generation},
    copyright = {http://www.springer.com/tdm},
    isbn = {978-1-4613-8645-2 978-1-4613-8643-8},
      doi = {10.1007/978-1-4613-8643-8},
    language = {en},
      publisher = {Springer New York},
    author = {Devroye, Luc},
    year = {1986},
}

@article{li2026JUQ,
author = {Li, Fengyi and Baptista, Ricardo and Marzouk, Youssef},
title = {Expected Information Gain Estimation via Density Approximations: Sample Allocation and Dimension Reduction},
journal = {SIAM/ASA Journal on Uncertainty Quantification},
volume = {14},
number = {3},
pages = {974-1007},
year = {2026},
doi = {10.1137/25M1740164},
}

@article{chen2025rotated,
  title={Rotated mean-field variational inference and iterative {G}aussianization},
  author={Chen, Yifan and Liu, Sifan},
  journal={arXiv preprint arXiv:2510.07732},
  year={2025}
}
